\documentclass[reprint,amsmath,amssymb,aps,prl,superscriptaddress]{revtex4-2}
\usepackage{amsmath}
\usepackage{graphicx}

\begin{document}
\def\rmd{\mathrm{d}}%
 
\def\rme{\mathrm{e}}%
 
\def\rmi{\mathrm{i}}%

\title{General theory for geometry-dependent non-Hermitian bands}
\author{Chenyang Wang}
\affiliation{State Key Laboratory of Low-Dimensional Quantum Physics, Department
of Physics, Tsinghua University, Beijing 100084, China}
\author{Jinghui Pi}
\affiliation{The Chinese University of Hong Kong Shenzhen Research Institute, Shenzhen 518057, China}
\author{Qinxin Liu}
\affiliation{State Key Laboratory of Low-Dimensional Quantum Physics, Department
of Physics, Tsinghua University, Beijing 100084, China}
\author{Yaohua Li}
\affiliation{State Key Laboratory of Low-Dimensional Quantum Physics, Department
of Physics, Tsinghua University, Beijing 100084, China}
\author{Yong-Chun Liu}
\email{ycliu@tsinghua.edu.cn}

\affiliation{State Key Laboratory of Low-Dimensional Quantum Physics, Department
of Physics, Tsinghua University, Beijing 100084, China}
\affiliation{Frontier Science Center for Quantum Information, Beijing 100084, China}
\begin{abstract}
In two- and higher-dimensional non-Hermitian lattices, systems can
exhibit geometry-dependent bands, where the spectrum and eigenstates
under open boundary conditions depend on the bulk geometry even in
the thermodynamic limit. Although geometry-dependent bands are widely
observed, the underlying mechanism for this phenomenon remains unclear.
In this work, we address this problem by establishing a higher-dimensional
non-Bloch band theory based on the concept of “strip generalized Brillouin
zones” (SGBZs), which describe the asymptotic behavior of non-Hermitian
bands when a lattice is extended sequentially along its linearly independent
axes. Within this framework, we demonstrate that geometry-dependent
bands arise from the incompatibility of SGBZs and, for the first time,
derive a general criterion for the geometry dependence of non-Hermitian
bands: non-zero area of the complex energy spectrum or the imaginary
momentum spectrum. Our work opens an avenue for future studies on
the interplay between geometric effects and non-Hermitian physics,
such as non-Hermitian band topology.
\end{abstract}
\maketitle
\textit{Introduction} --- The band structure of periodic lattices
is a cornerstone of modern condensed matter physics. In Hermitian
systems, the energy spectrum under open boundary conditions (OBCs)
is consistent with the Bloch bands in the thermodynamic limit. However,
non-Hermitian lattices can exhibit the non-Hermitian skin effect,
where the spectrum and eigenstates under OBC deviate from the Bloch
bands and Bloch wave functions \cite{leeAnomalousEdgeState2016,yaoEdgeStatesTopological2018}.
The deviation of the OBC bands from the Bloch bands has been observed
in various non-Hermitian classical \cite{helbigGeneralizedBulkBoundary2020,hofmannReciprocalSkinEffect2020,weidemannTopologicalFunnelingLight2020,wangGeneratingArbitraryTopological2021,zhangObservationHigherorderNonHermitian2021,zouObservationHybridHigherorder2021,liuComplexSkinModes2022,guTransientNonHermitianSkin2022,wangExtendedStateLocalized2022,shangExperimentalIdentificationSecondOrder2022,liuExperimentalObservationNonHermitian2023,wuEvidencingNonBlochDynamics2023,zhangElectricalCircuitRealization2023,zhuHigherRankChirality2023,liuLocalizationChiralEdge2024,liObservationGaugeField2025}
and quantum \cite{xiaoNonHermitianBulkBoundary2020,wangDetectingNonBlochTopological2021,xiaoObservationNonBlochParityTime2021,zhangObservationNonHermitianTopology2021,linObservationNonHermitianTopological2022,linTopologicalPhaseTransitions2022,chuBroadColossalEdge2023,ochkanNonHermitianTopologyMultiterminal2024,zhaoTwodimensionalNonHermitianSkin2025}
systems.

\begin{figure}
\centering

\includegraphics{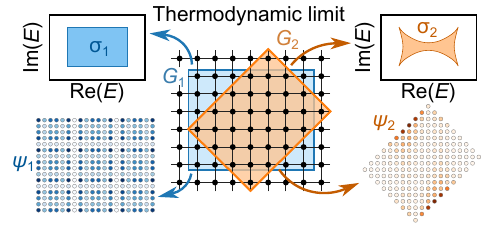}

\caption{Illustration of geometry-dependent non-Hermitian bands. For the same
system, different geometries ($G_{1}$ and $G_{2}$) yield distinct
spectra ($\sigma_{1}$ and $\sigma_{2}$) and eigenstates ($\psi_{1}$
and $\psi_{2}$), even in the thermodynamic limit.}
\label{fig:illustration-GD-bands} 
\end{figure}

For one-dimensional (1D) non-Hermitian systems, the thermodynamic
limit of OBC spectra and eigenstates is described by the non-Bloch
band theory using the generalized Brillouin zone (GBZ) \cite{yaoEdgeStatesTopological2018,yokomizoNonBlochBandTheory2019,yangNonHermitianBulkBoundaryCorrespondence2020},
which has been verified in numerous studies \cite{okumaTopologicalPhaseTransition2019,zhuPhotonicNonHermitianSkin2020,xiaoNonHermitianBulkBoundary2020,bartlettIlluminatingBulkboundaryCorrespondence2021,longhiNonHermitianTopologicalPhase2021,wangDetectingNonBlochTopological2021,xiaoObservationNonBlochParityTime2021}.
However, non-Hermitian bands in higher dimensions are not well understood
because of the geometry dependence of the energy bands \cite{zhangUniversalNonHermitianSkin2022}.
As schematically illustrated in Fig. \ref{fig:illustration-GD-bands},
for two- and higher-dimensional non-Hermitian lattices, different
geometries ($G_{1}$ and $G_{2}$) can result in different spectra
($\sigma_{1}$ and $\sigma_{2}$) and eigenstates ($\psi_{1}$ and
$\psi_{2}$) even in the thermodynamic limit. This remarkable effect
has been observed in various physical systems \cite{chengTruncationdependent$mathcalPT$Phase2022,fangGeometrydependentSkinEffects2022,wanObservationGeometrydependentSkin2023,zhouObservationGeometrydependentSkin2023,qinGeometrydependentSkinEffect2024},
yet its underlying mechanism is still not well understood, hindering
the development of a comprehensive band theory for higher-dimensional
non-Hermitian systems \cite{yaoNonHermitianChernBands2018,zhangDynamicalDegeneracySplitting2023,yokomizoNonBlochBandsTwodimensional2023,jiangDimensionalTransmutationNonHermiticity2023,wangAmoebaFormulationNonBloch2024,zhangEdgeTheoryNonHermitian2024,huTopologicalOriginNonHermitian2024,zhangAlgebraicNonHermitianSkin2025}.

To investigate the mechanism of geometry-dependent bands, we develop
a general formulation based on the strip generalized Brillouin zone
(SGBZ), which describes the energy bands when a non-Hermitian lattice
is extended sequentially along its linearly independent axes to infinity.
We demonstrate that the SGBZ of a non-Hermitian lattice can be dependent
of the sequence of axes, and the competition between incompatible
SGBZs results in geometry-dependent bands. Furthermore, through the
transformation of SGBZs, we derive a criterion that a non-Hermitian
system exhibits geometry-dependent bands if and only if its energy
spectrum or imaginary momentum spectrum has a non-zero area.

\begin{figure}
\centering

\includegraphics{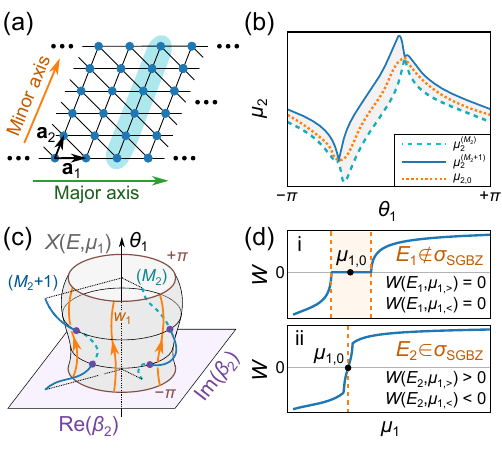}

\caption{Definition of the SGBZ. (a) Schematic diagram of the strip geometry
of a 2D non-Hermitian lattice, where the cyan region represents the
periodic unit of the strip. (b) Definition of the function $\mu_{2,0}$,
which is a periodic function between $\mu_{2}^{(M_{2})}$ and $\mu_{2}^{(M_{2}+1)}$.
(c) Definition of the base manifold $X(E,\mu_{1})$, where the blue
solid lines and cyan dashed lines represent the $M_{2}$-th and $(M_{2}+1)$-th
solutions of the eigenvalue equations, respectively, and the orange
solid curves represent the winding loops for $w_{1}(\theta_{2};E,\mu_{1})$.
(d) SGBZ bands and the strip winding number. When $W(E,\mu_{1})$
exhibits a plateau, the reference energy lies outside the SGBZ bands
(panel i). Otherwise, the reference energy belongs to the SGBZ bands
(panel ii).}
\label{fig:SGBZ} 
\end{figure}

\textit{SGBZ formulation for non-Hermitian bands} --- To obtain the
SGBZ, we first extend a lattice along a lattice vector $\mathbf{a}_{1}$
(named the ``major axis''), forming a strip geometry with finite
width, and then take the width to infinity. As illustrated in Fig.
\ref{fig:SGBZ}(a), the lattice confined in the strip geometry can
be viewed as a 1D lattice along $\mathbf{a}_{1}$, whose periodic
unit is a slice of sites along the other lattice vector $\mathbf{a}_{2}$
(named the ``minor axis''). Assume that the momentum-space Hamiltonian
of this 1D lattice is $\mathcal{H}_{L_{2}}(\rme^{\rmi k_{1}})$, where
$L_{2}$ is the width of the strip. When the length of the strip is
sufficiently large, the OBC spectrum tends to the GBZ bands defined
by $|\beta_{1}^{(M)}(E)|=|\beta_{1}^{(M+1)}(E)|$ \cite{yaoEdgeStatesTopological2018,yokomizoNonBlochBandTheory2019}.
Here, $\beta_{1}^{(j)}(E)$ is the $j$-th solution of the eigenvalue
equation $\det[E-\mathcal{H}_{L_{2}}(\beta_{1})]=0$, ordered by $|\beta_{1}^{(i)}|\leq|\beta_{1}^{(j)}|,\forall i<j$,
and $-M$ is the lowest degree of $\beta_{1}$ in $\det[E-\mathcal{H}_{L_{2}}(\beta_{1})]$
.

However, as $L_{2}\rightarrow\infty$, the 1D GBZ constraint $|\beta_{1}^{(M)}(E)|=|\beta_{1}^{(M+1)}(E)|$
becomes ill-defined due to the divergence of $M$. To address this
problem, we derive an equivalent formulation based on the ``strip
winding number'' $W(E,\mu_{1})$. Here, we summarize the main conclusion,
and the details are available in Sec. S1 of the Supplemental Material
(SM) \cite{supp-2SK}. Assuming the momentum-space Hamiltonian is
$h(\rme^{\rmi k_{1}},\rme^{\rmi k_{2}})$, where $k_{j}\equiv\theta_{j}-\rmi\mu_{j}\in\mathbb{C}$,
we consider the solutions $\beta_{2}^{(j)}(E,\beta_{1}),j=1,2,\dots,M_{2}+N_{2}$
of $\det[E-h(\beta_{1},\beta_{2})]=0$, where $-M_{2}$ and $N_{2}$
are the lowest and highest degrees of $\beta_{2}$ in $\det[E-h(\beta_{1},\beta_{2})]$,
respectively, and the solutions are ordered by $|\beta_{2}^{(j)}|\leq|\beta_{2}^{(k)}|,\forall j<k$.
As shown in Fig. \ref{fig:SGBZ}(b), for a given radius $|\beta_{1}|=\rme^{\mu_{1}}$,
we define the radius function $\mu_{2,0}(\theta_{1};E,\mu_{1})$ as,
\begin{equation}
\mu_{2}^{(M_{2})}\leq\mu_{2,0}\leq\mu_{2}^{(M_{2}+1)},\label{eq:def-radius-func}
\end{equation}
where $\mu_{2}^{(j)}(\theta_{1};E,\mu_{1})\equiv\ln|\beta_{2}^{(j)}(E,\rme^{\mu_{1}+\rmi\theta_{1}})|$
are the imaginary momentum components of the $j$-th solutions. We
require that $\mu_{2,0}$ is periodic in $\theta_{1}$, and the equality
in Eq. (\ref{eq:def-radius-func}) holds if and only if $\mu_{2}^{(M_{2})}(\theta_{1};E,\mu_{1})=\mu_{2,0}(\theta_{1};E,\mu_{1})=\mu_{2}^{(M_{2}+1)}(\theta_{1};E,\mu_{1})$.
Using the radius function, we define the base manifold $X(E,\mu_{1})$
as, 
\begin{align}
X(E,\mu_{1})\equiv & \left\{ \left(\beta_{1},\beta_{2}\right)\in\mathbb{C}^{2}\mid\beta_{1}=\rme^{\mu_{1}+\rmi\theta_{1}},\right.\nonumber \\
 & \left.\beta_{2}=\rme^{\mu_{2,0}(\theta_{1};E,\mu_{1})+\rmi\theta_{2}},\theta_{1},\theta_{2}\in\left[-\pi,\pi\right]\right\} .\label{eq:base-manifold}
\end{align}
As shown in Fig. \ref{fig:SGBZ}(c), in the three-dimensional space
defined by $\ln|\beta_{1}|=\mu_{1}$, the solutions $\beta_{2}^{(M_{2}+1)}$
(blue solid curves) lie outside $X(E,\mu_{1})$, while $\beta_{2}^{(M_{2})}$
(cyan dashed curves) are enclosed within $X(E,\mu_{1})$. On the base
manifold, we define $W(E,\mu_{1})$ as, 
\begin{equation}
W\left(E,\mu_{1}\right)\equiv\int_{-\pi}^{\pi}\frac{\rmd\theta_{2}}{2\pi}w_{1}\left(\theta_{2};E,\mu_{1}\right),
\end{equation}
where $w_{1}(\theta_{2};E,\mu_{1})$ is the winding number of $\det[E-h(\beta_{1},\beta_{2})]$
when $(\beta_{1},\beta_{2})$ traverses the loop defined by $\text{Arg}(\beta_{2})=\theta_{2}$
on the base manifold $X(E,\mu_{1})$, shown as the orange curves in
Fig. \ref{fig:SGBZ}(c). Using the strip winding number, the SGBZ
is defined as the point where the sign of $W(E,\mu_{1})$ changes.
Specifically, for a reference energy $E\in\mathbb{C}$ and a radius
$\mu_{1,0}\in\mathbb{R}$, if there exist $\mu_{1,<}\in(\mu_{1,0}-\epsilon,\mu_{1,0})$
and $\mu_{1,>}\in(\mu_{1,0},\mu_{1,0}+\epsilon)$ for every $\epsilon>0$,
such that, 
\begin{equation}
\begin{cases}
W\left(E,\mu_{1,<}\right)<0,\\
W\left(E,\mu_{1,>}\right)>0,
\end{cases}\label{eq:SGBZ-constraint}
\end{equation}
then, $E$ belongs to the SGBZ spectrum, and the solutions of $\det[E-h(\beta_{1},\beta_{2})]=0$
on the base manifold $X(E,\mu_{1,0})$ form the SGBZ. With above constructions,
it is verified that the energy bands defined on the SGBZ are consistent
with the GBZ bands of $\mathcal{H}_{L_{2}}(\rme^{\rmi k_{1}})$ as
$L_{2}\rightarrow\infty$ \cite{supp-2SK}. The SGBZ formulations
can also be extended to higher dimensions. For an $n$-dimensional
lattice, a strip is specified by a sequence of axes $(\mathbf{a}_{1},\mathbf{a}_{2},\dots,\mathbf{a}_{n})$,
and the strip winding numbers are defined recursively. A detailed
discussion is available in Sec. S1.D of the SM \cite{supp-2SK}.

It is proved that $W(E,\mu_{1})$ is a non-decreasing function of
$\mu_{1}$ (see Sec. S1.C of the SM \cite{supp-2SK}). Therefore,
as shown in the two panels of Fig. \ref{fig:SGBZ}(d), $W(E,\mu_{1})$
as a function of $\mu_{1}$ is either locally constant (panel i) or
increasing (panel ii) in the neighborhood of its zeros. In the former
case, as shown in panel i, $W(E,\mu_{1,<})<0$ and $W(E,\mu_{1,>})>0$
cannot hold simultaneously for any $\mu_{1,0}\in\mathbb{R}$. Thus,
the energy $E_{1}$ does not belong to the SGBZ spectrum $\sigma_{\text{SGBZ}}$.
Otherwise, as shown in panel ii, the constraint of Eq. (\ref{eq:SGBZ-constraint})
is satisfied at $\mu_{1,0}$, so that $E_{2}\in\sigma_{\text{SGBZ}}$.

\begin{figure*}
\centering

\includegraphics{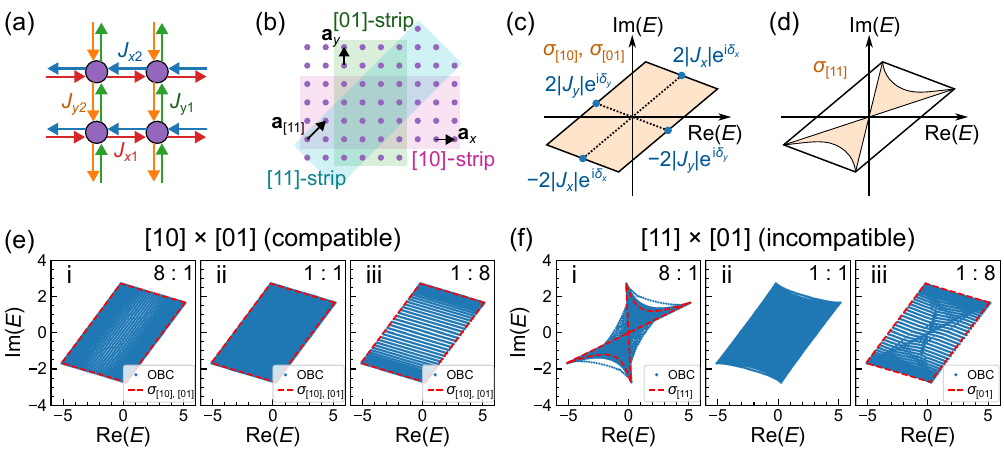}

\caption{Relation between SGBZs and geometry-dependent bands. (a) Illustration
of the 2D HN model. (b) Illustration of the {[}10{]}-strip, {[}01{]}-strip,
and {[}11{]}-strip, where the major axes are $\mathbf{a}_{x}=(1,0)$,
$\mathbf{a}_{y}=(0,1)$, and $\mathbf{a}_{[11]}=(1,1)$, respectively.
(c, d) Spectra of (c) $[10]$-SGBZ or $[01]$-SGBZ, and (d) $[11]$-SGBZ.
(e, f) Comparison between the SGBZ bands and the finite-size OBC spectra
in parallelogram regions with different aspect ratios, where (e) illustrates
the region with compatible SGBZs, and (f) illustrates the region with
incompatible SGBZs. The coupling coefficients are $J_{x1}=1+\protect\rmi$,
$J_{x2}=1.5+1.2\protect\rmi$, $J_{y1}=-1+\protect\rmi$, and $J_{y2}=-1.2-0.5\protect\rmi$.
In numerical calculations, the total number of sites is set to be
$12800$ (or the nearest integer to $12800$).}
\label{fig:OBC}
\end{figure*}

In the SGBZ formulation presented above, it is noted that the status
of $\beta_{1}$ and $\beta_{2}$ is not symmetric. This observation
implies that the SGBZs calculated under different major and minor
axes can be incompatible. In fact, the SGBZ is independent of the
minor axes (see Sec. S2 of the SM \cite{supp-2SK}), but depends on
the selection of the major axes. As an example, we consider the 2D
Hatano-Nelson (HN) model shown in Fig. \ref{fig:OBC}(a). The momentum-space
Hamiltonian of the model reads, 
\begin{equation}
h\left(\beta_{x},\beta_{y}\right)=J_{x1}\beta_{x}^{-1}+J_{x2}\beta_{x}+J_{y1}\beta_{y}^{-1}+J_{y2}\beta_{y}.\label{eq:HN}
\end{equation}
As illustrated in Fig. \ref{fig:OBC}(b), we consider three different
strips: the {[}10{]}-strip, {[}01{]}-strip, and {[}11{]}-strip, defined
by the major axes $\mathbf{a}_{x}=(1,0)$, $\mathbf{a}_{y}=(0,1)$,
and $\mathbf{a}_{[11]}=(1,1)$, respectively. In Sec. S3 of the SM,
we compute the SGBZs for these three distinct strips. The SGBZs of
the {[}10{]}-strip and {[}01{]}-strip are identical, which reads,
\begin{align}
\beta_{x} & =\exp\left(\gamma_{x}+\rmi\theta_{x}\right),\quad\beta_{y}=\exp\left(\gamma_{y}+\rmi\theta_{y}\right),\label{eq:xy-SGBZ-HN}
\end{align}
where $\gamma_{\alpha}\equiv\ln(|J_{\alpha1}/J_{\alpha2}|)/2$, $\alpha=x,y$.
The only distinction between the {[}10{]}-SGBZ and {[}01{]}-SGBZ lies
in whether $\mathbf{a}_{x}$ or $\mathbf{a}_{y}$ is the major axis.
For comparison, the SGBZ for the {[}11{]}-strip is given by, 
\begin{equation}
\begin{cases}
\tilde{\beta}_{[11]}=\rme^{\gamma_{x}+\gamma_{y}+\rmi\theta_{[11]}},\\
\tilde{\beta}_{y}=\rme^{\gamma_{y}+\rmi\theta_{y}}\sqrt{\left|\frac{J_{x}^{*}\rme^{\rmi\Delta_{xy}+\rmi\theta_{[11]}}+J_{y}}{J_{x}\rme^{\rmi\Delta_{xy}-\rmi\theta_{[11]}}+J_{y}^{*}}\right|},
\end{cases}\label{eq:11-SGBZ-HN}
\end{equation}
where $\Delta_{xy}\equiv\delta_{x}-\delta_{y}$, $\delta_{\alpha}\equiv\text{Arg}(J_{\alpha1}J_{\alpha2})/2$,
and $J_{\alpha}\equiv J_{\alpha1}/\exp(\gamma_{\alpha}+\rmi\delta_{\alpha})$
for $\alpha=x,y$. If the SGBZs of the three strips are compatible,
the coordinate transformation of momenta, which reads, 
\begin{equation}
\tilde{\beta}_{[11]}=\beta_{x}\beta_{y},\quad\tilde{\beta}_{y}=\beta_{y},\label{eq:trans}
\end{equation}
should hold for Eqs. (\ref{eq:xy-SGBZ-HN}) and (\ref{eq:11-SGBZ-HN}).
However, Eq. (\ref{eq:trans}) fails when $\sin\Delta_{xy}\neq0$,
indicating that the {[}11{]}-SGBZ differs from the {[}10{]}-SGBZ or
{[}01{]}-SGBZ in general. In fact, as shown in Fig. \ref{fig:OBC}(c)
and \ref{fig:OBC}(d), the spectrum of the {[}10{]}-SGBZ ($\sigma_{[10]}$)
or {[}01{]}-SGBZ ($\sigma_{[01]}$) deviates from that of the {[}11{]}-SGBZ
($\sigma_{[11]}$), except when $\rme^{\rmi\delta_{x}}$ is collinear
with $\rme^{\rmi\delta_{y}}$, i.e., $\sin\Delta_{xy}=0$.

To investigate the relation between the compatibility of SGBZs and
the geometry-dependent bands, we numerically calculate the OBC spectrum
in finite-size parallelogram regions with different aspect ratios.
For the compatible case, as shown in Fig. \ref{fig:OBC}(e), the parallelogram
region is spanned by the $[10]$-axis and $[01]$-axis, along which
the SGBZs are identical. For all aspect ratios, the OBC spectra match
well with the SGBZ bands. For the incompatible case, as shown in Fig.
\ref{fig:OBC}(f), the parallelogram region is spanned by the $[11]$-axis
and $[01]$-axis. In contrast to the compatible case, the OBC spectra
vary with the aspect ratios, and tend to the corresponding SGBZ bands
when the length of one side is much larger than the other side.

To understand this phenomenon, we return to the definition of the
SGBZ. Since the SGBZ is constructed by taking sequential limits, the
strip length remains much larger than the strip width during the process
of taking the width to infinity. Consequently, the SGBZ bands correspond
to the OBC spectra in the limit of both infinite sizes and extreme
aspect ratios. Therefore, when a lattice holds incompatible SGBZs,
the effects of lattice extensions along different axes compete with
each other, preventing the spectrum from converging as the system
size increases (see Sec. S4 of the SM \cite{supp-2SK}). Unlike Hermitian
systems, where the OBC spectra with different boundary geometries
converge to the Bloch bands, in non-Hermitian systems with incompatible
SGBZs, there is no uniform thermodynamic limit of OBC spectra, resulting
in the geometry-dependent bands. 

\textit{Criterion for geometry-dependent bands} --- Because the geometry-dependent
bands result from the incompatible SGBZs, a universal criterion for
geometry-dependent bands can be derived by checking the compatibility
of SGBZs. Assume that a non-Hermitian system exhibits uniform bands,
i.e., the OBC spectra of the system converge to a uniform thermodynamic
limit. Under the coordinate transformation, 
\begin{equation}
\begin{pmatrix}\tilde{\mathbf{a}}_{1} & \tilde{\mathbf{a}}_{2}\end{pmatrix}=\begin{pmatrix}\mathbf{a}_{1} & \mathbf{a}_{2}\end{pmatrix}\mathbf{P},
\end{equation}
the complex momenta of the SGBZs corresponding to $(\tilde{\mathbf{a}}_{1},\tilde{\mathbf{a}}_{2})$
and $(\mathbf{a}_{1},\mathbf{a}_{2})$ must satisfy,
\begin{equation}
\begin{pmatrix}\ln\tilde{\beta}_{1} & \ln\tilde{\beta}_{2}\end{pmatrix}=\begin{pmatrix}\ln\beta_{1} & \ln\beta_{2}\end{pmatrix}\mathbf{P},\label{eq:trans-beta}
\end{equation}
where $\mathbf{P}\in\mathbb{Z}^{2\times2}$ is an arbitrary transformation
matrix. 

\begin{figure}
\centering

\includegraphics{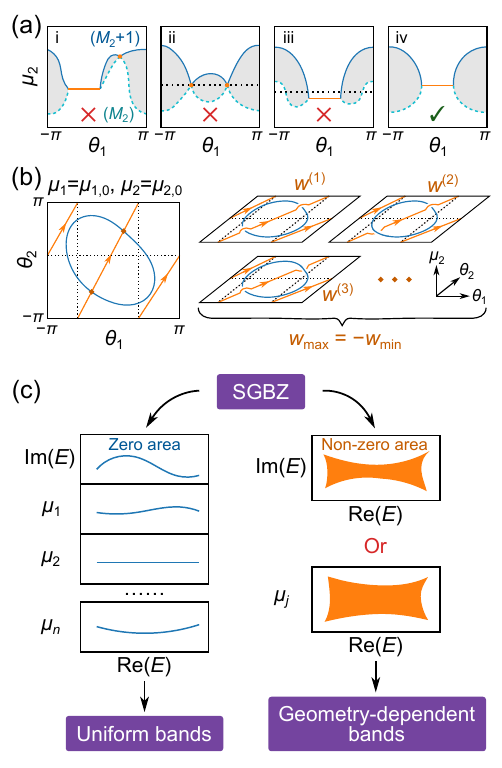}

\caption{Criterion for uniform or geometry-dependent bands. (a) Cases with
or without uniform bands, where the blue solid curve and cyan dashed
curve represent $\mu_{2}^{(M_{2}+1)}(\theta_{1};E,\mu_{1,0})$ and
$\mu_{2}^{(M_{2})}(\theta_{1};E,\mu_{1,0})$, respectively, and the
black dotted lines in ii and iii are reference lines. In cases i--iii,
uniform bands are not allowed, while in case iv, uniform bands are
allowed. (b) Requirements on the winding numbers when a system exhibits
a uniform band. The blue curve represents the solutions of $\det[E-h(\beta_{1},\beta_{2})]=0$
on the plane $(\mu_{1},\mu_{2})=(\mu_{1,0},\mu_{2,0})$, and the orange
curve represents the winding loop. (c) Illustration of the criterion
for uniform or geometry-dependent bands. For an arbitrary SGBZ, if
the complex energy spectrum and all the imaginary momentum spectra
have zero area, the bands are uniform. Otherwise, if the energy spectrum
or at least one of the imaginary momentum spectra has nonzero area,
the bands are geometry-dependent.}
\label{fig:uniform-cond}
\end{figure}

Under the basis $(\mathbf{a}_{1},\mathbf{a}_{2})$, as illustrated
in Fig. \ref{fig:uniform-cond}(a), we consider the curves $\mu_{2}^{(j)}(\theta_{1};E,\mu_{1,0})$
for $j=M_{2}$ and $M_{2}+1$, where $\mu_{1,0}$ is the critical
point at which $W(E,\mu_{1})$ changes sign. By definition, the common
points of the two curves correspond to SGBZ points with eigenenergy
$E$. First, we consider the permutation of two axes, i.e., $P_{11}=P_{22}=0$
and $P_{12}=P_{21}=1$, where $P_{ij}$ is the matrix element of $\mathbf{P}$.
By the definition of $X(E,\mu_{1,0})$, $\ln|\beta_{1}|=\mu_{1,0}$
is a constant in the original SGBZ. To ensure the transformed points
also satisfy the SGBZ constraint, $\ln|\tilde{\beta}_{1}|=\ln|\beta_{2}|$
should also be constant. Therefore, for uniform bands, the common
points between $\mu_{2}^{(M_{2})}$ and $\mu_{2}^{(M_{2}+1)}$ should
have constant value, which rules out the case i in Fig. \ref{fig:uniform-cond}(a).

Second, for a given value of $\beta_{1}$ on the SGBZ, both solutions
$(\beta_{1},\beta_{2}^{(M_{2})})$ and $(\beta_{1},\beta_{2}^{(M_{2}+1)})$
belong to the SGBZ, so the SGBZ points occur in pairs. However, the
transformation in Eq. (\ref{eq:trans-beta}) may disrupt this pairing.
To ensure the transformed points also form pairs, for each SGBZ point
$(\beta_{1},\beta_{2})$, there must exist another SGBZ point $(\beta_{1}^{\prime},\beta_{2}^{\prime})$
at the same eigenenergy such that $\tilde{\beta}_{1}=\tilde{\beta}_{1}^{\prime}$,
that is, 
\begin{equation}
\left(\frac{\beta_{1}}{\beta_{1}^{\prime}}\right)^{P_{11}}=\left(\frac{\beta_{2}^{\prime}}{\beta_{2}}\right)^{P_{21}}.\label{eq:beta-pairing}
\end{equation}
Since the solutions of Eq. (\ref{eq:11-SGBZ-HN}) for $(\beta_{1}^{\prime},\beta_{2}^{\prime})$
depend on $P_{11}/P_{21}$, when $\mathbf{P}$ ranges over all possible
transformation matrices, there must exist infinitely many SGBZ points
paired with $(\beta_{1},\beta_{2})$. Therefore, if the SGBZ points
form a finite set for some eigenenergy, as illustrated in panel ii
of Fig. \ref{fig:uniform-cond}(a), the system cannot exhibit uniform
bands.

Third, as illustrated in panel iii of Fig. \ref{fig:uniform-cond}(a),
if a horizontal line (black dotted line) passes through both the upper
region of $\mu_{2}^{(M_{2}+1)}$ and the lower region of $\mu_{2}^{(M_{2})}$,
the system cannot exhibit uniform bands either. According to Sec.
S5 of the SM \cite{supp-2SK}, this case can be transformed into the
case i under a certain coordinate transformation.

As discussed above, only when an SGBZ has infinitely many SGBZ points
with constant values of $\mu_{1}$ and $\mu_{2}$ for every eigenenergy
$E$, as illustrated in panel iv of Fig. \ref{fig:uniform-cond}(a),
can the system exhibit uniform non-Hermitian bands. In this case,
the base manifold $X(E,\mu_{1,0})$ becomes the (hyper)plane $(\mu_{1},\mu_{2})=(\mu_{1,0},\mu_{2,0})$.
In Sec. S5 of the SM \cite{supp-2SK}, we show that the uniformity
of bands requires the winding numbers of the closed loops to satisfy
the conditions shown in Fig. \ref{fig:uniform-cond}(b): For an arbitrary
loop on the base manifold (orange lines), if the loop intersects the
SGBZ points (blue curve), the winding number is ill-defined. In this
case, as illustrated in the right panel, we can increase or decrease
$\mu_{2}$ (or equivalently $\mu_{1}$) to avoid the intersections.
For all possible perturbations, uniform bands require that the maximum
and minimum winding numbers satisfy $w_{\text{max}}=-w_{\text{min}}$.
The condition in Fig. \ref{fig:uniform-cond}(b) is also sufficient
for uniform bands.

Moreover, the criterion for uniform or geometry-dependent bands also
manifests in the spectrum and the imaginary momentum spectrum. On
one hand, the SGBZ points corresponding to a fixed energy form 1D
curves, so the dimensionality of a 2D uniform spectrum is less than
that of the 2D SGBZ. Therefore, the uniform spectrum should have zero
area. On the other hand, for uniform bands, the SGBZ points corresponding
to a fixed energy should have constant $\mu_{1}$ and $\mu_{2}$,
so the imaginary momentum spectra, i.e., the plots of $\text{Re}(E)$-$\mu_{j}$,
should have zero area. The conclusion can be generalized to higher
dimensions (see Sec. S5 of the SM \cite{supp-2SK}). For an $n$-dimensional
system, as shown in Fig. \ref{fig:uniform-cond}(c), if and only if
both the spectrum and all the imaginary momentum spectra have zero
area, the system exhibits uniform bands. It is noted that zero area
of the SGBZ spectrum is not sufficient for uniform bands; an example
is available in Sec. S6 of the SM \cite{supp-2SK}.

\textit{Conclusion} --- In this work, we develop the SGBZ formulation
for the non-Bloch bands of 2D and higher-dimensional lattices, corresponding
to the limit of OBC spectrum when a lattice is extended sequentially
along its axes. Using the SGBZ formulation, we reveal the mechanism
behind geometry-dependent non-Hermitian bands. That is, the effects
of the extensions along different axes on the OBC spectrum may compete
with each other, causing the thermodynamic limit to vary with shape.
Crucially, we establish a universal criterion for geometry-dependent
non-Hermitian bands, where the geometry dependence is characterized
by the non-zero area of the energy spectrum or the imaginary momentum
spectrum. Our work provides the necessary tools to describe the interplay
between boundary geometries and novel non-Hermitian effects, such
as non-Hermitian band topologies, which is crucial for the discovery
of new physical phenomena in the future.

\nocite{yaoEdgeStatesTopological2018} \nocite{yokomizoNonBlochBandTheory2019}
\nocite{yangNonHermitianBulkBoundaryCorrespondence2020} \nocite{zhangCorrespondenceWindingNumbers2020}
\nocite{hatcherAlgebraicTopology2001} \nocite{verscheldeAlgorithm795PHCpack1999}
\begin{acknowledgments}
This work is supported by the National Key R\textbackslash\&D Program
of China (Grant No. 2023YFA1407600), and the National Natural Science
Foundation of China (NSFC) (Grants No. 12275145, No. 92050110, No.
91736106, No. 11674390, and No. 91836302).
\end{acknowledgments}

\bibliography{2D-skin-effect}

\end{document}


\def\rme{\mathrm{e}}%
 
\def\rmi{\mathrm{i}}%
 
\def\rmd{\mathrm{d}}%

\title{Supplemental Material for \\
 ``General theory for geometry-dependent non-Hermitian bands''}
\author{Chenyang Wang}
\affiliation{State Key Laboratory of Low-Dimensional Quantum Physics, Department
of Physics, Tsinghua University, Beijing 100084, China}
\author{Jinghui Pi}
\affiliation{The Chinese University of Hong Kong Shenzhen Research Institute, Shenzhen 518057, China}
\author{Qinxin Liu}
\affiliation{State Key Laboratory of Low-Dimensional Quantum Physics, Department
of Physics, Tsinghua University, Beijing 100084, China}
\author{Yaohua Li}
\affiliation{State Key Laboratory of Low-Dimensional Quantum Physics, Department
of Physics, Tsinghua University, Beijing 100084, China}
\author{Yong-Chun Liu}
\email{ycliu@tsinghua.edu.cn}

\affiliation{State Key Laboratory of Low-Dimensional Quantum Physics, Department
of Physics, Tsinghua University, Beijing 100084, China}
\affiliation{Frontier Science Center for Quantum Information, Beijing 100084, China}

\maketitle
\onecolumngrid
\setcounter{section}{0} 
\global\long\def\thesection{S\arabic{section}}
 \setcounter{figure}{0} 
\global\long\def\thefigure{S\arabic{figure}}
 \setcounter{equation}{0} 
\global\long\def\theequation{S\arabic{equation}}
 \setcounter{table}{0} 
\global\long\def\thetable{S\arabic{table}}
\makeatletter
\renewcommand{\p@subsection}{\thesection.}
\makeatother
\numberwithin{equation}{section}

\tableofcontents{}

\section{The formulation of strip generalized Brillouin zone }

\subsection{Introduction to 1D non-Bloch band theory}\label{subsec:GBZ-1D}

The non-Bloch band theory is a generalization of the Bloch band theory
by extending the lattice momenta to complex numbers \cite{yaoEdgeStatesTopological2018,yokomizoNonBlochBandTheory2019,yangNonHermitianBulkBoundaryCorrespondence2020}.
However, not all complex values are permissible as lattice momenta.
To maintain the same dimensionality in momentum space as in real space,
$n$ real-valued constraints are required for an $n$-dimensional
($n$D) non-Hermitian lattice.

For one-dimensional (1D) lattices, such real-valued constraints can
be derived from the open boundary condition (OBC) in the thermodynamic
limit. In general, consider the 1D Hamiltonian in the following form,
\begin{equation}
H=\sum_{r\in\mathbb{Z}}\sum_{t=-t_{R}}^{t_{R}}\sum_{\mu,\nu=1}^{m}\mathcal{T}_{t,\mu,\nu}c_{r+t,\mu}^{\dagger}c_{r,\nu},
\end{equation}
where $c_{r,\mu}$ is the annihilation operator at the site with coordinate
$r$ and sublattice index $\mu$, and $\mathcal{T}_{t,\mu,\nu}$ is
the coupling coefficient with maximum coupling range $t_{R}$. Using
a Fourier transformation, we get the momentum-space Hamiltonian,
\begin{equation}
h_{\mu,\nu}\left(\rme^{\rmi k}\right)=\sum_{t=-t_{R}}^{t_{R}}\mathcal{T}_{t,\mu,\nu}\rme^{-\rmi kt}.
\end{equation}
By substituting $\beta=\rme^{\rmi k}$, the elements of $h(\beta)$
are all Laurent polynomials in $\beta$. The corresponding characteristic
polynomial reads, 
\begin{align}
f\left(E,\beta\right) & \equiv\det\left[E-h\left(\beta\right)\right],\nonumber \\
 & =\begin{vmatrix}E-\sum\limits_{t=-t_{R}}^{t_{R}}\mathcal{T}_{t,1,1}\beta^{-t} & -\sum\limits_{t=-t_{R}}^{t_{R}}\mathcal{T}_{t,1,2}\beta^{-t} & \cdots & -\sum\limits_{t=-t_{R}}^{t_{R}}\mathcal{T}_{t,1,m}\beta^{-t}\\
-\sum\limits_{t=-t_{R}}^{t_{R}}\mathcal{T}_{t,2,1}\beta^{-t} & E-\sum\limits_{t=-t_{R}}^{t_{R}}\mathcal{T}_{t,2,2}\beta^{-t} & \cdots & -\sum\limits_{t=-t_{R}}^{t_{R}}\mathcal{T}_{t,2,m}\beta^{-t}\\
\vdots & \vdots & \ddots & \vdots\\
-\sum\limits_{t=-t_{R}}^{t_{R}}\mathcal{T}_{t,m,1}\beta^{-t} & -\sum\limits_{t=-t_{R}}^{t_{R}}\mathcal{T}_{t,m,2}\beta^{-t} & \cdots & E-\sum\limits_{t=-t_{R}}^{t_{R}}\mathcal{T}_{t,m,m}\beta^{-t}
\end{vmatrix}.\label{seq:det-1D-ChP}
\end{align}
According to Eq. (\ref{seq:det-1D-ChP}), when $\mathcal{T}_{t,\mu,\nu}$
are non-zero for any $\mu,\nu=1,2,\dots,m$, the highest and lowest
degrees of $\beta$ in $f\left(E,\beta\right)$ are $-M=-mt_{R}$
and $N=mt_{R}$, respectively. Therefore, for given values of $E$,
there are $M+N=2mt_{R}$ zeros for the eigenvalue equation $f\left(E,\beta\right)=0$.
We sort the zeros by $\left|\beta^{(1)}\right|\leq\left|\beta^{(2)}\right|\leq\cdots\leq\left|\beta^{(2mt_{R})}\right|$.
The corresponding ``non-Bloch'' waves in momentum space can be calculated
by, 
\begin{equation}
h\left(\beta^{(j)}\right)\tilde{\phi}^{(j)}=E\tilde{\phi}^{(j)},\label{seq:non-Bloch-wave-reciprocal}
\end{equation}
where $\tilde{\phi}^{(j)}=(\tilde{\phi}_{1}^{(j)},\tilde{\phi}_{2}^{(j)},\dots,\tilde{\phi}_{m}^{(j)})\in\mathbb{C}^{m}$,
and the corresponding real-space expression is given by, 
\begin{align}
\phi_{\mu}^{(j)}\left(r\right) & \equiv\langle r,\mu|\phi^{(j)}\rangle=\tilde{\phi}_{\mu}^{(j)}\left(\beta^{(j)}\right)^{r},
\end{align}
where $\left|r,\mu\right>\equiv c_{r,\mu}^{\dagger}\left|0\right>$
is the single-particle basis.

Next, we consider a finite-size open chain of length $L$. We assume
that the eigenstates under open boundary conditions (OBCs) are superpositions
of non-Bloch waves, which read, 
\begin{equation}
\left|\psi\right>=\sum_{j=1}^{2mt_{R}}C_{j}\left|\phi^{(j)}\right>.\label{seq:formal-OBC-eigenstate}
\end{equation}
Substituting Eqs. (\ref{seq:non-Bloch-wave-reciprocal}-\ref{seq:formal-OBC-eigenstate})
into the eigenvalue equation $H\left|\psi\right>=E\left|\psi\right>$,
the equations for $C_{j}$ read, 
\begin{align}
\psi_{\mu}\left(-l\right) & \equiv\sum_{j=1}^{2mt_{R}}C_{j}\tilde{\phi}_{\mu}^{(j)}\left(\beta^{(j)}\right)^{-l}=0,\label{seq:eq-C-1}\\
\psi_{\mu}\left(L+1+l\right) & \equiv\sum_{j=1}^{2mt_{R}}C_{j}\tilde{\phi}_{\mu}^{(j)}\left(\beta^{(j)}\right)^{L+1+l}=0,\label{seq:eq-C-2}
\end{align}
for $l=0,1,\dots,t_{R}-1$ and $\mu=1,\dots,m$. Equations (\ref{seq:eq-C-1})
and (\ref{seq:eq-C-2}) are homogeneous linear equations in $C_{j}$,
where $j=1,2,\dots,mt_{R}$. Therefore, the condition for non-zero
solutions of $\left|\psi\right>$ requires the determinant of the
coefficients to vanish, i.e., 
\begin{equation}
\begin{vmatrix}\tilde{\phi}^{(1)} & \tilde{\phi}^{(2)} & \cdots & \tilde{\phi}^{(2mt_{R})}\\
\tilde{\phi}^{(1)}\left(\beta^{(1)}\right)^{-1} & \tilde{\phi}^{(2)}\left(\beta^{(2)}\right)^{-1} & \cdots & \tilde{\phi}^{(2mt_{R})}\left(\beta^{(2mt_{R})}\right)^{-1}\\
\vdots & \vdots &  & \vdots\\
\tilde{\phi}^{(1)}\left(\beta^{(1)}\right)^{-t_{R}+1} & \tilde{\phi}^{(2)}\left(\beta^{(2)}\right)^{-t_{R}+1} & \cdots & \tilde{\phi}^{(2mt_{R})}\left(\beta^{(2mt_{R})}\right)^{-t_{R}+1}\\
\tilde{\phi}^{(1)}\left(\beta^{(1)}\right)^{L+1} & \tilde{\phi}^{(2)}\left(\beta^{(2)}\right)^{L+1} & \cdots & \tilde{\phi}^{(2mt_{R})}\left(\beta^{(2mt_{R})}\right)^{L+1}\\
\tilde{\phi}^{(1)}\left(\beta^{(1)}\right)^{L+2} & \tilde{\phi}^{(2)}\left(\beta^{(2)}\right)^{L+2} & \cdots & \tilde{\phi}^{(2mt_{R})}\left(\beta^{(2mt_{R})}\right)^{L+2}\\
\vdots & \vdots &  & \vdots\\
\tilde{\phi}^{(1)}\left(\beta^{(1)}\right)^{L+t_{R}} & \tilde{\phi}^{(2)}\left(\beta^{(2)}\right)^{L+t_{R}} & \cdots & \tilde{\phi}^{(2mt_{R})}\left(\beta^{(2mt_{R})}\right)^{L+t_{R}}
\end{vmatrix}=0.\label{seq:det-of-C}
\end{equation}
It is noted that $\tilde{\phi}^{(j)}$ in Eq. (\ref{seq:det-of-C})
is an $m$-vector; thus, the determinant is of size $2mt_{R}\times2mt_{R}$.
According to the definition of $\tilde{\phi}^{(j)}$, $\tilde{\phi}^{(j)}$
is independent of $L$, so the determinant in Eq. (\ref{seq:det-of-C})
can be expanded in the general form, which reads,
\begin{equation}
\sum_{1\leq j_{1}<j_{2}<\cdots<j_{M_{}}\leq2M}\left(\beta^{(j_{1})}\beta^{(j_{2})}\cdots\beta^{(j_{M})}\right)^{L}g_{j_{1}j_{2}\dots j_{M}}\left(E,\beta^{(1)},\dots,\beta^{(2M)}\right)=0,\label{seq:new-eq-C}
\end{equation}
where the functions $g_{j_{1}j_{2}\dots j_{M}}\left(E,\beta^{(1)},\dots,\beta^{(2M)}\right)$
are independent of $L$. As $L$ tends to infinity, dividing both
sides of Eq. (\ref{seq:new-eq-C}) by $\left(\beta^{(M+1)}\beta^{(M+2)}\cdots\beta^{(2M)}\right)^{L}$
yields, 
\begin{equation}
g_{M+1,\dots,2M}\left(E,\beta^{(1)},\dots,\beta^{(2M)}\right)+\left(\frac{\beta^{(M)}}{\beta^{(M+1)}}\right)^{L}g_{M,M+2,\dots,2M}\left(E,\beta^{(1)},\dots,\beta^{(2M)}\right)+o\left[\left(\frac{\beta^{(M)}}{\beta^{(M+1)}}\right)^{L}\right]=0.\label{seq:simplified-eq}
\end{equation}
Since $\beta^{(j)},j=1,2,\dots,2M$ depend on $E$ through $f\left(E,\beta\right)=0$,
the functions $g_{j_{1}j_{2}\dots j_{M_{2}}}$ are univariate functions
of $E$. If $\left|\beta^{(M)}\right|\neq\left|\beta^{(M+1)}\right|$,
all terms in Eq. (\ref{seq:simplified-eq}) vanish except for $g_{M+1,\dots,2M}\left(E,\beta^{(1)},\dots,\beta^{(2M)}\right)$,
so only a finite number of solutions (independent of $L$) can be
obtained from Eq. (\ref{seq:simplified-eq}). Otherwise, if $\left|\beta^{(M)}\right|=\left|\beta^{(M+1)}\right|$,
the first two terms in Eq. (\ref{seq:simplified-eq}) are preserved,
and the number of solutions increases as $O(L)$ when $L\rightarrow\infty$.

Based on the above discussion, the thermodynamic limit under OBC imposes
a real-valued constraint ($|\beta^{(M)}|=|\beta^{(M+1)}|$) on the
complex momentum, which restricts $\beta$ to a 1D closed loop in
the complex plane. This 1D closed loop is defined as the generalized
Brillouin zone (GBZ) of the 1D non-Hermitian lattice. Similar to the
Brillouin zone (BZ) in Hermitian systems, the bands on the GBZ correspond
to the OBC spectrum of a 1D non-Hermitian lattice when the system
is sufficiently large.

Next, we consider the density of states on the GBZ. Define the relative
phase $\exp\left(\rmi\phi\right)\equiv\beta^{(M+1)}/\beta^{(M)}$.
For a fixed $\phi$, the values of $E$ are determined by solving
the equations, 
\begin{equation}
\begin{cases}
f\left(E,\beta\right)=0,\\
f\left(E,\beta\rme^{\rmi\phi}\right)=0.
\end{cases}
\end{equation}
Consequently, all functions $g_{j_{1}j_{2}\dots j_{M}}$ are univariate
functions of $\exp\left(\rmi\phi\right)$. When $L$ is sufficiently
large, Eq. (\ref{seq:simplified-eq}) becomes, 
\begin{equation}
\phi=\frac{\rmi}{L}\ln\frac{g_{M,M+2,\dots,2M}\left(\rme^{\rmi\phi}\right)}{g_{M+1,\dots,2M}\left(\rme^{\rmi\phi}\right)}+\frac{2n\pi}{L},n=0,\pm1,\pm2,\dots.\label{seq:phi-eq}
\end{equation}
Next, we define $\tilde{g}\left(\rme^{\rmi\phi}\right)\equiv\rmi\ln\left[g_{M_{2},M_{2}+2,\dots,2M_{2}}\left(\rme^{\rmi\phi}\right)/g_{M_{2}+1,\dots,2M_{2}}\left(\rme^{\rmi\phi}\right)\right]$,
and consider two adjacent solutions, 
\begin{align}
\phi_{1} & =\frac{1}{L}\tilde{g}\left(\rme^{\rmi\phi_{1}}\right)+\frac{2n\pi}{L},\\
\phi_{2} & =\frac{1}{L}\tilde{g}\left(\rme^{\rmi\phi_{2}}\right)+\frac{2\left(n+1\right)\pi}{L},
\end{align}
then, the difference between the two solutions is given by, 
\begin{align}
\phi_{2}-\phi_{1} & =\frac{2\pi}{L}+\frac{1}{L}\frac{\rmd\tilde{g}\left(\phi_{1}\right)}{\rmd\phi}\left(\phi_{2}-\phi_{1}\right)+O\left(\frac{1}{L^{3}}\right)=\frac{2\pi}{L}+O\left(\frac{1}{L^{2}}\right),
\end{align}
where the second equation holds because the order of $\phi_{2}-\phi_{1}$
is $O(1/L)$. Therefore, if the relative phase changes by $\Delta\phi$,
the number of solutions is given by, 
\begin{align}
N_{\text{sols}} & =\frac{\left|\Delta\phi\right|}{\frac{2\pi}{L}+O\left(\frac{1}{L^{2}}\right)}=\frac{L}{2\pi}\left|\Delta\phi\right|+O\left(1\right),\label{eq:sols-and-argument}
\end{align}
indicating that the number of OBC eigenstates on a segment of the
GBZ is proportional to the change in relative phases in the thermodynamic
limit.

\subsection{Basic idea of the SGBZ}

In general, we consider an $n$-dimensional ($n$D) non-Hermitian
lattice. Using a set of lattice vectors $(\mathbf{a}_{1},\mathbf{a}_{2},\dots,\mathbf{a}_{n})$,
the real-space Hamiltonian can be expressed in the general form, 
\begin{equation}
H=\sum_{\mathbf{r}}\sum_{t_{1},t_{2},\dots,t_{n}}\sum_{\mu,\nu=1}^{m}\mathcal{T}_{\mathbf{t},\mu,\nu}c_{\mathbf{r}+\mathbf{t},\mu}^{\dagger}c_{\mathbf{r},\nu},\label{eq:non-interacting-Hamiltonian}
\end{equation}
where $\mathcal{T}_{\mathbf{t},\mu,\nu}\in\mathbb{C}$ is the coupling
coefficient, and $c_{\mathbf{r},\nu}$ is the annihilation operator
at position $\mathbf{r}$ and sublattice index $\nu$. The subscript
$\mathbf{r}=\sum_{j=1}^{n}r_{j}\mathbf{a}_{j}$ denotes the position
vector, and $\mathbf{t}=\sum_{j=1}^{n}t_{j}\mathbf{a}_{j}$ is the
coupling vector within the range $|t_{j}|\leq t_{Rj}$, $j=1,2,\dots,n$,
where $r_{j},t_{j}\in\mathbb{Z}$, and $t_{Rj}$, $j=1,2,\dots,n$
are positive integers. We first discuss the SGBZ for 2D lattices and
then extend it to higher dimensions.

\begin{figure*}
\centering

\includegraphics{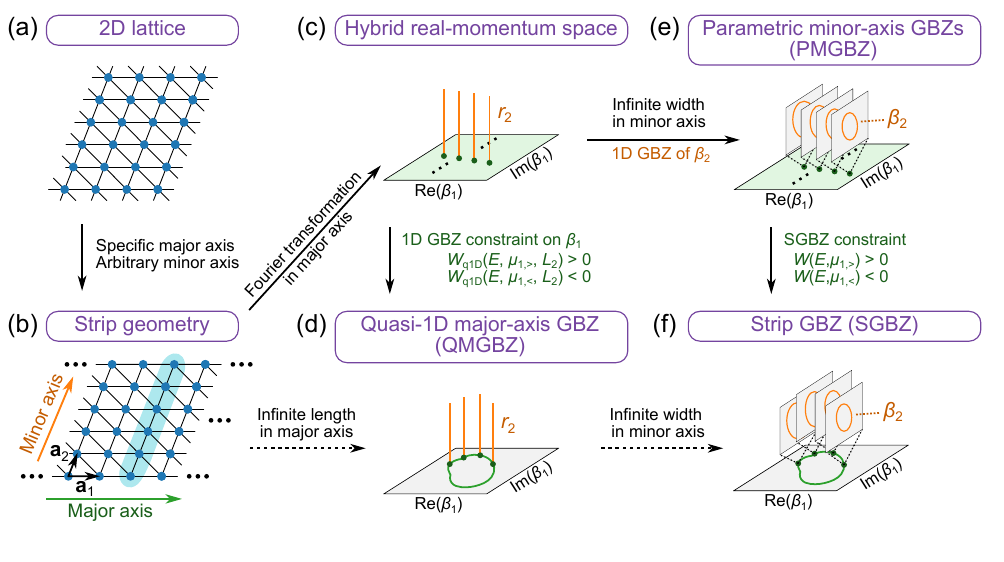}

\caption{Schematics of the SGBZ. (a) Sketch of a 2D lattice. (b) Schematic
diagram of the strip geometry, where $\mathbf{a}_{1}$ and $\mathbf{a}_{2}$
denote the major and minor axes, respectively, and the cyan region
represents the supercell. (c) The hybrid real-momentum space formed
by the 2D complex plane of $\beta_{1}$ and the 1D real space of $r_{2}$.
(d) The QMGBZ, defined as the 1D GBZ for the strip geometry. (e) The
PMGBZ, defined as the parametric GBZ along the minor axis with parameter
$\beta_{1}$. (f) The SGBZ, defined as the limit of the QMGBZ as the
width approaches infinity. }
\label{fig:main-idea}
\end{figure*}

The concept of the SGBZ is illustrated in Fig. \ref{fig:main-idea}.
For a 2D lattice {[}Fig. \ref{fig:main-idea}(a){]}, a strip geometry
is defined by selecting a major axis and a minor axis, then extending
the structure along the major axis. As shown in Fig. \ref{fig:main-idea}(b),
the lattice in a strip geometry is periodic along the major axis,
with its unit cell (referred to as the quasi-1D supercell) comprising
a slice of lattice points parallel to the minor axis (cyan region).
According to the 1D non-Bloch band theory, as the length along the
major axis approaches infinity, the OBC eigensystem of the strip geometry
converges to the eigensystem of its 1D GBZ, termed the quasi-1D major-axis
GBZ (QMGBZ). Assuming the real-space Hamiltonian takes the form of
Eq. (\ref{eq:non-interacting-Hamiltonian}), applying Fourier transformation
along the major axis yields a hybrid real-momentum Hamiltonian, given
by 
\begin{align}
\mathcal{H}_{L_{2}}\left(\beta_{1}\right)= & \sum_{t_{1},t_{2}}\sum_{r_{2}=1}^{L_{2}}\sum_{\mu,\nu=1}^{m}\mathcal{T}_{t_{1},t_{2},\mu,\nu}\beta_{1}^{-t_{1}}c_{\beta_{1},r_{2}+t_{2};\mu}^{\dagger}c_{\beta_{1},r_{2};\nu},\label{eq:hybrid-non-Bloch-Hamiltonian}
\end{align}
where $L_{2}$ is the width, and $c_{\beta_{1},r_{2};\mu}$ is the
annihilator of the non-Bloch wave with complex momentum $\exp(\rmi k_{1})\equiv\beta_{1}\in\mathbb{C}$,
coordinate $r_{2}$, and sublattice index $\mu$. As illustrated in
Fig. \ref{fig:main-idea}(c), $\beta_{1}$ and $r_{2}$ form a 3D
hybrid real-momentum space. To obtain the QMGBZ {[}Fig. \ref{fig:main-idea}(d){]},
the following 1D GBZ constraint, 
\begin{equation}
|\beta_{1}^{(M)}(E)|=|\beta_{1}^{(M+1)}(E)|,\label{eq:quasi-1D-GBZ-constraint}
\end{equation}
is imposed on the 3D hybrid space, where $\beta_{1}^{(j)}(E)$ is
the $j$-th solution of $F_{L_{2}}(E,\beta_{1})\equiv\det[E-\mathcal{H}_{L_{2}}(\beta_{1})]=0$
ordered by $|\beta_{1}^{(i)}(E)|\leq|\beta_{1}^{(j)}(E)|,\forall i<j$,
and $-M,N$ are the lowest and highest degrees of $\beta_{1}$ in
$F_{L_{2}}(E,\beta_{1})$, respectively.

Next, to obtain the SGBZ, we take $L_{2}$ to infinity and define
the SGBZ as the limit of the QMGBZ. Since the QMGBZ is defined as
the set of points in the hybrid real-momentum space restricted by
the 1D GBZ constraint, the SGBZ can be derived by taking the limit
of both the hybrid real-momentum space and the 1D GBZ constraint on
$\beta_{1}$ as $L_{2}\rightarrow\infty$.

For the hybrid real-momentum space, treating $\beta_{1}$ as a parameter,
the hybrid Hamiltonian $\mathcal{H}_{L_{2}}(\beta_{1})$ can be regarded
as a parametric Hamiltonian of a 1D open chain along the minor axis.
As $L_{2}\rightarrow\infty$, the eigensystem of $\mathcal{H}_{L_{2}}(\beta_{1})$
tends to its GBZ, referred to as the parametric minor-axis GBZ (PMGBZ).
As illustrated in Fig. \ref{fig:main-idea}(e), each $\beta_{1}\in\mathbb{C}$
defines a 1D GBZ of $\beta_{2}$. When $\beta_{1}$ varies over $\mathbb{C}$,
all these 1D GBZs form a 3D space.

For the GBZ constraint, since the degree $M$ in Eq. (\ref{eq:quasi-1D-GBZ-constraint})
diverges as the width tends to infinity, we employ the winding number
formulation for the 1D GBZ constraint \cite{zhangCorrespondenceWindingNumbers2020},
which is based on the quasi-1D winding number defined as, 
\begin{equation}
W_{\text{q1D}}\left(E,\mu_{1},L_{2}\right)=\frac{1}{2\pi\rmi}\int_{-\pi}^{\pi}\rmd\theta_{1}\frac{\partial\ln F_{L_{2}}\left(E,\rme^{\mu_{1}+\rmi\theta_{1}}\right)}{\partial\theta_{1}}.\label{eq:definition-winding-number-quasi-1D}
\end{equation}
When the width tends to infinity, we prove that $W_{\text{q1D}}(E,\mu_{1},L_{2})/L_{2}$
converges and define the ``strip winding number'' $W(E,\mu_{1})$
as the limit of $W_{\text{q1D}}(E,\mu_{1},L_{2})/L_{2}$. Substituting
$W_{\text{q1D}}$ with $W$, we obtain the SGBZ constraint. 

In the following text, we will give the explicit expressions for the
SGBZ. Frequently used notations are listed in Tab. \ref{tab:notation}.
Additionally, in this work, some functions have arguments separated
into two groups by a semicolon, such as $f|_{X}(\theta_{1},\theta_{2};E,\mu_{1})$,
where the arguments to the right of the semicolon are treated as parameters.
In some cases, the parameter part is omitted to keep the expressions
concise.

\begin{table}
\centering

\caption{ Frequently used notations}\label{tab:notation}

\begin{tabular}{|c|c|}
\hline 
Notation & Definition\tabularnewline
\hline 
\hline 
$k_{j}$ & Complex momentum conjugate to $\mathbf{a}_{j}$, $j=1,2,\dots,n$\tabularnewline
\hline 
$\beta_{j},\mu_{j},\theta_{j}$ & $\beta_{j}\equiv\rme^{\rmi k_{j}}$, $\theta_{j}\equiv\text{Re}k_{j}$
and $\mu_{j}\equiv-\text{Im}k_{j}$\tabularnewline
\hline 
$H$ & Hamiltonian in real space \tabularnewline
\hline 
$\mathcal{H}_{L_{2}}(\beta_{1})$ & Hamiltonian in hybrid real-momentum space\tabularnewline
\hline 
$h(\beta_{1},\beta_{2})$ & Hamiltonian in momentum space\tabularnewline
\hline 
$F_{L_{2}}(E,\beta_{1})$ & Characteristic polynomial of $\mathcal{H}_{L_{2}}$, i.e., $F_{L_{2}}(E,\beta_{1})\equiv\det\left[E-\mathcal{H}_{L_{2}}(\beta_{1})\right]$\tabularnewline
\hline 
$f(E,\beta_{1},\beta_{2})$ & Characteristic polynomial of $h$, i.e., $f(E,\beta_{1},\beta_{2})\equiv\det\left[E-h(\beta_{1},\beta_{2})\right]$\tabularnewline
\hline 
$M$, $N$ & $-M$, $N$ are lowest and highest degrees of $\beta_{1}$ in $F_{L_{2}}$\tabularnewline
\hline 
$M_{j}$, $N_{j}$ & $-M_{j}$, $N_{j}$ are lowest and highest degrees of $\beta_{j}$
in $f$\tabularnewline
\hline 
$W_{\text{q1D}}$ & Quasi-1D winding number\tabularnewline
\hline 
$W$ & Strip winding number\tabularnewline
\hline 
\end{tabular}
\end{table}

\subsection{Derivation of the SGBZ formulation}\label{subsec:derivation-SGBZ}

\begin{figure}
\centering

\includegraphics{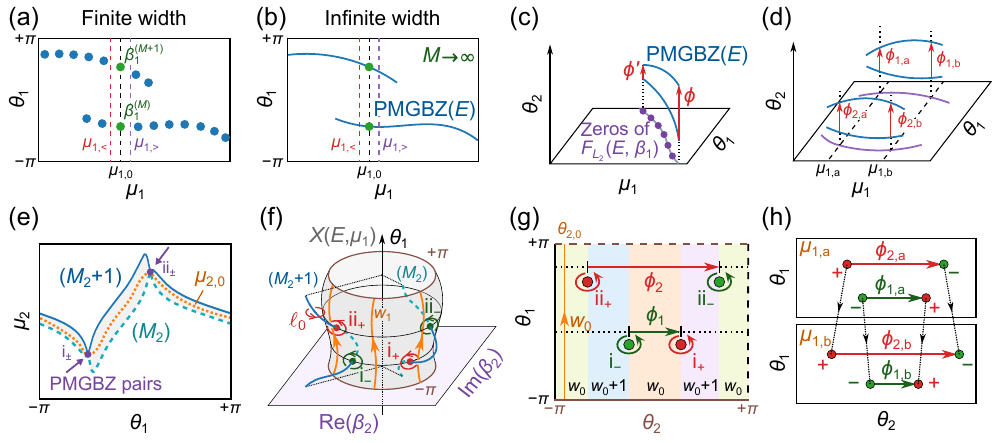}

\caption{Distribution of the zeros of $F_{L_{2}}$ and properties of the strip
winding number. (a, b) Illustrations of the zeros of $F_{L_{2}}(E,\beta_{1})$
when the strip width is (a) finite and (b) infinite. (c) Relation
between the density of zeros on a segment of $\text{PMGBZ}(E)$ and
the relative phases $\phi$ and $\phi^{\prime}$. (d) Relation between
the increment of $W_{\text{q1D}}(E,\mu_{1},L_{2})$ and the differences
of relative phases. (e) Definition of the radius function $\mu_{2,0}$.
(f) Definition of the base manifold $X(E,\mu_{1})$ and the strip
winding number, where the cyan dashed line and the blue solid line
denote the $M_{2}$-th and $(M_{2}+1)$-th zeros of $f(E,\protect\rme^{\mu_{1}+\protect\rmi\theta_{1}},\beta_{2})$,
respectively. The red and green dots represent PMGBZ points with positive
and negative charges, respectively, and the orange curves are the
winding loops for $w_{1}(\theta_{2};E,\mu_{1})$. (g) Unfolded view
of $X(E,\mu_{1})$. The red and green dots correspond to PMGBZ points.
The manifold is divided into slices with constant $w_{1}$. Given
the winding number around a specific loop (orange line), $w_{1}(\theta_{1};E,\mu_{1})$
is determined for any $\theta_{1}$ based on the topological charges
of the PMGBZ points. (h) Movement of PMGBZ points as $\mu_{1}$ increases.
When $\mu_{1}$ increases from $\mu_{1,a}$ to $\mu_{1,b}>\mu_{1,a}$,
the distance from the positive charge to the negative charge increases,
and vice versa.}
\label{fig:SGBZ-formulation}
\end{figure}

According to the Cauchy argument principle, when $F_{L_{2}}(E,\beta_{1})$
does not vanish on the circle $|\beta_{1}|=\rme^{\mu_{1}}$, $W_{\text{q1D}}(E,\mu_{1},L_{2})$
is well-defined and is related to the $\beta_{1}$-zeros of $F_{L_{2}}(E,\beta_{1})$
by, 
\begin{equation}
W_{\text{q1D}}\left(E,\mu_{1},L_{2}\right)=N_{\text{zeros}}-M,\label{eq:Cauchy-argument}
\end{equation}
where $N_{\text{zeros}}$ is the number of zeros satisfying $\ln|\beta_{1}^{(j)}|<\mu_{1}$.
Therefore, the QMGBZ constraint in Eq. (\ref{eq:quasi-1D-GBZ-constraint})
is equivalent to the following condition: For some imaginary momentum
$\mu_{1,0}$, if $\forall\epsilon>0$, there exist $\mu_{1,<}\in(\mu_{1,0}-\epsilon,\mu_{1,0})$
and $\mu_{1,>}\in(\mu_{1,0},\mu_{1,0}+\epsilon)$ such that, 
\begin{equation}
\begin{cases}
W_{\text{q1D}}\left(E,\mu_{1,<},L_{2}\right)<0,\\
W_{\text{q1D}}\left(E,\mu_{1,>},L_{2}\right)>0,
\end{cases}\label{eq:QMGBZ-constraint}
\end{equation}
then, the zeros of $F_{L_{2}}(E,\beta_{1})$ satisfying $|\beta_{1}|=\rme^{\mu_{1,0}}$
are QMGBZ points. The geometric interpretation of Eq. (\ref{eq:QMGBZ-constraint})
is illustrated in Fig. \ref{fig:SGBZ-formulation}(a), where the dots
represent the zeros of $F_{L_{2}}(E,\beta_{1})$. According to Eq.
(\ref{eq:Cauchy-argument}), there are at most $M-1$ zeros to the
left of $\mu_{1,<}$ (red dashed line), and at least $M+1$ zeros
to the left of $\mu_{1,>}$ (purple dashed line). Therefore, $\beta_{1}^{(M)}(E)$
and $\beta_{1}^{(M+1)}(E)$ must be located between $\mu_{1,<}$ and
$\mu_{1,>}$. Because Eq. (\ref{eq:QMGBZ-constraint}) is satisfied
for every $\epsilon>0$, $|\beta_{1}^{(M)}(E)|=|\beta_{1}^{(M+1)}(E)|=\rme^{\mu_{1,0}}$
must hold, which is the QMGBZ constraint in the form of Eq. (\ref{eq:quasi-1D-GBZ-constraint}).

When the width $L_{2}$ approaches infinity, the degrees $M$ and
$N$ both diverge, making Eq. (\ref{eq:quasi-1D-GBZ-constraint})
ill-defined. As illustrated in Fig. \ref{fig:SGBZ-formulation}(b),
the zeros of $F_{L_{2}}(E,\beta_{1})$ form 1D curves in the complex
plane, which are the $\beta_{1}$ components of the PMGBZ points with
eigenenergy $E$, denoted as $\text{PMGBZ}(E)$. The set $\text{PMGBZ}(E)$
can be determined by first solving the following auxiliary GBZ (aGBZ)
equations \cite{yangNonHermitianBulkBoundaryCorrespondence2020},
\begin{equation}
\begin{cases}
f\left(E,\beta_{1},\beta_{2}\right)=0\\
f\left(E,\beta_{1},\beta_{2}\rme^{\rmi\phi}\right)=0
\end{cases},\phi\in\left[0,\pi\right),
\end{equation}
then checking the GBZ constraint $|\beta_{2}^{(M_{2})}(E,\beta_{1})|=|\beta_{2}^{(M_{2}+1)}(E,\beta_{1})|$,
where $\beta_{2}^{(j)}(E,\beta_{1})$ is the $j$-th zero of $f(E,\beta_{1},\beta_{2})$
ordered by modulus. For each fixed $\phi$, the aGBZ equations have
a finite number of solutions for $(\beta_{1},\beta_{2})$. Hence,
as $\phi$ traverses $[0,\pi)$, the solutions form 1D curves.

In Sec. \ref{subsec:GBZ-1D}, we have shown that the number of states
on a GBZ is proportional to the relative phase between a GBZ pair
by Eq. (\ref{eq:sols-and-argument}). For $\text{PMGBZ}(E)$, the
reasoning in Sec. \ref{subsec:GBZ-1D} still holds by replacing $E$
with $\beta_{1}$. As shown in Fig. \ref{fig:SGBZ-formulation}(c),
the number of $\beta_{1}$-zeros on a segment of $\text{PMGBZ}(E)$
is proportional to the change in relative phases at both endpoints
when $L_{2}$ is large enough, i.e., $N_{\text{q1D}}=L_{2}|\phi^{\prime}-\phi|/2\pi+O(1)$.
Therefore, as illustrated in Fig. \ref{fig:SGBZ-formulation}(d),
the increment of $W_{\text{q1D}}(E,\mu_{1},L_{2})$ when $\mu_{1}$
increases from $\mu_{1,a}$ to $\mu_{1,b}$ ($\mu_{1,b}>\mu_{1,a}$)
equals, 
\begin{equation}
W_{\text{q1D}}\left(E,\mu_{1,b},L_{2}\right)-W_{\text{q1D}}\left(E,\mu_{1,a},L_{2}\right)=\frac{L_{2}}{2\pi}\sum_{j}\left|\phi_{j,a}-\phi_{j,b}\right|+O(1),\label{eq:increment-Wq1D}
\end{equation}
where $\phi_{j,a}$ and $\phi_{j,b}$ are the relative phases of the
$j$-th PMGBZ pairs at $\mu_{1,a}$ and $\mu_{1,b}$, respectively.

Equation (\ref{eq:increment-Wq1D}) shows that the increment of $W_{\text{q1D}}(E,\mu_{1},L_{2})/L_{2}$
converges as $L_{2}\rightarrow\infty$. Therefore, the QMGBZ constraint
in Eq. (\ref{eq:QMGBZ-constraint}) remains valid when $L_{2}\rightarrow\infty$,
provided $W_{\text{q1D}}(E,\mu_{1},L_{2})$ is replaced by the limit
of $W_{\text{q1D}}(E,\mu_{1},L_{2})/L_{2}$. We define the strip winding
number $W(E,\mu_{1})$ as the limit of $W_{\text{q1D}}(E,\mu_{1},L_{2})/L_{2}$.
In the following text, we will first derive the expression for $W(E,\mu_{1})$,
then prove that $W(E,\mu_{1})=\lim_{L_{2}\rightarrow\infty}W_{\text{q1D}}(E,\mu_{1},L_{2})/L_{2}$.

For fixed $E$ and $\mu_{1}$, the zeros $\beta_{2}^{(j)}(\theta_{1};E,\mu_{1})$,
$j=1,2,\dots,M_{2}$ of $f(E,\rme^{\mu_{1}+\rmi\theta_{1}},\beta_{2})$
can be regarded as functions of $\theta_{1}$. If the zeros are ordered
by $|\beta_{2}^{(j)}|\leq|\beta_{2}^{(k)}|$, $\forall j<k$, $\beta_{2}^{(j)}$
is not always continuous in $\theta_{1}$, but its modulus $\mu_{2}^{(j)}(\theta_{1};E,\mu_{1})\equiv\ln|\beta_{2}^{(j)}(\theta_{1};E,\mu_{1})|$
is continuous. As illustrated in Fig. \ref{fig:SGBZ-formulation}(e),
when $\mu_{2}^{(M_{2}+1)}$ intersects with $\mu_{2}^{(M_{2})}$,
the intersection points satisfy the PMGBZ constraint.

To define the base manifold for the strip winding number, we define
the radius function $\mu_{2,0}(\theta_{1};E,\mu_{1})$ as a periodic
function of $\theta_{1}$ between $\mu_{2}^{(M_{2})}$ and $\mu_{2}^{(M_{2}+1)}$,
i.e., 
\begin{equation}
\mu_{2}^{(M_{2})}\left(\theta_{1};E,\mu_{1}\right)\leq\mu_{2,0}\left(\theta_{1};E,\mu_{1}\right)\leq\mu_{2}^{(M_{2}+1)}\left(\theta_{1};E,\mu_{1}\right),\label{eq:radius-function}
\end{equation}
where the equality holds if and only if $\mu_{2}^{(M_{2})}=\mu_{2}^{(M_{2}+1)}$.
In Fig. \ref{fig:SGBZ-formulation}(e), the radius function $\mu_{2,0}$
is represented by the orange dotted line. Using the radius function
$\mu_{2,0}$, we define the base manifold $X(E,\mu_{1})$, 
\begin{equation}
X\left(E,\mu_{1}\right)\equiv\left\{ \left(\rme^{\mu_{1}+\rmi\theta_{1}},\beta_{2}\right)\mid\left|\beta_{2}\right|=\rme^{\mu_{2,0}\left(\theta_{1};E,\mu_{1}\right)},\theta_{1}\in\left[-\pi,\pi\right]\right\} ,\label{eq:base-manifold}
\end{equation}
which is illustrated by the gray surface in Fig. \ref{fig:SGBZ-formulation}(f).
Due to the periodicity of $\mu_{2,0}$, $X(E,\mu_{1})$ has the topology
of a torus. We can use the arguments $\theta_{1}\equiv\text{Arg}\beta_{1}$
and $\theta_{2}\equiv\text{Arg}\beta_{2}$ as a global coordinate
system for $X(E,\mu_{1})$.

On the base manifold, the restriction of $f$ to $X(E,\mu_{1})$ reads,
\begin{equation}
f|_{X}\left(\theta_{1},\theta_{2};E,\mu_{1}\right)\equiv f\left(E,\rme^{\mu_{1}+\rmi\theta_{1}},\rme^{\mu_{2,0}(\theta_{1})+\rmi\theta_{2}}\right).
\end{equation}
By the definition of $X(E,\mu_{1})$, $f|_{X}$ vanishes only at the
PMGBZ pairs (green and red dots marked by $\text{i}_{\pm}$ and $\text{ii}_{\pm}$).
We consider the winding number around loops with constant $\theta_{2}$,
i.e., 
\begin{equation}
w_{1}\left(\theta_{2};E,\mu_{1}\right)\equiv\int_{-\pi}^{\pi}\frac{\rmd\theta_{1}}{2\pi\rmi}\frac{\partial\ln\left[f|_{X}\left(\theta_{1},\theta_{2};E,\mu_{1}\right)\right]}{\partial\theta_{1}},
\end{equation}
illustrated by the orange curves in Fig. \ref{fig:SGBZ-formulation}(f).
Then, we define the strip winding number as, 
\begin{equation}
W\left(E,\mu_{1}\right)\equiv\int_{-\pi}^{\pi}\frac{\rmd\theta_{2}}{2\pi}w_{1}\left(\theta_{2};E,\mu_{1}\right).\label{eq:def-strip-winding}
\end{equation}

The strip winding number can be simplified by the topological invariance
of $w_{1}$. In general, $f|_{X}$ is a continuous map from $X(E,\mu_{1})\backslash\text{PMGBZ}(E)$
to $\mathbb{C}\backslash\{0\}$, which induces a homomorphism between
the fundamental groups $w_{\text{loop}}:\pi_{1}\left(X(E,\mu_{1})\backslash\text{PMGBZ}(E)\right)\rightarrow\pi_{1}\left(\mathbb{C}\backslash\{0\}\right)=\mathbb{Z}$,
representing the winding number of a closed loop in $X(E,\mu_{1})\backslash\text{PMGBZ}(E)$
\cite{hatcherAlgebraicTopology2001}. For two closed loops $\ell_{1}$
and $\ell_{2}$ in $X(E,\mu_{1})\backslash\text{PMGBZ}(E)$, the homomorphism
implies that $w_{\text{loop}}\left(\ell_{1}\circ\ell_{2}\right)=w_{\text{loop}}\left(\ell_{1}\right)+w_{\text{loop}}\left(\ell_{2}\right)$.
Moreover, since $\pi_{1}\left(\mathbb{C}\backslash\{0\}\right)$ is
abelian, $w_{\text{loop}}$ also defines a homomorphism between the
homology groups $H_{1}\left(X(E,\mu_{1})\backslash\text{PMGBZ}(E)\right)$
and $H_{1}\left(\mathbb{C}\backslash\{0\}\right)=\mathbb{Z}$.

The winding number $w_{1}(\theta_{2};E,\mu_{1})$ equals $w_{\text{loop}}(\ell_{\theta_{2}})$,
where $\ell_{\theta_{2}}$ is the loop at constant $\theta_{2}$ in
$X(E,\mu_{1})$. Due to topological invariance, $w_{1}(\theta_{2})$
remains constant when $\ell_{\theta_{2}}$ does not cross any PMGBZ
points. However, if $\ell_{\theta_{2}}$ passes through a PMGBZ point,
such as the red and green dots in Fig. \ref{fig:SGBZ-formulation}(f),
$w_{1}(\theta_{2})$ changes, and the increment or decrement depends
on the homology difference between the winding loops on the two sides
of the PMGBZ point. As shown in Fig. \ref{fig:SGBZ-formulation}(f),
when $\theta_{2}$ increases across a PMGBZ point, the change in $w_{1}(\theta_{2})$
equals the winding number of $f|_{X}$ around an infinitesimal loop
indicated by the green and red circles. We define the winding numbers
around these infinitesimal loops as the topological charges of the
PMGBZ points. To compute these topological charges, we topologically
deform the infinitesimal loops into horizontal loops parallel to the
$\beta_{2}$ complex plane, illustrated as loop $\ell_{0}$ in Fig.
\ref{fig:SGBZ-formulation}(f). In general, the loop $\ell_{0}$ has
the following parametric form, 
\begin{equation}
\ell_{0}:\begin{cases}
\theta_{1}(t)=\theta_{1,0},\\
\beta_{2}(t)=\beta_{2}^{(j)}\left(\theta_{1,0};E,\mu_{1}\right)+\epsilon\rme^{\rmi t},
\end{cases},t\in\left[0,2\pi\right],
\end{equation}
where $\beta_{2}^{(j)}(E,\rme^{\mu_{1}+\rmi\theta_{1,0}}),j=M_{2},M_{2}+1$
is the $\beta_{2}$-zero of $f(E,\beta_{1},\beta_{2})$. The winding
number around $\ell_{0}$ is given by, 
\begin{align}
w_{\text{loop}}\left(\ell_{0}\right) & =\int_{0}^{2\pi}\frac{\rmd t}{2\pi\rmi}\frac{\partial\ln f\left(E,\rme^{\mu_{1}+\rmi\theta_{1}(t)},\beta_{2}(t)\right)}{\partial t}.
\end{align}
For infinitesimal $\epsilon$, the characteristic polynomial is given
by, 
\begin{equation}
f\left(E,\rme^{\mu_{1}+\rmi\theta_{1}(t)},\beta_{2}(t)\right)=\epsilon\frac{\partial f}{\partial\beta_{2}}|_{\left(\theta_{1,0},\beta_{2}^{(j)}\right)}\rme^{\rmi t}+O\left(\epsilon^{2}\right),
\end{equation}
where $\partial f/\partial\beta_{2}$ is evaluated at $\beta_{1}=\rme^{\mu_{1}+\rmi\theta_{1,0}}$
and $\beta_{2}=\beta_{2}^{(j)}(\theta_{1,0};E,\mu_{1})$. Therefore,
the winding number around $\ell_{0}$ reads, 
\begin{align}
w_{\text{loop}}\left(\ell_{0}\right) & =\int_{0}^{2\pi}\frac{\rmd t}{2\pi\rmi}\frac{\partial\left[\ln\left(\epsilon\frac{\partial f}{\partial\beta_{2}}|_{\left(\theta_{1,0},\beta_{2}^{(j)}\right)}+O\left(\epsilon^{2}\right)\right)+\rmi t\right]}{\partial t},\nonumber \\
 & =1.
\end{align}
According to Fig. \ref{fig:SGBZ-formulation}(f), the infinitesimal
loops around the PMGBZ points are either homologous to $\ell_{0}$
or homologous to its inverse. As $\theta_{1}$ increases, if the zeros
of $f$ (blue and cyan curves) move outward from $X(E,\mu_{1})$ at
the PMGBZ point, the infinitesimal loop is homologous to $\ell_{0}$,
and the PMGBZ point has a topological charge of $+1$ (shown as $\text{i}_{+}$
and $\text{ii}_{+}$). Otherwise, as shown for $\text{i}_{-}$ and
$\text{ii}_{-}$, the topological charge is $-1$.

With the topological charges of the PMGBZ points, the expression for
$W(E,\mu_{1})$ can be simplified. As illustrated in Fig. \ref{fig:SGBZ-formulation}(g),
on the unfolded view of $X(E,\mu_{1})$, we select an arbitrary winding
loop $\theta_{2}=\theta_{2,0}$ and compute its winding number $w_{1}(\theta_{2,0};E,\mu_{1})=w_{0}$.
Then, the value of $w_{1}(\theta_{2};E,\mu_{1})$ for any $\theta_{2}\in[-\pi,\pi]$
is determined based on the topological charges, as indicated by the
colored stripes and the labels at the bottom of each stripe. For each
pair of PMGBZ points, labeled $\text{i}_{\pm}$ and $\text{ii}_{\pm}$
in Fig. \ref{fig:SGBZ-formulation}(g), we define $\phi_{j}$ as the
relative phase of the segment that does not intersect the winding
loop $\theta_{2}=\theta_{2,0}$. Then, the strip winding number is
given by, 
\begin{equation}
W\left(E,\mu_{1}\right)=w_{0}+\sum_{j}\frac{\left(-1\right)^{\tau_{j}}}{2\pi}\phi_{j},
\end{equation}
where the summation is over all PMGBZ pairs on $X(E,\mu_{1})$, and
$\tau_{j}=0,1$ is determined by the topological charge of the PMGBZ
pair. If the arrow of $\phi_{j}$ {[}red and green arrows in Fig.
\ref{fig:SGBZ-formulation}(g){]} starts from a positive charge, then
$\tau_{j}=0$; otherwise, $\tau_{j}=1$.

When $\mu_{1}$ increases by a small value, the increment of the winding
number reads, 
\begin{equation}
W\left(E,\mu_{1,b}\right)-W\left(E,\mu_{1,a}\right)=\sum_{j}\frac{\left(-1\right)^{\tau_{j}}}{2\pi}\left(\phi_{j,b}-\phi_{j,a}\right),
\end{equation}
where $\mu_{1,a}$ and $\mu_{1,b}$ ($\mu_{1,b}>\mu_{1,a}$) are two
nearby values of $\mu_{1}$, and $\phi_{j,a}$ and $\phi_{j,b}$ are
the values of $\phi_{j}$ at $\mu_{1}=\mu_{1,a}$ and $\mu_{1}=\mu_{1,b}$,
respectively. Next, we will prove that the sign of $\phi_{j,b}-\phi_{j,a}$
is related to the topological charge. In general, consider the $\beta_{2}$-zero
of $f(E,\beta_{1},\beta_{2})$ as a function of $\beta_{1}$. Because
$f(E,\beta_{1},\beta_{2})$ is a holomorphic function on $\mathbb{C}\setminus\{0\}$
for both $\beta_{1}$ and $\beta_{2}$, the function $\beta_{2}(E,\beta_{1})$
is locally analytic, as is $\ln\beta_{2}$ as a function of $\ln\beta_{1}$.
According to the Cauchy-Riemann equations, the following relations
hold, 
\begin{equation}
\frac{\partial\mu_{2}}{\partial\mu_{1}}=\frac{\partial\theta_{2}}{\partial\theta_{1}},\quad\frac{\partial\theta_{2}}{\partial\mu_{1}}=-\frac{\partial\mu_{2}}{\partial\theta_{1}}.\label{eq:Cauchy-Riemann}
\end{equation}
For a pair of PMGBZ points, such as $\text{i}_{\pm}$ or $\text{ii}_{\pm}$
in Fig. \ref{fig:SGBZ-formulation}(f), the value of $\partial\mu_{2}/\partial\theta_{1}$
at the positive charge is larger than that at the negative charge.
Consequently, the difference between the values of $\theta_{2}$ at
the positive and negative charges decreases as $\mu_{1}$ increases.
Therefore, as illustrated in Fig. \ref{fig:SGBZ-formulation}(g),
for the relative phase $\phi_{j}$ starting from the positive charge,
$\phi_{j}$ increases when $\mu_{1}$ increases, and vice versa. By
the definition of $\tau_{j}$, the increment of $W(E,\mu_{1})$ is
given by, 
\begin{equation}
W\left(E,\mu_{1,b}\right)-W\left(E,\mu_{1,a}\right)=\sum_{j}\frac{\left|\phi_{j,b}-\phi_{j,a}\right|}{2\pi}.
\end{equation}
Compared with Eq. (\ref{eq:increment-Wq1D}), the increment of $W(E,\mu_{1})$
is equal to the increment of $W_{\text{q1D}}(E,\mu_{1},L_{2})/L_{2}$
for every small increment, except for a remainder of order $O(1/L_{2})$.

Since the increment of $W$ equals $W_{\text{q1D}}/L_{2}$ as $L_{2}\rightarrow\infty$,
we only need to verify whether the two quantities are equal for some
specific value of $\mu_{1}$. Here, we consider the case when $\mu_{1}\rightarrow\pm\infty$.
On one hand, when $\mu_{1}\rightarrow-\infty$, i.e., $|\beta_{1}|\rightarrow0$,
the term with $\beta_{1}^{-M}$ dominates in $F_{L_{2}}(E,\beta_{1})$,
so that $W_{\text{q1D}}(E,-\infty,L_{2})=-M$. Similarly, $W_{\text{q1D}}(E,+\infty,L_{2})=N$.
On the other hand, the term containing $\beta_{1}^{-M_{1}}$ dominates
in $f(E,\beta_{1},\beta_{2})$ when $\mu_{1}\rightarrow-\infty$,
so that $w_{1}(\theta_{2};E,-\infty)=-M_{1}$ for arbitrary $\theta_{2}$,
and $W(E,-\infty)=-M_{1}$. Because the quasi-1D supercell consists
of $L_{2}$ copies of the unit cell, the degrees of $F_{L_{2}}(E,\beta_{1})$
are related to the degrees of $f(E,\beta_{1},\beta_{2})$ by, 
\begin{align}
M & =L_{2}M_{1}+O(1),\label{eq:relation-M}\\
N & =L_{2}N_{1}+O(1).\label{eq:relation-N}
\end{align}
Equations (\ref{eq:relation-M}) and (\ref{eq:relation-N}) can also
be rigorously proved using the mathematical forms of $h(\beta_{1},\beta_{2})$
and $\mathcal{H}_{L_{2}}(\beta_{1})$ (see Sec. \ref{subsec:rigor-proof-degree}
for details). Therefore, the relation $W(E,\pm\infty)=W_{\text{q1D}}(E,\pm\infty,L_{2})/L_{2}+O(1/L_{2})$
holds. Consequently, we have proved the relation, 
\begin{equation}
W\left(E,\mu_{1}\right)=W_{\text{q1D}}\left(E,\mu_{1},L_{2}\right)/L_{2}+O\left(1/L_{2}\right).\label{eq:relation-winding}
\end{equation}

\begin{figure}
\centering

\includegraphics{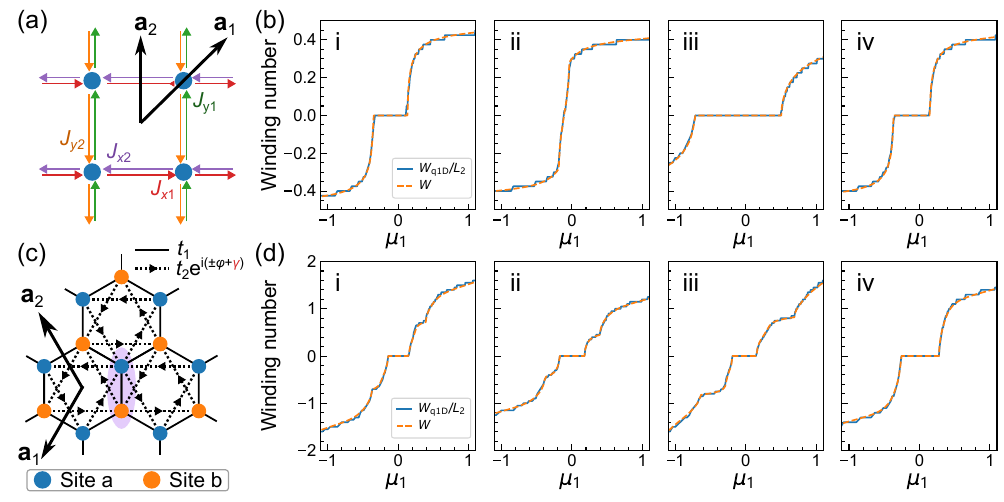}

\caption{Comparison of $W(E,\mu_{1})$ and $W_{\text{q1D}}(E,\mu_{1},L_{2})/L_{2}$
in the 2D Hatano-Nelson model and the non-Hermitian Haldane model.
(a) Schematic of the 2D Hatano-Nelson model with parameters $J_{x1}=1+\protect\rmi$,
$J_{x2}=1.5+1.2\protect\rmi$, $J_{y1}=-1+\protect\rmi$, and $J_{y2}=-1.2-0.5\protect\rmi$.
The major (minor) axis is indicated by the lattice vector $\mathbf{a}_{1}$
($\mathbf{a}_{2}$). (b) Numerical results for $W_{\text{q1D}}(E,\mu_{1},L_{2})/L_{2}$
(blue solid lines) and $W(E,\mu_{1})$ (orange dashed lines) in the
2D Hatano-Nelson model along the major axis $\mathbf{a}_{1}$. The
system width is $L_{2}=40$, and the reference energies $E$ are (i)
$1.00296-0.21641\protect\rmi$, (ii) $1.55832+0.91741\protect\rmi$,
(iii) $-2.57608+1.13451\protect\rmi$, and (iv) $-1.57168+0.02125\protect\rmi$.
(c) Schematic of the non-Hermitian Haldane model with parameters $t_{1}=0.70502$,
$t_{2}=-1.32760$, $\gamma=2.15618$, $\varphi=0.05877$, and $m=-0.64569$.
The unit cell is highlighted by a purple ellipse, and the major (minor)
axis is denoted by the lattice vector $\mathbf{a}_{1}$ ($\mathbf{a}_{2}$).
(d) Numerical results for $W_{\text{q1D}}(E,\mu_{1},L_{2})/L_{2}$
(blue solid lines) and $W(E,\mu_{1})$ (orange dashed lines) in the
non-Hermitian Haldane model along the major axis $\mathbf{a}_{1}$.
The width is $L_{2}=20$ (corresponding to 40 sites in the supercell),
and the reference energies $E$ are (i) $0.01162+0.68736\protect\rmi$,
(ii) $0.54100-1.99811\protect\rmi$, (iii) $0.59046+1.89903\protect\rmi$,
and (iv) $0.10591-0.34594\protect\rmi$.}
\label{fig:numerical-winding}
\end{figure}

The relation between $W(E,\mu_{1})$ and $W_{\text{q1D}}(E,\mu_{1},L_{2})$
is also verified numerically. Figure \ref{fig:numerical-winding}(a)
shows the 2D HN model with complex coupling coefficients, defined
as,
\begin{equation}
h\left(\beta_{x},\beta_{y}\right)=J_{x1}\beta_{x}^{-1}+J_{x2}\beta_{x}+J_{y1}\beta_{y}^{-1}+J_{y2}\beta_{y},
\end{equation}
which is the same as the example in the main text. In the numerical
calculation, the coefficients are $J_{x1}=1+\rmi$, $J_{x2}=1.5+1.2\rmi$,
$J_{y1}=-1+\rmi$, and $J_{y2}=-1.2-0.5\rmi$. The lattice vector
$\mathbf{a}_{1}=(1,1)$ is selected as the main axis and $\mathbf{a}_{2}=(0,1)$
as the minor axis. To calculate the quasi-1D winding number, the supercell
is constructed by selecting successive $L_{2}$ unit cells along the
minor axis $\mathbf{a}_{2}$. Figure \ref{fig:numerical-winding}(b)
illustrates the numerical results of $W(E,\mu_{1})$ and $W_{\text{q1D}}(E,\mu_{1},L_{2})/L_{2}$
for the 2D HN model with randomly generated reference energies. In
the numerical calculation, $L_{2}$ is set to $40$, and the reference
energies are (i) $1.00296-0.21641\rmi$, (ii) $1.55832+0.91741\rmi$,
(iii) $-2.57608+1.13451\rmi$, and (iv) $-1.57168+0.02125\rmi$. As
shown in Fig. \ref{fig:numerical-winding}(b), the curves of $W(E,\mu_{1})$
versus $\mu_{1}$ (orange dashed curves) are piecewise smooth, while
the curves of $W_{\text{q1D}}(E,\mu_{1},L_{2})/L_{2}$ (blue solid
curves) exhibit plateaus due to the finite $L_{2}$. For all four
samples in Fig. \ref{fig:numerical-winding}(b), $W_{\text{q1D}}(E,\mu_{1},L_{2})/L_{2}$
agrees well with $W(E,\mu_{1})$.

To demonstrate the generality of our conclusion, we also compare $W(E,\mu_{1})$
and $W_{\text{q1D}}(E,\mu_{1},L_{2})$ in a non-Hermitian version
of the Haldane model. As shown in Fig. \ref{fig:numerical-winding}(c),
the nearest-neighbor coupling is $t_{1}\in\mathbb{R}$, and the next-nearest-neighbor
coupling has a nonreciprocal phase; that is, the coupling coefficient
is $t_{2}\rme^{\rmi\left(\varphi+\gamma\right)}$ along the direction
of the arrow and $t_{2}\rme^{-\rmi\left(\varphi-\gamma\right)}$ against
the arrow. Under the basis $\{\mathbf{a}_{1},\mathbf{a}_{2}\}$ shown
in Fig. \ref{fig:numerical-winding}(c), the Hamiltonian reads, 
\begin{align}
H= & m\sum_{r_{1},r_{2}}\left(c_{r_{1},r_{2},\mathrm{a}}^{\dagger}c_{r_{1},r_{2},\mathrm{a}}-c_{r_{1},r_{2},\mathrm{b}}^{\dagger}c_{r_{1},r_{2},\mathrm{b}}\right)+\nonumber \\
 & +t_{1}\sum_{r_{1},r_{2}}\left(c_{r_{1},r_{2},\mathrm{b}}^{\dagger}c_{r_{1},r_{2},\mathrm{a}}+c_{r_{1},r_{2}+1,\mathrm{b}}^{\dagger}c_{r_{1},r_{2},\mathrm{a}}+c_{r_{1}-1,r_{2},\mathrm{b}}^{\dagger}c_{r_{1},r_{2},\mathrm{a}}+\text{h.c.}\right)+\nonumber \\
 & +t_{2}\rme^{\rmi\gamma}\sum_{r_{1},r_{2}}\left[\rme^{\rmi\varphi}\left(c_{r_{1}-1,r_{2},\mathrm{a}}^{\dagger}c_{r_{1},r_{2},\mathrm{a}}+c_{r_{1},r_{2}-1,\mathrm{a}}^{\dagger}c_{r_{1},r_{2},\mathrm{a}}+c_{r_{1}+1,r_{2}+1,\mathrm{a}}^{\dagger}c_{r_{1},r_{2},\mathrm{a}}\right)+\right.\nonumber \\
 & \left.+\rme^{-\rmi\varphi}\left(c_{r_{1}-1,r_{2},\mathrm{b}}^{\dagger}c_{r_{1},r_{2},\mathrm{b}}+c_{r_{1},r_{2}-1,\mathrm{b}}^{\dagger}c_{r_{1},r_{2},\mathrm{b}}+c_{r_{1}+1,r_{2}+1,\mathrm{b}}^{\dagger}c_{r_{1},r_{2},\mathrm{b}}\right)+\text{h.c.}\right],
\end{align}
where $c_{r_{1},r_{2},\mu}$, with $r_{1}\in\mathbb{Z}$, $r_{2}\in\mathbb{Z}$,
and $\mu=\mathrm{a},\mathrm{b}$, is the annihilation operator at
sublattice $\mu$ in the unit cell located at coordinate $r_{1}\mathbf{a}_{1}+r_{2}\mathbf{a}_{2}$,
and $m\in\mathbb{R}$ is the detuning between site $\mathrm{a}$ and
site $\mathrm{b}$. In the numerical calculations, the parameters
$t_{1}$, $t_{2}$, $\gamma$, $\varphi$, and $m$ are randomly generated
as $t_{1}=0.70502$, $t_{2}=-1.32760$, $\gamma=2.15618$, $\varphi=0.05877$,
and $m=-0.64569$. As shown in Fig. \ref{fig:numerical-winding}(d),
we compute $W(E,\mu_{1})$ and $W_{\text{q1D}}(E,\mu_{1},L_{2})$
as functions of $\mu_{1}$ along the $\mathbf{a}_{1}$ direction for
four random reference energies: (i) $0.01162+0.68736\rmi$, (ii) $0.54100-1.99811\rmi$,
(iii) $0.59046+1.89903\rmi$, and (iv) $0.10591-0.34594\rmi$. For
the quasi-1D winding number $W_{\text{q1D}}(E,\mu_{1},L_{2})$, $L_{2}$
is set to $20$ (corresponding to $40$ sites in the supercell). Similar
to the case of the 2D HN model, the curves of $W_{\text{q1D}}(E,\mu_{1},L_{2})/L_{2}$
align well with $W(E,\mu_{1})$.

Based on the above analysis, the strip winding number $W(E,\mu_{1})$
is exactly the limit of $W_{\text{q1D}}(E,\mu_{1},L_{2})/L_{2}$ as
$L_{2}\rightarrow\infty$. Therefore, we obtain the SGBZ constraint
by substituting Eq. (\ref{eq:relation-winding}) into Eq. (\ref{eq:QMGBZ-constraint}),
which reads, 
\begin{equation}
\begin{cases}
W\left(E,\mu_{1,<}\right)<0,\\
W\left(E,\mu_{1,>}\right)>0.
\end{cases}\label{eq:SGBZ-constraint}
\end{equation}

\subsection{SGBZ in higher dimensions}

\begin{figure}
\centering

\includegraphics{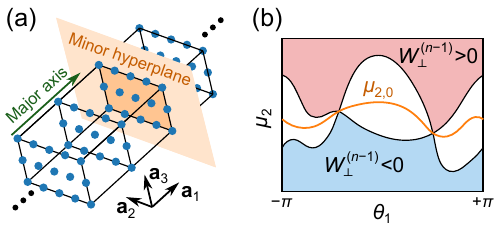}

\caption{SGBZ in higher dimensions. (a) Illustration of an $n$D strip geometry,
where the strip is extended along the major axis ($\mathbf{a}_{1}$)
and truncated in the minor hyperplane (spanned by $\mathbf{a}_{2},\mathbf{a}_{3},\dots$).
(b) Definition of the radius function $\mu_{2,0}(\theta_{1};E,\mu_{1})$
(orange curve), which is a periodic function in the region where $W_{\perp}^{(n-1)}(E,\protect\rme^{\mu_{1}+\protect\rmi\theta_{1}},\mu_{2})=0$.}
\label{fig:high-dim-SGBZ}
\end{figure}

The SGBZ formulation, which is based on the concept of sequential
thermodynamic limits, can be naturally extended to arbitrary $n$D
non-Hermitian lattices. Similar to 2D lattices, we construct the $n$D
SGBZ using a quasi-1D strip geometry. As illustrated in Fig. \ref{fig:high-dim-SGBZ}(a),
the strip geometry is extended along a major axis $\mathbf{a}_{1}$
and truncated in the minor hyperplane spanned by the remaining axes
($\mathbf{a}_{2},\mathbf{a}_{3},\dots,\mathbf{a}_{n}$). We denote
the momentum-space characteristic polynomial as $f^{(n)}(E,\beta_{1},\beta_{2},\dots,\beta_{n})$,
where $\beta_{1},\dots,\beta_{n}$ are the complex momenta conjugate
to $\mathbf{a}_{1},\dots,\mathbf{a}_{n}$, respectively.

The strip winding number, denoted as $W^{(n)}(E,\mu_{1})$, can be
defined inductively. First, for 1D lattices, we define, 
\begin{align}
X^{(1)}\left(E,\mu_{1}\right) & \equiv\left\{ \beta_{1}\in\mathbb{C}\mid\ln\left|\beta_{1}\right|=\mu_{1}\right\} ,\\
W^{(1)}\left(E,\mu_{1}\right) & \equiv\int_{-\pi}^{\pi}\frac{\rmd\theta_{1}}{2\pi\rmi}\frac{\partial\ln\left[f^{(1)}|_{X}\left(\theta_{1};E,\mu_{1}\right)\right]}{\partial\theta_{1}}=\int_{-\pi}^{\pi}\frac{\rmd\theta_{1}}{2\pi\rmi}\frac{\partial\ln\left[f^{(1)}\left(E,\rme^{\mu_{1}+\rmi\theta_{1}}\right)\right]}{\partial\theta_{1}}.
\end{align}
Then, supposing that the base manifold $X^{(n-1)}$ and strip winding
number $W^{(n-1)}$ are well defined for an arbitrary $(n-1)$D momentum-space
characteristic polynomial $f^{(n-1)}(E,\beta_{2},\beta_{3},\dots,\beta_{n})$.
For $f^{(n)}$, because each $\beta_{1}$ specifies a parametric $(n-1)$D
lattice by taking $\beta_{1}$ as a parameter, we define the base
manifold and strip winding number corresponding to the parametric
$(n-1)$D lattice as the ``transverse'' base manifold and strip
winding number, denoted as $X_{\perp}^{(n-1)}(E,\beta_{1},\mu_{2})$
and $W_{\perp}^{(n-1)}(E,\beta_{1},\mu_{2})$, respectively. To define
the base manifold $X^{(n)}(E,r_{1})$, as illustrated in Fig. \ref{fig:high-dim-SGBZ}(b),
we define the radius function $\mu_{2,0}(\theta_{1};E,\mu_{1})$ as
the periodic function of $\theta_{1}$ located in the region where
$W_{\perp}^{(n-1)}(E,\rme^{\mu_{1}+\rmi\theta_{1}},r_{2})=0$. Then,
the base manifold is defined as, 
\begin{align}
X^{(n)}\left(E,\mu_{1}\right)= & \left\{ \boldsymbol{\beta}\in\mathbb{C}^{n}\mid\beta_{1}=\rme^{\mu_{1}+\rmi\theta_{1}},\theta_{1}\in\left[-\pi,\pi\right],(\beta_{2},\dots,\beta_{n})\in X_{\perp}^{(n-1)}\left(\theta_{1};E,\mu_{1}\right)\right\} ,\label{eq:n-base-manifold}
\end{align}
where, 
\begin{equation}
X_{\perp}^{(n-1)}\left(\theta_{1};E,\mu_{1}\right)\equiv X_{\perp}^{(n-1)}\left(E,\rme^{\mu_{1}+\rmi\theta_{1}},\mu_{2,0}(\theta_{1};E,\mu_{1})\right).
\end{equation}
When $n=2$, Eq. (\ref{eq:n-base-manifold}) is equivalent to the
2D base manifold defined in Eq. (\ref{eq:base-manifold}).

Using mathematical induction, we will show that the base manifold
$X^{(n)}(E,r_{1})$ is homeomorphic to an $n$D torus, with the arguments
$\boldsymbol{\theta}\equiv(\theta_{1},\theta_{2},\dots,\theta_{n})\equiv(\text{Arg}\beta_{1},\text{Arg}\beta_{2},\dots,\text{Arg}\beta_{n})$
as global coordinates. First, $X^{(1)}(E,\mu_{1})$ is a 1D torus
with coordinate $\theta_{1}\equiv\text{Arg}\beta_{1}$. Next, assuming
that the points $(\beta_{2},\beta_{3},\dots,\beta_{n})\in X^{(n-1)}(E,\mu_{2})$
are uniquely determined by $(\theta_{2},\theta_{3},\dots,\theta_{n})$
and $\mu_{2}$, and since $\mu_{2,0}(\theta_{1};E,\mu_{1})$ is determined
by $\theta_{1}$ for fixed $E$ and $\mu_{1}$, the points $\boldsymbol{\beta}_{\perp}\in X_{\perp}^{(n-1)}(\theta_{1};E,\mu_{1})$
are functions of $\boldsymbol{\theta}$. We denote this function as
$\boldsymbol{\beta}_{\perp}(\boldsymbol{\theta})$. According to Eq.
(\ref{eq:n-base-manifold}), the points $\boldsymbol{\beta}=(\rme^{\mu_{1}+\rmi\theta_{1}},\boldsymbol{\beta}_{\perp}(\boldsymbol{\theta}))\in X^{(n)}(E,\mu_{1})$
are uniquely determined by $\boldsymbol{\theta}$. Therefore, the
$n$D base manifold has the topology of an $n$D torus.

With the above preparations, the $n$D strip winding number can be
defined as, 
\begin{equation}
W^{(n)}\left(E,\mu_{1}\right)=\int_{T^{n-1}}\frac{\rmd^{n-1}\boldsymbol{\theta}_{\perp}}{(2\pi)^{n-1}}w_{1}\left(\boldsymbol{\theta}_{\perp};E,\mu_{1}\right),
\end{equation}
where $T^{n-1}$ denotes the $(n-1)$D torus spanned by the transverse
angular coordinates $\boldsymbol{\theta}_{\perp}=(\theta_{2},\theta_{3},\dots,\theta_{n})$.
The winding number $w_{1}(\boldsymbol{\theta}_{\perp};E,\mu_{1})$
is defined as, 
\begin{equation}
w_{1}\left(\boldsymbol{\theta}_{\perp};E,\mu_{1}\right)=\int_{-\pi}^{\pi}\frac{\rmd\theta_{1}}{2\pi\rmi}\frac{\partial\ln\left[f^{(n)}|_{X^{(n)}}\left(\boldsymbol{\theta};E,\mu_{1}\right)\right]}{\partial\theta_{1}},\label{eq:w-loop-nD}
\end{equation}
where $f^{(n)}|_{X^{(n)}}(\boldsymbol{\theta};E,\mu_{1})$ denotes
the restriction of $f^{(n)}(E,\boldsymbol{\beta})$ to $X^{(n)}(E,\mu_{1})$,
with $\boldsymbol{\beta}$ determined by the coordinate $\boldsymbol{\theta}$.
Thus, by applying the SGBZ constraint {[}Eq. (\ref{eq:SGBZ-constraint}){]}
to $W^{(n)}(E,\mu_{1})$, the spectrum $E$ and the corresponding
critical value $\mu_{1,0}$ are determined, and the $n$D SGBZ is
defined as the zeros of $f^{(n)}|_{X^{(n)}}$ on $X^{(n)}(E,\mu_{1,0})$.

In Sec. \ref{subsec:3D-HN-model}, we will illustrate our scheme with
the three-dimensional Hatano-Nelson model as an example.

\subsection{Rigorous proof of Eqs. (\ref{eq:relation-M}) and (\ref{eq:relation-N})}\label{subsec:rigor-proof-degree}

First, we expand the 2D non-Bloch Hamiltonian $h(\beta_{1},\beta_{2})$
as a polynomial in $\beta_{2}$, that is, 
\begin{equation}
h\left(\beta_{1},\beta_{2}\right)=\sum_{j=-M_{2}}^{N_{2}}h^{(j)}\left(\beta_{1}\right)\beta_{2}^{j},\label{seq:h-in-polynomial}
\end{equation}
where $h^{(j)}(\beta_{1})$ is an $m\times m$ matrix. Then, the hybrid
non-Bloch Hamiltonian can be expanded into a block Toeplitz form under
the single-particle basis, i.e., 
\begin{equation}
\mathcal{H}_{L_{2}}\left(\beta_{1}\right)=\underbrace{\left(\begin{matrix}h^{(0)} & h^{(-1)} & \cdots & h^{(-M_{2})}\\
h^{(1)} & h^{(0)} & h^{(-1)} & \cdots & \ddots\\
\vdots & h^{(1)} & h^{(0)} & h^{(-1)} & \cdots & h^{(-M_{2})}\\
h^{(N_{2})} & \vdots & h^{(1)} & h^{(0)} & \ddots & \vdots\\
 & \ddots & \vdots & \ddots & \ddots & h^{(-1)}\\
 &  & h^{(N_{2})} & \cdots & h^{(1)} & h^{(0)}
\end{matrix}\right)}_{L_{2}\text{ blocks}}.\label{seq:quasi-1D-Toeplitz}
\end{equation}

We first consider the lowest degrees; the case of the highest degrees
can be proved by the same reasoning. Note that the matrix elements
of $h^{(j)}(\beta_{1})$ are all Laurent polynomials in $\beta_{1}$.
Suppose the lowest degree of $\beta_{1}$ in $h_{\mu\nu}(\beta_{1},\beta_{2})$
is $-M_{\mu\nu}$, and the degree of $\beta_{2}$ in the term with
the lowest degree of $\beta_{1}$ is $t_{\mu\nu}$. Then, the elements
of $f(E,\beta_{1},\beta_{2})$ can be expressed as the leading term
$a_{ij}\beta_{1}^{-M_{ij}}\beta_{2}^{t_{ij}}$ plus a remainder in
which the degrees of $\beta_{1}$ are greater than $-M_{ij}$, which
reads,
\begin{align}
f\left(E,\beta_{1},\beta_{2}\right) & =\det\left[E-h\left(\beta_{1},\beta_{2}\right)\right],\nonumber \\
 & =\begin{vmatrix}a_{11}\beta_{1}^{-M_{11}}\beta_{2}^{t_{11}}+\cdots & a_{12}\beta_{1}^{-M_{12}}\beta_{2}^{t_{12}}+\cdots & \cdots & a_{1m}\beta_{1}^{-M_{1m}}\beta_{2}^{t_{1m}}+\cdots\\
a_{21}\beta_{1}^{-M_{21}}\beta_{2}^{t_{21}}+\cdots & a_{22}\beta_{1}^{-M_{22}}\beta_{2}^{t_{22}}+\cdots &  & \vdots\\
\vdots &  & \ddots\\
a_{m1}\beta_{1}^{-M_{m1}}\beta_{2}^{t_{m1}}+\cdots & \cdots &  & a_{mm}\beta_{1}^{-M_{mm}}\beta_{2}^{t_{mm}}+\cdots
\end{vmatrix},\label{seq:expansion-f}
\end{align}
where $a_{ij}\in\mathbb{C}$ are the coefficients. In the determinant
of Eq. (\ref{seq:expansion-f}), only the leading terms contribute
to the lowest degree of $f(E,\beta_{1},\beta_{2})$. Therefore, the
term in $f(E,\beta_{1},\beta_{2})$ with the lowest degree in $\beta_{1}$
can be constructed by the following steps: 
\begin{enumerate}
\item Find $i_{\max}$ and $j_{\max}$ such that $M_{i_{\max}j_{\max}}=\max_{i,j}\{M_{ij}\}$.
\item Obtain the $(i_{\max},j_{\max})$-cofactor of $f(E,\beta_{1},\beta_{2})$,
denoted as $f_{1}(E,\beta_{1},\beta_{2})$, and then factorize $f(E,\beta_{1},\beta_{2})$
as follows: 
\begin{equation}
f\left(E,\beta_{1},\beta_{2}\right)=a_{i_{\max}j_{\max}}\beta_{1}^{-M_{i_{\max}j_{\max}}}\beta_{2}^{t_{i_{\max}j_{\max}}}f_{1}\left(E,\beta_{1},\beta_{2}\right)+\dots
\end{equation}
\item Substitute the determinant $f$ with the cofactor $f_{1}$, and repeat
the two steps above to obtain $f_{2}$, $f_{3}$, $\dots$, until
the order of the cofactor is reduced to $1$.
\end{enumerate}
After this procedure, we obtain a sequence of indices $(i_{\max}^{(1)},j_{\max}^{(1)}),(i_{\max}^{(2)},j_{\max}^{(2)}),\dots,(i_{\max}^{(m)},j_{\max}^{(m)})$.
By construction, $i_{\max}^{(1)},i_{\max}^{(2)},\dots,i_{\max}^{(m)}$
and $j_{\max}^{(1)},j_{\max}^{(2)},\dots,j_{\max}^{(m)}$ are permutations
of $1,2,\dots,m$. Therefore, we can reorder $M_{i_{\max}^{(1)}j_{\max}^{(1)}},M_{i_{\max}^{(2)}j_{\max}^{(2)}},\dots,M_{i_{\max}^{(m)}j_{\max}^{(m)}}$
as $M_{1\nu_{1}},M_{2\nu_{2}},\dots,M_{m\nu_{m}}$. Then, the minimum
negative degree of $f(E,\beta_{1},\beta_{2})$ is given by, 
\begin{equation}
M_{1}=\sum_{j=1}^{m}M_{j\nu_{j}}.
\end{equation}

Returning to the hybrid non-Bloch Hamiltonian $\mathcal{H}_{L_{2}}(\beta_{1})$,
Eq. (\ref{seq:h-in-polynomial}) indicates that the term $a_{j\nu_{j}}\beta_{1}^{-M_{j\nu_{j}}}$
appears in the matrix element $h_{j\nu_{j}}^{(t_{j\nu_{j}})}(\beta_{1})$.
Consequently, in each row of the blocks in Eq. (\ref{seq:quasi-1D-Toeplitz}),
we can select the terms $a_{1\nu_{1}}\beta_{1}^{-M_{1\nu_{1}}},a_{2\nu_{2}}\beta_{1}^{-M_{2\nu_{2}}},\dots,a_{m\nu_{m}}\beta_{1}^{-M_{m\nu_{m}}}$,
except for the first $N_{2}$ or last $M_{2}$ rows. For distinct
rows of the blocks, the column indices of the chosen terms do not
overlap, since $\nu_{1},\nu_{2},\dots,\nu_{m}$ is a permutation of
$1,2,\dots,m$. Next, examining the determinant $F_{L_{2}}(E,\beta_{1})\equiv\det[E-\mathcal{H}_{L_{2}}(\beta_{1})]$,
aside from the first $N_{2}$ and last $M_{2}$ rows of the blocks,
the lowest degree of $\beta_{1}$ contributed by each row is given
by, 
\[
\beta_{1}^{-M_{1\nu_{1}}}\beta_{1}^{-M_{2\nu_{2}}}\cdots\beta_{1}^{-M_{m\nu_{m}}}=\beta_{1}^{-M_{1}},
\]
and the total contribution of the $L_{2}-M_{2}-N_{2}$ rows is, 
\[
\beta_{1}^{-M_{1}\left(L_{2}-M_{2}-N_{2}\right)}=\beta_{1}^{-M_{1}L_{2}+O(1)}.
\]
Since the contributions of $\beta_{1}$ in the first $N_{2}$ and
last $M_{2}$ rows can alter the total degree of $\beta_{1}$ by only
a finite amount (i.e., independent of $L_{2}$), the lowest degree
of $F_{L_{2}}(E,\beta_{1})$ satisfies Eq. (\ref{eq:relation-M}).
Similarly, we can also verify Eq. (\ref{eq:relation-N}) using the
same reasoning.

\section{Details about selections of minor axes}

In the schematic diagram of Fig. \ref{fig:main-idea}, we claimed
that SGBZ is independent of the minor axes. In this section, we will
prove this observation.

\begin{figure}
\centering

\includegraphics{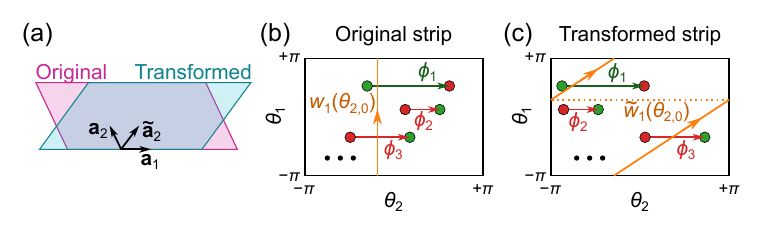}

\caption{Coordinate transformations of minor axes. (a) Illustration of the
transformation. (b) Unfolded base manifold $X(E,\mu_{1})$ in the
original strip. (c) Unfolded base manifold $\tilde{X}(E,\mu_{1})$
in the transformed strip.}
\label{fig:minor-transformation}
\end{figure}

As illustrated in Fig. \ref{fig:minor-transformation}(a), we consider
the coordinate transformation in the following form, 
\begin{equation}
\begin{pmatrix}\tilde{\mathbf{a}}_{1} & \tilde{\mathbf{a}}_{2}\end{pmatrix}=\begin{pmatrix}\mathbf{a}_{1} & \mathbf{a}_{2}\end{pmatrix}\begin{pmatrix}1 & \alpha\\
0 & 1
\end{pmatrix},
\end{equation}
where $\alpha$ is an arbitrary integer. Under the coordinate transformations,
the characteristic polynomials satisfy, 
\begin{equation}
f\left(E,\beta_{1},\beta_{2}\right)=\tilde{f}\left(E,\beta_{1},\beta_{1}^{\alpha}\beta_{2}\right),\label{eq:ChP-transformation-major-axis}
\end{equation}
where $f$ and $\tilde{f}$ are the characteristic polynomials under
the original and the transformed bases, respectively. Assuming that
the $\beta_{2}$-zeros of $f$ are $\beta_{2}^{(j)}(\theta_{1};E,\mu_{1})$,
$j=1,2,\dots,M_{2}+N_{2}$, ordered by $\left|\beta_{2}^{(j)}\right|\leq\left|\beta_{2}^{(k)}\right|$,
$\forall j<k$, in the transformed strip, the transformed minor-axis
momenta, 
\begin{equation}
\tilde{\beta}_{2}^{(j)}(\theta_{1};E,\mu_{1})=\beta_{1}^{\alpha}\beta_{2}^{(j)}(\theta_{1};E,\mu_{1}),
\end{equation}
are also zeros of $\tilde{f}\left(E,\beta_{1},\tilde{\beta}_{2}\right)$,
and the ordering $|\tilde{\beta}_{2}^{(j)}|\leq|\tilde{\beta}_{2}^{(k)}|$
for all $j<k$ holds.

For the base manifold $X(E,\mu_{1})$, by definition, the function
$\mu_{2,0}(\theta_{1};E,\mu_{1})$ satisfies $\ln\left|\beta_{2}^{(M_{2})}\right|\leq\mu_{2,0}\leq\left|\beta_{2}^{(M_{2}+1)}\right|$.
Then, the image of $X(E,\mu_{1})$ under the coordinate transformation,
i.e., 
\begin{equation}
\tilde{X}\left(E,\mu_{1}\right)=\left\{ \left(\rme^{\mu_{1}+\rmi\theta_{1}},\tilde{\beta}_{2}\right)\mid\ln\left|\tilde{\beta}_{2}\right|=\alpha\mu_{1}+\mu_{2,0}(\theta_{1};E,\mu_{1})\right\} ,
\end{equation}
is also a valid base manifold in the transformed strip. That is, the
function $\tilde{\mu}_{2,0}\left(\theta_{1};E,\mu_{1}\right)\equiv\alpha\mu_{1}+\mu_{2,0}(\theta_{1};E,\mu_{1})$
satisfies the relation $\ln|\tilde{\beta}_{2}^{(M_{2})}|\leq\tilde{\mu}_{2,0}\leq\ln|\tilde{\beta}_{2}^{(M_{2}+1)}|$.

For the strip winding number in the original and transformed strips,
denoted $W(E,r)$ and $\tilde{W}(E,r)$, we consider the transformation
of the winding loop of $w_{1}(\theta_{2};E,\mu_{1})$ in $X(E,\mu_{1})$
into the closed loop on $\tilde{X}(E,\mu_{1})$. As illustrated in
Fig. \ref{fig:minor-transformation}(b) and \ref{fig:minor-transformation}(c),
for the winding loop $\theta_{2}=\theta_{2,0}$ on $X$ with winding
number $w_{1}(\theta_{2,0};E,\mu_{1})$, the transformed winding loop
reads, 
\begin{equation}
\tilde{\theta}_{2,0}\left(\theta_{1}\right)=\alpha\theta_{1}+\theta_{2,0},\label{eq:transformed-theta2}
\end{equation}
and the characteristic polynomials satisfy, 
\begin{equation}
f|_{X}\left(\theta_{1},\theta_{2,0};E,\mu_{1}\right)=\tilde{f}|_{\tilde{X}}\left(\theta_{1},\tilde{\theta}_{2,0}\left(\theta_{1}\right);E,\mu_{1}\right),\label{eq:relation-ChP}
\end{equation}
for arbitrary $\theta_{1}\in\left[-\pi,\pi\right]$. Therefore, in
the transformed strip, the winding number $\tilde{w}_{1}(\theta_{2,0};E,\mu_{1})$
around the loop of Eq. (\ref{eq:transformed-theta2}) equals $w_{1}(\theta_{2,0};E,\mu_{1})$.
With an equivalent form of the strip winding number (see Sec. \ref{subsec:Equivalent-forms}),
$\tilde{W}(E,\mu_{1})$ is equal to the integral of $\tilde{w}_{1}(\theta_{2,0};E,\mu_{1})$,
and consequently equal to $W(E,\mu_{1})$ in the original strip.

Because the strip winding numbers $W(E,\mu_{1})$ and $\tilde{W}(E,\mu_{1})$
are equal for arbitrary $E$ and $\mu_{1}$, every SGBZ point in the
original strip is transformed into an SGBZ point in the transformed
strip. Therefore, the SGBZs with the same major axis and different
minor axes are compatible with each other.

\section{SGBZs of 2D and higher-dimensional Hatano-Nelson model}\label{sec:SGBZs-HN}

In this section, we discuss the details of the SGBZs for the 2D and
higher-dimensional Hatano-Nelson (HN) model. We first present equivalent
forms of the strip winding number and then calculate the SGBZs for
the 2D and 3D HN models.

\subsection{Equivalent forms of strip winding number}\label{subsec:Equivalent-forms}

In some cases, the winding number $w_{1}(\theta_{2};E,\mu_{1})$ is
difficult to calculate. Nevertheless, by virtue of the topological
invariance of the winding number, the strip winding number can also
be computed using the winding number around a loop with non-constant
$\theta_{2}$. In this section, we will show that the winding number
$w_{1}(\theta_{2};E,\mu_{1})$ can be replaced by $\tilde{w}_{1}(s;E,\mu_{1})$,
defined as, 
\begin{equation}
\tilde{w}_{1}\left(s;E,\mu_{1}\right)=\frac{1}{2\pi\rmi}\int_{-\pi}^{\pi}\rmd\theta_{1}\frac{\partial\ln f|_{X}\left(\theta_{1},s+\delta_{2}\left(\theta_{1}\right);E,\mu_{1}\right)}{\partial\theta_{1}},\label{seq:def-tilde-w}
\end{equation}
where $\delta_{2}(\theta_{1})$, $\theta_{1}\in[-\pi,\pi]$ is an
arbitrary function of $\theta_{1}$ satisfying $\exp[\rmi\delta_{2}(-\pi)]=\exp[\rmi\delta_{2}(\pi)]$,
and the strip winding number is equal to, 
\begin{equation}
W(E,\mu_{1})=\int_{-\pi}^{\pi}\frac{\rmd s}{2\pi}\tilde{w}_{1}\left(s;E,\mu_{1}\right).\label{eq:new-strip-winding}
\end{equation}

\begin{figure}
\centering

\includegraphics{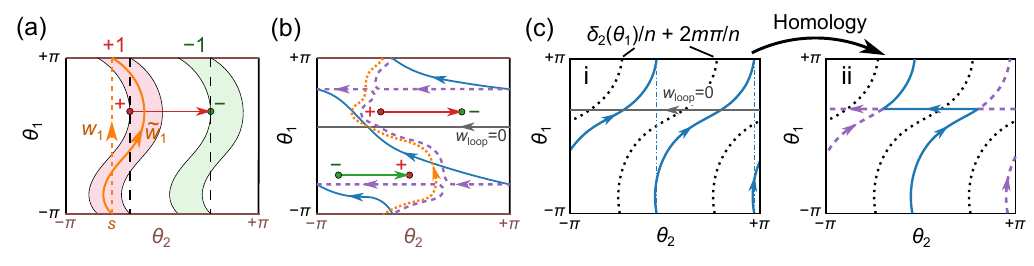}

\caption{Geometrical illustrations of equivalent formulations for the strip
winding number. (a) Relation between $w_{1}$ and $\tilde{w}_{1}$.
The red (green) regions indicate where $\tilde{w}_{1}$ is greater
(less) than $w_{1}$ by 1. (b) Cases with non-zero winding number
around the $\theta_{2}$ axis. The original winding loop (blue curve)
is homologous to the purple dashed curve, which in turn is homologous
to the sum of the orange dotted curve and two horizontal loops. (c)
Cases with non-zero winding number around both axes. An $n$-fold
winding loop can be decomposed into $n$ $1$-fold loops via homology.}
\label{fig:winding-general-loop}
\end{figure}

As defined in Eq. (\ref{seq:def-tilde-w}), for each constant value
$s$, $\tilde{w}(s;E,\mu_{1})$ equals the winding number of the loop
defined by $\theta_{2}(\theta_{1})=\delta_{2}(\theta_{1})+s$ and
$\theta_{1}=2\pi t,t\in[0,1]$. We first consider the case where $\delta_{2}(-\pi)=\delta_{2}(\pi)$,
that is, the loop does not wind around the $\theta_{2}$-axis. Without
loss of generality, we suppose $\delta_{2}(-\pi)=\delta_{2}(\pi)=0$.
As shown in Fig. \ref{fig:winding-general-loop}(a), for each constant
value $s$, the difference between $\tilde{w}_{1}$ and $w_{1}$ equals
the total topological charges enclosed by the loop of $\tilde{w}_{1}$
{[}orange solid line in Fig. \ref{fig:winding-general-loop}(a){]}
and the reverse of the loop of $w_{1}$ {[}orange dashed line in Fig.
\ref{fig:winding-general-loop}(a){]}. For a pair of PMGBZ points,
as illustrated by the light red and light green regions in Fig. \ref{fig:winding-general-loop}(a),
the contributions of the positive and negative charges cancel each
other, so that the integral of $\tilde{w}_{1}$ equals the integral
of $w_{1}$, which proves Eq. (\ref{eq:new-strip-winding}).

Next, we consider the general case where the loop defined by $\theta_{2}=\delta_{2}\left(\theta_{1}\right)$
is allowed to wind around the $\theta_{2}$-axis, shown as the blue
solid line in Fig. \ref{fig:winding-general-loop}(b). As illustrated
in the figure, the blue loop is homotopic to the purple dashed loop,
and the purple dashed loop is homologous to the sum of the orange
dotted loop, which satisfies the condition $\delta_{2}\left(-\pi\right)=\delta_{2}\left(\pi\right)$,
and the gray loop, around which the winding number vanishes. Therefore,
for each winding loop (solid blue line), the winding number around
the loop equals the winding number of a special loop satisfying $\delta_{2}\left(-\pi\right)=\delta_{2}\left(\pi\right)$
(dotted orange line), which is the case shown in Fig. \ref{fig:winding-general-loop}(a).

Furthermore, in some cases, the form of the characteristic polynomial
is complicated, but the product, 
\begin{equation}
g\left(\theta_{1},\theta_{2};E,\mu_{1}\right)=\prod_{m=1}^{n}f|_{X}\left(\theta_{1},\theta_{2}+\frac{2m\pi}{n};E,\mu_{1}\right),
\end{equation}
has a simple form, such as the $[11]$-SGBZ in the 2D HN model discussed
in the main text. Then, $W$ can be calculated by, 
\begin{equation}
W\left(E,\mu_{1}\right)=\int_{-\pi/n}^{\pi/n}\frac{\rmd s}{2\pi}\tilde{w}_{g}\left(s;E,\mu_{1}\right),\label{eq:W-strip-g}
\end{equation}
where, 
\begin{equation}
\tilde{w}_{g}\left(s;E,\mu_{1}\right)=\int_{-\pi}^{\pi}\frac{\rmd\theta_{1}}{2\pi\rmi}\frac{\partial\ln\left[g\left(\theta_{1},s+\delta_{2}(\theta_{1})/n;E,\mu_{1}\right)\right]}{\partial\theta_{1}},\label{seq:w-g}
\end{equation}
is the winding number of $g$ on the loop $s+\delta_{2}(\theta_{1})/n$.
Here, we only require $\delta_{2}(\theta_{1})$ to satisfy the periodicity
condition $\exp[\rmi\delta_{2}(-\pi)]=\exp[\rmi\delta_{2}(\pi)]$
rather than $\delta_{2}(\theta_{1})/n$. Now, we will prove Eq. (\ref{eq:W-strip-g}).

First, $g(\theta_{1},s+\delta_{2}(\theta_{1})/n;E,\mu_{1})$ is a
periodic function of $\theta_{1}$. When $\theta_{1}$ increases by
$2\pi$, $\delta_{2}(\theta_{1})/n$ increases by an integer multiple
of $2\pi/n$. Supposing $\delta_{2}(\pi)/n=\delta_{2}(-\pi)/n+2\pi m_{0}/n\mod{2\pi}$,
we get, 
\begin{align}
g\left(\pi,s+\delta_{2}(\pi)/n;E,\mu_{1}\right) & =\prod_{m=1}^{n}f\left(E,\rme^{\mu_{1}+\rmi\pi},\rme^{\mu_{2,0}(\pi)+\rmi\left(s+\delta_{2}(\pi)/n+2m\pi/n\right)}\right),\nonumber \\
 & =\prod_{m=1}^{n}f\left(E,\rme^{\mu_{1}-\rmi\pi},\rme^{\mu_{2,0}(-\pi)+\rmi\left(s+\delta_{2}(-\pi)/n+2(m+m_{0})\pi/n\right)}\right),\nonumber \\
 & =g\left(-\pi,s+\delta_{2}(-\pi)/n;E,\mu_{1}\right).
\end{align}
Therefore, the number $\tilde{w}_{g}(s;E,\mu_{1})$ defined by Eq.
(\ref{seq:w-g}) is a valid winding number.

Next, we consider the relation between $\tilde{w}_{g}(s;E,\mu_{1})$
and the winding number of the characteristic polynomial. When $\delta_{2}(-\pi)/n=\delta_{2}(\pi)/n\mod{2\pi}$,
the curve $\theta_{2}=s+\delta_{2}(\theta_{1})/n$ forms a closed
loop on $X(E,\mu_{1})$. Therefore, the following relation holds,
\begin{align}
\tilde{w}_{g}\left(s;E,\mu_{1}\right) & =\sum_{m=1}^{n}\int_{-\pi}^{\pi}\frac{\rmd\theta_{1}}{2\pi\rmi}\frac{\partial\ln\left[f|_{X}\left(\theta_{1},s+\frac{\delta_{2}(\theta_{1})}{n}+\frac{2m\pi}{n};E,\mu_{1}\right)\right]}{\partial\theta_{1}},\nonumber \\
 & =\sum_{m=1}^{n}\tilde{w}_{1}\left(s+\frac{2m\pi}{n};E,\mu_{1}\right),\label{seq:relation-winding-numbers}
\end{align}
where $\tilde{w}_{1}(s;E,\mu_{1})$ is defined by the loop $\delta_{2}^{\prime}(\theta_{1})\equiv\delta_{2}(\theta_{1})/n$.
By Eq. (\ref{eq:new-strip-winding}), the integral in Eq. (\ref{eq:W-strip-g})
reads, 
\begin{align}
\int_{-\pi/n}^{\pi/n}\frac{\rmd s}{2\pi}\tilde{w}_{g}\left(s;E,\mu_{1}\right)= & \sum_{m=1}^{n}\int_{-\pi/n}^{\pi/n}\frac{\rmd s}{2\pi}\tilde{w}_{1}\left(s+\frac{2m\pi}{n};E,\mu_{1}\right),\nonumber \\
= & \int_{\pi/n}^{2\pi+\pi/n}\frac{\rmd s}{2\pi}\tilde{w}_{1}\left(s;E,\mu_{1}\right),
\end{align}
which equals $W(E,\mu_{1})$.

When $\delta_{2}(-\pi)/n=\delta_{2}(\pi)/n+2\pi m_{0}/n\mod{2\pi}$
with $m_{0}\neq0$, the curve $\theta_{2}=s+\delta_{2}(\theta_{1})/n$
for $\theta_{1}\in[-\pi,\pi]$ is not closed, rendering the winding
number $\tilde{w}_{1}$ in Eq. (\ref{seq:relation-winding-numbers})
ill-defined. However, through homology transformations, $\tilde{w}_{g}$
can be transformed into the sum of winding numbers around $n$ congruent
loops with a phase shift of $2\pi/n$ along the $\theta_{2}$-axis.
Taking $n=2,m_{0}=1$ as an example, as shown in panel i of Fig. \ref{fig:winding-general-loop}(c),
the curve $\delta_{2}(\theta_{1})/2$ is braided with $\delta_{2}(\theta_{1})/2+\pi$
(black dotted curves), forming a twofold winding loop around the $\theta_{1}$-axis
(solid blue curves). The winding number $\tilde{w}_{g}(s;E,\mu_{1})$
equals the winding number of the characteristic polynomial around
this twofold loop. Due to the homology invariance of the winding number,
a winding loop parallel to the $\theta_{2}$-axis can be added to
the twofold loop without altering the winding number, shown as the
gray horizontal line in panel i of Fig. \ref{fig:winding-general-loop}(c).
Then, the sum of the twofold loop and the horizontal loop is homologous
to the sum of two onefold loops, shown as the blue solid curve and
the purple dashed curve in panel ii of Fig. \ref{fig:winding-general-loop}(c).
Thus, we return to the case with $m_{0}=0$. In general, using the
same method, an arbitrary $n$-fold loop can be split into $n$ onefold
loops by homology, so Eq. (\ref{eq:W-strip-g}) holds for arbitrary
$n$ and $m_{0}$.

With the above preparations, we will calculate the SGBZs of the 2D
HN model in the following sections.

\subsection{Calculations of the SGBZs}\label{subsec:HN-2D}

\begin{figure}
\centering

\includegraphics{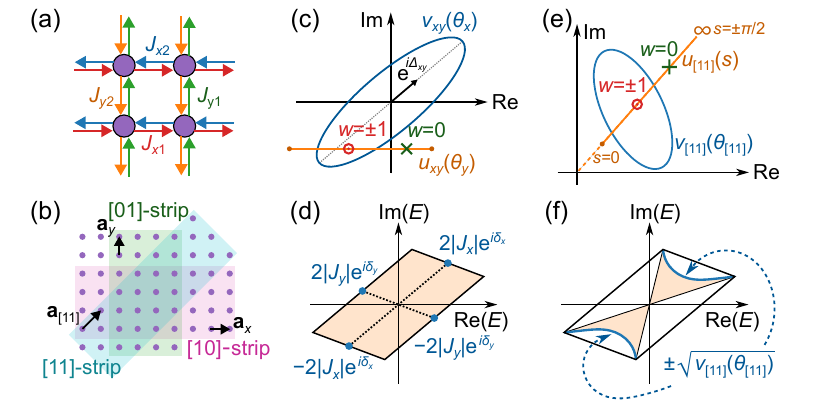}

\caption{ Schematic diagrams of the 2D HN model, three special strips and
the SGBZ bands. (a) Illustration of the 2D HN model defined as Eq.
(\ref{eq:HN-model}). (b) Illustration of the $[10]$-strip, $[01]$-strip
and $[11]$-strip of the 2D HN model. (c) Sketches for the calculations
of $[10]$($[01]$)-SGBZ. (d) Illustration of the $[10]$($[01]$)-SGBZ
bands. (e) Sketches for the calculation of $[11]$-SGBZ. (f) Illustration
of the $[11]$-SGBZ bands.}\label{fig:2D-HN-and-stripes}

\end{figure}

In this section, we calculate the SGBZs of the 2D HN model discussed
in the main text. As illustrated in Fig. \ref{fig:2D-HN-and-stripes}(a),
the Hamiltonian of the system is given by 
\begin{equation}
h_{xy}\left(\beta_{x},\beta_{y}\right)=J_{x1}\beta_{x}^{-1}+J_{x2}\beta_{x}+J_{y1}\beta_{y}^{-1}+J_{y2}\beta_{y},\label{eq:HN-model}
\end{equation}
where $\beta_{x}$ and $\beta_{y}$ are the components conjugate to
the lattice vectors $\mathbf{a}_{x}=(1,0)$ and $\mathbf{a}_{y}=(0,1)$.
$J_{x1}$, $J_{x2}$, $J_{y1}$, and $J_{y2}$ are arbitrary complex
numbers. To simplify the expressions, we factorize the coupling coefficients
into Hermitian and non-Hermitian parts, which reads, 
\begin{equation}
J_{\alpha1}=\rme^{\gamma_{\alpha}+\rmi\delta_{\alpha}}J_{\alpha},\quad J_{\alpha2}=\rme^{-\gamma_{\alpha}+\rmi\delta_{\alpha}}J_{\alpha}^{*},\quad\alpha=x,y,
\end{equation}
where $J_{\alpha}\in\mathbb{C}$ is the Hermitian part, and $\gamma_{\alpha},\delta_{\alpha}\in\mathbb{R}$
are the non-Hermitian parts. As illustrated in Fig. \ref{fig:2D-HN-and-stripes}(b),
we consider three different strips: the $[10]$-strip, $[01]$-strip,
and $[11]$-strip, defined by the major axes $\mathbf{a}_{x}$, $\mathbf{a}_{y}$,
and $\mathbf{a}_{[11]}=(1,1)$, respectively.

For the $[10]$- and $[01]$-directional SGBZs, the 2D momentum-space
characteristic polynomial under the basis $(\mathbf{a}_{x},\mathbf{a}_{y})$
reads, 
\begin{align}
f_{xy}\left(E,\beta_{x},\beta_{y}\right)= & E-J_{x1}\beta_{x}^{-1}-J_{x2}\beta_{x}-J_{y1}\beta_{y}^{-1}-J_{y2}\beta_{y}.
\end{align}
We first construct the base manifold $X(E,\mu_{x})$. By Vieta's formulas,
the two $\beta_{y}$-zeros of $f_{xy}$ satisfy, 
\begin{equation}
\beta_{y}^{(1)}\left(E,\beta_{x}\right)\beta_{y}^{(2)}\left(E,\beta_{x}\right)=\frac{J_{y1}}{J_{y2}},\label{eq:Vieta-theorem-xy}
\end{equation}
where the two solutions are ordered by $|\beta_{y}^{(1)}(E,\beta_{x})|\leq|\beta_{y}^{(2)}(E,\beta_{x})|$.
For any radius $\ln|\beta_{x}|=\mu_{x}$, if we choose $\mu_{y,0}=\ln\sqrt{|J_{y1}/J_{y2}|}=\gamma_{y}$,
the relation $\ln|\beta_{y}^{(1)}|\leq\mu_{y,0}\leq\ln|\beta_{y}^{(2)}|$
is satisfied, and equality holds if and only if the two solutions
have the same modulus. Therefore, $\mu_{y,0}=\gamma_{y}$ is a valid
radius function for $X(E,\mu_{x})$.

To compute $w_{x}(\theta_{y};E,\mu_{x})$, we consider the value of
$f_{xy}$ restricted to $X(E,\mu_{x})$, i.e., 
\begin{align}
f_{xy}|_{X}\left(\theta_{x},\theta_{y};E,\mu_{x}\right)\equiv & f_{xy}\left(E,\rme^{\mu_{x}+\rmi\theta_{x}},\rme^{\mu_{y,0}(\theta_{x})+\rmi\theta_{y}}\right),\nonumber \\
= & E-J_{x1}\rme^{-\mu_{x}-\rmi\theta_{x}}-J_{x2}\rme^{\mu_{x}+\rmi\theta_{x}}-J_{y1}\rme^{-\gamma_{y}-\rmi\theta_{y}}-J_{y2}\rme^{\gamma_{y}+\rmi\theta_{y}}.\nonumber \\
= & \rme^{\rmi\delta_{y}}\left[u_{xy}\left(\theta_{y};E\right)-v_{xy}\left(\theta_{x};\mu_{x}\right)\right]
\end{align}
where, 
\begin{align}
u_{xy}\left(\theta_{y};E\right) & =E\rme^{-\rmi\delta_{y}}-2\text{Re}\left(J_{y}^{*}\rme^{\rmi\theta_{y}}\right),\label{eq:uxy}\\
v_{xy}\left(\theta_{x};\mu_{x}\right) & =\rme^{\rmi\Delta_{xy}}\left(J_{x}\rme^{\gamma_{x}-\mu_{x}-\rmi\theta_{x}}-J_{x}^{*}\rme^{\mu_{x}-\gamma_{x}+\rmi\theta_{x}}\right),\label{eq:vxy}
\end{align}
and $\Delta_{xy}\equiv\delta_{x}-\delta_{y}$. By definition, $w_{x}(\theta_{y};E,\mu_{x})$
is the winding number of $f_{xy}$ when $\theta_{y}$ remains constant
and $\theta_{x}$ winds around $2\pi$, so it equals the winding number
of $v_{xy}(\theta_{x};\mu_{x})$ around $u_{xy}(\theta_{y};E)$ on
the complex plane.

As illustrated in Fig. \ref{fig:2D-HN-and-stripes}(c), the curve
of $u_{xy}$ is a horizontal line segment. The curve of $v_{xy}$
is an ellipse with its major axis parallel to $\exp(\rmi\Delta_{xy})$.
The length of its major semi-axis is $2\cosh(\mu_{x}-\gamma_{x})$,
and the length of its minor semi-axis is $|2\sinh(\mu_{x}-\gamma_{x})|$.
Furthermore, when $\mu_{x}<\gamma_{x}$, $v_{xy}(\theta_{x})$ rotates
clockwise as $\theta_{x}$ increases, and vice versa. Therefore, when
$u_{xy}(\theta_{y};E)$ is enclosed inside the ellipse of $v_{xy}$,
shown as the red circle in Fig. \ref{fig:2D-HN-and-stripes}(c), $w_{x}(\theta_{y};E,\mu_{x})=\pm1$,
where the sign of $w_{x}(\theta_{y};E,\mu_{x})$ is the same as the
sign of $\mu_{x}-\gamma_{x}$. Otherwise, as shown by the green cross
in Fig. \ref{fig:2D-HN-and-stripes}(c), $w_{x}(\theta_{y};E,\mu_{x})=0$.
As a result, $\mu_{x}=\gamma_{x}$ satisfies the SGBZ constraint.
The corresponding SGBZ points are the points on $X(E,\gamma_{x})$
satisfying $u_{xy}(\theta_{y};E)=v_{xy}(\theta_{x};\mu_{x})$. Thus,
we obtain the SGBZ for the $x$-strip, that is, 
\begin{equation}
\beta_{x}=\exp\left(\gamma_{x}+\rmi\theta_{x}\right),\quad\beta_{y}=\exp\left(\gamma_{y}+\rmi\theta_{y}\right).\label{eq:SGBZ-xy}
\end{equation}
where $\theta_{x},\theta_{y}\in[-\pi,\pi]$, and the corresponding
eigenenergy is given by, 
\begin{equation}
E\left(\theta_{x},\theta_{y}\right)=2\rme^{\rmi\delta_{x}}\text{Re}\left(J_{x}^{*}\rme^{\rmi\theta_{x}}\right)+2\rme^{\rmi\delta_{y}}\text{Re}\left(J_{y}^{*}\rme^{\rmi\theta_{y}}\right).\label{eq:SGBZ-xy-E}
\end{equation}
On the complex plane, the spectrum forms a parallelogram spanned by
$\pm2|J_{x}|\rme^{\rmi\delta_{x}}$ and $\pm2|J_{y}|\rme^{\rmi\delta_{y}}$,
as shown in Fig. \ref{fig:2D-HN-and-stripes}(d). Using the same method,
we can also calculate the SGBZ for the $y$-strip, which is the same
as Eq. (\ref{eq:SGBZ-xy}).

For the $[11]$-directional SGBZ, the 2D characteristic polynomial
under the basis $(\mathbf{a}_{[11]},\mathbf{a}_{y})$ reads, 
\begin{align}
f_{[11]}\left(E,\beta_{[11]},\beta_{y}\right)= & E-J_{x1}\beta_{[11]}^{-1}\beta_{y}-J_{x2}\beta_{[11]}\beta_{y}^{-1}-J_{y1}\beta_{y}^{-1}-J_{y2}\beta_{y},
\end{align}
where $\beta_{[11]}$ and $\beta_{y}$ are conjugate to $\mathbf{a}_{[11]}$
and $\mathbf{a}_{y}$, respectively. Similar to the $x$-directional
case, the two $\beta_{y}$-zeros satisfy, 
\begin{equation}
\beta_{y}^{(1)}\left(E,\beta_{[11]}\right)\beta_{y}^{(2)}\left(E,\beta_{[11]}\right)=\frac{J_{x2}\beta_{[11]}+J_{y1}}{J_{x1}\beta_{[11]}^{-1}+J_{y2}}.\label{seq:11-product-beta2-zeros}
\end{equation}
Therefore, the function $\mu_{y,0}(\theta_{[11]};E,\mu_{[11]})$,
defined as, 
\begin{equation}
\rme^{\mu_{y,0}(\theta_{[11]};E,\mu_{[11]})}=\sqrt{\left|\frac{J_{x2}\rme^{\mu_{[11]}+\rmi\theta_{[11]}}+J_{y1}}{J_{x1}\rme^{-\mu_{[11]}-\rmi\theta_{[11]}}+J_{y2}}\right|},\label{eq:rho-theta11-11}
\end{equation}
is a valid radius function for $X(E,\mu_{[11]})$. The restriction
of $f_{[11]}$ to $X(E,\mu_{[11]})$ reads, 
\begin{align}
f_{[11]}|_{X}\left(\theta_{[11]},\theta_{y};E,\mu_{[11]}\right)\equiv & f_{[11]}\left(E,\rme^{\mu_{[11]}+\rmi\theta_{[11]}},\rme^{\mu_{y,0}(\theta_{[11]})+\rmi\theta_{y}}\right),\nonumber \\
= & E-2\rme^{\rmi\bar{\varphi}(\theta_{[11]})}\sqrt{\left|v_{[11]}\left(\theta_{[11]};\mu_{[11]}\right)\right|}\cos\left(\theta_{y}-\frac{\Delta\varphi(\theta_{[11]})}{2}\right),\label{eq:ChP-on-X-11}
\end{align}
where, 
\begin{align}
v_{[11]}\left(\theta_{[11]};\mu_{[11]}\right)\equiv & J_{x2}J_{y2}\rme^{\mu_{[11]}+\rmi\theta_{[11]}}+J_{x1}J_{y1}\rme^{-\mu_{[11]}-\rmi\theta_{[11]}}+J_{x1}J_{x2}+J_{y1}J_{y2},\label{eq:v-11}
\end{align}
and 
\begin{align}
\varphi_{1}\left(\theta_{[11]};\mu_{[11]}\right) & \equiv\text{Arg}\left(J_{x1}\rme^{-\mu_{[11]}-\rmi\theta_{[11]}}+J_{y2}\right),\\
\varphi_{2}\left(\theta_{[11]};\mu_{[11]}\right) & \equiv\text{Arg}\left(J_{x2}\rme^{\mu_{[11]}+\rmi\theta_{[11]}}+J_{y1}\right),\\
\bar{\varphi}\left(\theta_{[11]};\mu_{[11]}\right) & \equiv\frac{\varphi_{1}\left(\theta_{[11]};\mu_{[11]}\right)+\varphi_{2}\left(\theta_{[11]};\mu_{[11]}\right)}{2},\\
\Delta\varphi\left(\theta_{[11]};\mu_{[11]}\right) & \equiv\varphi_{2}\left(\theta_{[11]};\mu_{[11]}\right)-\varphi_{1}\left(\theta_{[11]};\mu_{[11]}\right).\label{eq:Delta-varphi}
\end{align}
It is noted that $f_{[11]}$ is continuous in $\theta_{[11]}$ because
$\exp[\rmi\bar{\varphi}(\theta_{[11]})]$ and $\cos(\theta_{y}-\Delta\varphi(\theta_{[11]})/2)$
in Eq. (\ref{eq:ChP-on-X-11}) flip their signs simultaneously when
$\theta_{[11]}$ passes the branch cut of $\varphi_{1}$ or $\varphi_{2}$.
To simplify Eq. (\ref{eq:ChP-on-X-11}), we compute the product, 
\begin{align}
g\left(\theta_{[11]},\theta_{y};E,\mu_{[11]}\right)\equiv & f_{[11]}|_{X}\left(\theta_{[11]},\theta_{y};E,\mu_{[11]}\right)f_{[11]}|_{X}\left(\theta_{[11]},\theta_{y}+\pi;E,\mu_{[11]}\right),\nonumber \\
= & 4\cos^{2}\left(\theta_{y}-\frac{\Delta\varphi(\theta_{[11]})}{2}\right)\left[u_{[11]}\left(\theta_{y}-\frac{\Delta\varphi(\theta_{[11]})}{2};E\right)-v_{[11]}\left(\theta_{[11]};\mu_{[11]}\right)\right],
\end{align}
where $u_{[11]}\left(s;E\right)\equiv E^{2}/4\cos^{2}s$. According
to Eq. (\ref{eq:W-strip-g}), $W(E,\mu_{[11]})$ is given by, 
\begin{equation}
W\left(E,\mu_{[11]}\right)=\int_{-\pi/n}^{\pi/n}\frac{\rmd s}{2\pi}\tilde{w}_{g}\left(s;E,\mu_{[11]}\right),\label{eq:W-strip-expression-11}
\end{equation}
where, 
\begin{equation}
\tilde{w}_{g}\left(s;E,\mu_{[11]}\right)\equiv\int_{-\pi}^{\pi}\frac{\rmd\theta_{[11]}}{2\pi\rmi}\frac{\partial\ln\left[g\left(\theta_{[11]},\frac{\Delta\varphi(\theta_{[11]})}{2}+s;E,\mu_{[11]}\right)\right]}{\partial\theta_{[11]}},
\end{equation}
is the winding number of $g$ around the loop $\theta_{y}=\Delta\varphi(\theta_{[11]})/2+s$,
which equals the winding number of $v_{[11]}(\theta_{[11]};\mu_{[11]})$
around the fixed point $u_{[11]}(s;E)$. As illustrated in Fig. \ref{fig:2D-HN-and-stripes}(e),
the curve of $v_{[11]}$ is an ellipse, and the curve of $u_{[11]}$
is a ray collinear with the origin. When $\mu_{[11]}<\gamma_{x}+\gamma_{y}$,
$v_{[11]}(\theta_{[11]})$ rotates clockwise, and vice versa. When
the point $u_{[11]}(s;E)$ is enclosed by the ellipse, shown as the
red circle in Fig. \ref{fig:2D-HN-and-stripes}(e), $\tilde{w}_{g}(s)=\pm1$,
where the sign is the same as the sign of $\mu_{[11]}-\gamma_{x}-\gamma_{y}$.
Otherwise, when $u_{[11]}(s;E)$ lies outside the ellipse, $\tilde{w}_{g}(s)=0$.
Therefore, according to Eq. (\ref{eq:W-strip-expression-11}), the
SGBZ constraint is satisfied if and only if $\mu_{[11]}=\gamma_{x}+\gamma_{y}$
and $u_{[11]}$ intersects with $v_{[11]}$. The corresponding SGBZ
points are solutions of $f_{[11]}|_{X}(\theta_{[11]},\theta_{y})=0$
on $X(E,\gamma_{x}+\gamma_{y})$, which reads, 
\begin{equation}
\begin{cases}
\tilde{\beta}_{[11]}=\rme^{\gamma_{x}+\gamma_{y}+\rmi\theta_{[11]}},\\
\tilde{\beta}_{y}=\rme^{\gamma_{y}+\rmi\theta_{y}}\sqrt{\left|\frac{J_{x}^{*}\rme^{\rmi\Delta_{xy}+\rmi\theta_{[11]}}+J_{y}}{J_{x}\rme^{\rmi\Delta_{xy}-\rmi\theta_{[11]}}+J_{y}^{*}}\right|},
\end{cases}\label{eq:11-SGBZ-HN}
\end{equation}
where $\theta_{[11]},\theta_{y}\in[-\pi,\pi]$. Here, the tildes on
$\tilde{\beta}_{[11]}$ and $\tilde{\beta}_{y}$ serve to distinguish
them from the $[10]$-SGBZ or $[01]$-SGBZ. The corresponding eigenenergy
is, 
\begin{align}
E\left(\theta_{[11]},\theta_{y}\right)= & 2\rme^{\rmi\bar{\varphi}(\theta_{[11]})}\sqrt{\left|v_{[11]}\left(\theta_{[11]};\gamma_{x}+\gamma_{y}\right)\right|}\cos\left(\theta_{y}-\frac{\Delta\varphi(\theta_{[11]};\gamma_{x}+\gamma_{y})}{2}\right).\label{eq:SGBZ-E-11}
\end{align}
It is noted that the curves $\pm\rme^{\rmi\bar{\varphi}}\sqrt{|v_{[11]}|}$
are the two square roots of $v_{[11]}$. Therefore, as illustrated
in Fig. \ref{fig:2D-HN-and-stripes}(f), the spectrum of the $[11]$-SGBZ
is the region swept by the line segments connecting the two square
roots of $v_{[11]}(\theta_{[11]};\gamma_{x}+\gamma_{y})$ for $\theta_{[11]}\in[-\pi,\pi]$.

\subsection{Verification with numerical calculations of QMGBZs}\label{subsec:verification-numerical}

By definition, the SGBZ is the limit of the QMGBZ as the width of
the strip approaches infinity. To verify this observation, we numerically
compute the QMGBZs for the three strips and compare them with the
SGBZ solutions.

\begin{figure}
\centering

\includegraphics{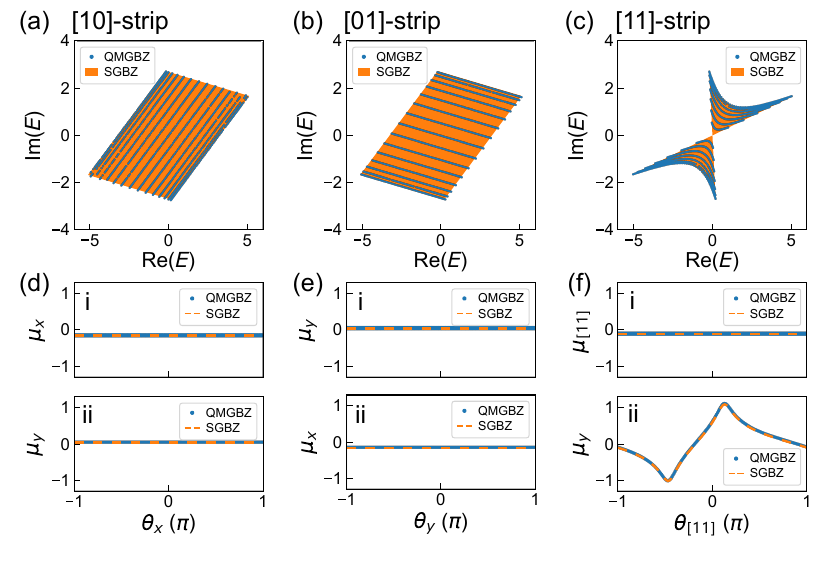}

\caption{Comparison between SGBZs and QMGBZs. (a--c) Theoretical results for
the SGBZ spectra and numerical results for the QMGBZ spectra for the
(a) $[10]$-strip, (b) $[01]$-strip, and (c) $[11]$-strip. (d--f)
Complex momenta from the theoretical SGBZ results and the numerical
QMGBZ results for the (d) $[10]$-strip, (e) $[01]$-strip, and (f)
$[11]$-strip, where panels i and ii in each plot display the major
and minor components, respectively. The parameters of the 2D HN model
are $J_{x1}=1+\protect\rmi$, $J_{x2}=1.5+1.2\protect\rmi$, $J_{y1}=-1+\protect\rmi$,
and $J_{y2}=-1.2-0.5\protect\rmi$.}
\label{fig:quasi-1D}
\end{figure}

By definition, the SGBZ is the thermodynamic limit of the QMGBZ. To
verify this, we numerically calculate the QMGBZs of the 2D HN model
in the $x$-strip, the $y$-strip, and the $[11]$-strip, then compare
the numerical results with the analytical results given above.

In numerical calculations, the width for each QMGBZ is set to $15$.
The quasi-1D GBZ is solved by first solving the auxiliary GBZ equations
\cite{yangNonHermitianBulkBoundaryCorrespondence2020}, 
\begin{equation}
\begin{cases}
F_{L_{2}}\left(E,\beta_{1}\right)=0\\
F_{L_{2}}\left(E,\beta_{1}\rme^{\rmi\phi}\right)=0
\end{cases},\label{eq:numerical-aGBZ-equation}
\end{equation}
where $E$ and $\beta_{1}$ are the unknowns, and $\phi$ is the relative
phase ranging from $[0,\pi]$. We then check the 1D GBZ constraint,
i.e., $|\beta_{1}^{(M)}(E)|=|\beta_{1}^{(M+1)}(E)|$. The numerical
solution of Eq. (\ref{eq:numerical-aGBZ-equation}) is obtained using
the Python package ``phcpy''\cite{verscheldeAlgorithm795PHCpack1999},
which implements a polynomial homotopy continuation algorithm. The
phase $\phi$ is discretized as $\phi=j\pi/N_{\phi}$, for $j=0,1,\dots,N_{\phi}$,
with $N_{\phi}=49$.

By solving Eq. (\ref{eq:numerical-aGBZ-equation}), the eigenenergy
$E$ and the complex momentum along the major axis are directly computed.
The complex momentum in the minor axis is computed with the profile
of the non-Bloch waves. For a given pair $(E,\beta_{1})$, the non-Bloch
wave $\psi_{E,\beta_{1}}$ is obtained by solving $\left(E-\mathcal{H}_{L_{2}}(\beta_{1})\right)\psi_{E,\beta_{1}}=0$.
We expand the non-Bloch state in the hybrid space basis, i.e., 
\begin{equation}
\psi_{E,\beta_{1}}=\sum_{j}\psi_{E,\beta_{1}}^{(j)}\left|\beta_{1},j\right>,\label{eq:expand-non-Bloch-on-a2}
\end{equation}
where $|\beta_{1},j\rangle\equiv c_{\beta_{1},j}^{\dagger}|0\rangle$
is the state in which one particle occupies the site with coordinate
$j\mathbf{a}_{2}$ in the supercell. Then, we assume that the non-Bloch
state $\psi_{E,\beta_{1}}^{(j)}$ has the form, 
\begin{equation}
\left|\psi_{E,\beta_{1}}^{(j)}\right|^{2}=C\rho^{2j}\cos\left(kx+\varphi\right),\label{eq:fit-supercell}
\end{equation}
where $C$, $\rho$, $k$, and $\varphi$ are fitting parameters.
By definition, $\mu_{2}$ equals the fitting parameter $\ln\rho$
in Eq. (\ref{eq:fit-supercell}).

For generality, the coupling terms are four arbitrary complex numbers
chosen as $J_{x1}=1+\rmi$, $J_{x2}=1.5+1.2\rmi$, $J_{y1}=-1+\rmi$,
and $J_{y2}=-1.2-0.5\rmi$. Figure \ref{fig:quasi-1D} shows a comparison
between the analytical solutions of the SGBZ and the numerical solutions
of the QMGBZ for the $[10]$-strip, $[01]$-strip, and $[11]$-strip.
The spectra of the three strips are displayed in Fig. \ref{fig:quasi-1D}(a--c).
In all cases, the QMGBZ spectra (blue dots) agree well with the SGBZ
spectra (orange patches). For the $[10]$-strip and $[01]$-strip,
as shown in Figs. \ref{fig:quasi-1D}(a) and \ref{fig:quasi-1D}(b),
the QMGBZ spectra form parallel line segments aligned with one pair
of sides of the SGBZ spectrum. The difference between the QMGBZ spectra
of the $[10]$-strip and $[01]$-strip lies in the orientation of
these parallel segments. For the $[11]$-strip, as shown in Fig. \ref{fig:quasi-1D}(c),
the quasi-1D spectrum consists of curves with constant $\theta_{y}-\Delta\varphi(\theta_{[11]})/2$.

Figure \ref{fig:quasi-1D}(d--f) shows the complex momenta of the
QMGBZs and SGBZs for (d) the $[10]$-strip, (e) the $[01]$-strip,
and (f) the $[11]$-strip, where panels i display the imaginary momenta
of the major components, and panels ii show those of the minor components.
For both the $[10]$-strip and $[01]$-strip, $\mu_{x}$ and $\mu_{y}$
are constant, with values $\mu_{x}=\gamma_{x}\approx-0.1531$ and
$\mu_{y}=\gamma_{y}\approx0.0421$. For the $[11]$-strip, $\mu_{[11]}$
remains constant at $\mu_{[11]}=\gamma_{x}+\gamma_{y}\approx-0.1110$,
while $\mu_{y}$ varies with $\theta_{[11]}$. In all three cases,
the numerical results agree well with the theoretical analysis.

In conclusion, for a 2D non-Hermitian lattice, both the spectrum and
the eigenstates of the QMGBZ match those of the SGBZ when the strip
width is sufficiently large. This finding supports the idea that the
QMGBZ converges to the SGBZ as the width increases.

\subsection{3D HN model with complex coupling coefficients}\label{subsec:3D-HN-model}

\begin{figure}
\centering

\includegraphics{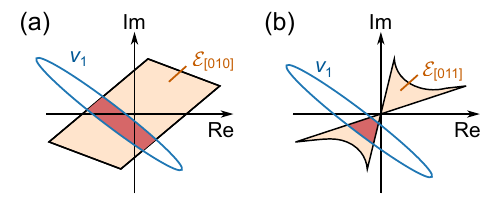}

\caption{Illustration of strip winding numbers in the 3D HN model for (a) $\mathbf{a}_{1}=(1,0,0)$,
$\mathbf{a}_{2}=(0,1,0)$, $\mathbf{a}_{3}=(0,0,1)$, and (b) $\mathbf{a}_{1}=(1,0,0)$,
$\mathbf{a}_{2}=(0,1,1)$, $\mathbf{a}_{3}=(0,0,1)$.}
\label{fig:higher-dim-HN}
\end{figure}

As an example of higher-dimensional SGBZs, we consider the 3D HN model
defined as, 
\begin{align}
h_{\text{HN}}\left(\beta_{x},\beta_{y},\beta_{z}\right)= & \sum_{\alpha=x,y,z}\left(J_{\alpha1}\beta_{\alpha}^{-1}+J_{\alpha2}\beta_{\alpha}\right),
\end{align}
where $J_{\alpha1}$, $J_{\alpha2}$, $\alpha=x,y,z$ are arbitrary
complex numbers. Similar to the 2D HN model, the coefficients are
factorized into, 
\begin{equation}
J_{\alpha1}=\rme^{\gamma_{\alpha}+\rmi\delta_{\alpha}}J_{\alpha},\quad J_{\alpha2}=\rme^{-\gamma_{\alpha}+\rmi\delta_{\alpha}}J_{\alpha}^{*},
\end{equation}
for $\alpha=x,y,z$. We first calculate the 3D SGBZ with the axes
$\mathbf{a}_{1}=\mathbf{a}_{x}\equiv(1,0,0)$, $\mathbf{a}_{2}=\mathbf{a}_{y}\equiv(0,1,0)$,
and $\mathbf{a}_{3}=\mathbf{a}_{z}\equiv(0,0,1)$. Under the selected
basis, the momentum-space characteristic polynomial is given by, 
\begin{align}
f\left(E,\beta_{x},\beta_{y},\beta_{z}\right)= & E-\sum_{\alpha=x,y,z}\left(J_{\alpha1}\beta_{\alpha}^{-1}+J_{\alpha2}\beta_{\alpha}\right).
\end{align}
Taking $\beta_{x}$ as the parameter, the 3D momentum-space characteristic
polynomial is equivalent to the characteristic polynomial of the 2D
HN model with $E^{\prime}=E-J_{\alpha1}\beta_{x}^{-1}-J_{\alpha2}\beta_{x}$
under the basis $\mathbf{a}_{y}$ and $\mathbf{a}_{z}$. According
to the results of the $x$($y$)-SGBZ in Sec. \ref{subsec:HN-2D},
the radius function of $\beta_{y}$ is $\mu_{y}=\gamma_{y}$, and
the corresponding transverse base manifold is, 
\begin{equation}
X_{\perp}^{(2)}\left(\theta_{x};E,\mu_{x}\right)=\left\{ \left(\beta_{y},\beta_{z}\right)\in\mathbb{C}^{2}\mid\left|\beta_{y}\right|=\rme^{\gamma_{y}},\left|\beta_{z}\right|=\rme^{\gamma_{z}}\right\} .
\end{equation}
Therefore, the restriction of $f$ to $X^{(3)}(E,\mu_{x})$ is, 
\begin{equation}
f|_{X^{(3)}}\left(E,\theta_{x},\theta_{y},\theta_{z}\right)=v_{1}\left(E,\theta_{x}\right)-\mathcal{E}_{[010]}\left(\theta_{y},\theta_{z}\right),\label{eq:3D-ChP-1}
\end{equation}
where, 
\begin{equation}
\mathcal{E}_{[010]}\left(\theta_{y},\theta_{z}\right)=2\rme^{\rmi\delta_{y}}\text{Re}\left(J_{y}^{*}\rme^{\rmi\theta_{y}}\right)+2\rme^{\rmi\delta_{z}}\text{Re}\left(J_{z}^{*}\rme^{\rmi\theta_{z}}\right),
\end{equation}
is the 2D SGBZ spectra of the strip along $\mathbf{a}_{y}$ in the
parametric 2D HN model, and the function $v_{1}(E,\theta_{x})$ is
given by, 
\begin{equation}
v_{1}\left(\theta_{x};E,\mu_{x}\right)\equiv E-J_{x1}\rme^{-\mu_{x}-\rmi\theta_{x}}-J_{x2}\rme^{\mu_{x}+\rmi\theta_{x}}.
\end{equation}
When $\theta_{x}$ runs over $\left[-\pi,\pi\right]$, the trajectory
of $v_{1}(\theta_{x};E,\mu_{x})$ forms an ellipse on the complex
plane. The lengths of the semi-major and semi-minor axes are $2\cosh(\mu_{x}-\gamma_{x})$
and $2|\sinh(\mu_{x}-\gamma_{x})|$, respectively. When $\mu_{x}<\gamma_{x}$,
$v_{1}(\theta_{x};E,\mu_{x})$ rotates clockwise as $\theta_{x}$
increases, and vice versa. Figure \ref{fig:higher-dim-HN}(a) illustrates
the geometric representation of $\mathcal{E}_{[010]}(\theta_{y},\theta_{z})$
(orange region) and $v_{1}(\theta_{x};E,\mu_{x})$ (blue ellipse)
on the complex plane. According to Eqs. (\ref{eq:w-loop-nD}) and
(\ref{eq:3D-ChP-1}), the winding number $w_{x}(\theta_{y},\theta_{z};E,\mu_{x})$
is non-vanishing if and only if $\mathcal{E}_{[010]}(\theta_{y},\theta_{z})$
is enclosed by the ellipse of $v_{1}(\theta_{x};E,\mu_{x})$, as shown
in the red region of Fig. \ref{fig:higher-dim-HN}(a), and the sign
of $w_{x}(\theta_{y},\theta_{z};E,\mu_{x})$ in this region matches
the sign of $r_{x}-\gamma_{x}$. Therefore, the 3D SGBZ constraint
is satisfied at $\mu_{x}=\gamma_{x}$, and the SGBZ is given by, 
\begin{equation}
\beta_{x}=\rme^{\gamma_{x}+\rmi\theta_{x}},\beta_{y}=\rme^{\gamma_{y}+\rmi\theta_{y}},\beta_{z}=\rme^{\gamma_{z}+\rmi\theta_{z}},
\end{equation}
and the corresponding spectrum reads, 
\begin{equation}
E=2\rme^{\rmi\delta_{x}}\text{Re}\left(J_{x}^{*}\rme^{\rmi\theta_{y}}\right)+\mathcal{E}_{[010]}\left(\theta_{y},\theta_{z}\right).
\end{equation}

To investigate the effects of the selection of $\mathbf{a}_{2}$,
we choose $\mathbf{a}_{2}=(0,1,1)$ instead of $\mathbf{a}_{y}$ and
calculate the corresponding 3D SGBZ. By the same reasoning, the parametric
2D system in this case is equivalent to the $[11]$-strip of the 2D
HN model. Therefore, the transverse base manifold reads, 
\begin{align}
X_{\perp}^{(2)}\left(\theta_{x};E,\mu_{x}\right)= & \left\{ \left(\beta_{2},\beta_{z}\right)\in\mathbb{C}^{2}\mid\beta_{2}=\rme^{\gamma_{y}+\gamma_{z}+\rmi\theta_{2}},\right.\nonumber \\
 & \left.\left|\beta_{z}\right|=\rme^{\mu_{z,0}(\theta_{2})},\theta_{2}\in[-\pi,\pi]\right\} ,
\end{align}
where, 
\begin{equation}
\mu_{z,0}(\theta_{2})=\gamma_{z}+\ln\sqrt{\left|\frac{J_{y}^{*}\rme^{\rmi\Delta_{yz}+\rmi\theta_{2}}+J_{z}}{J_{y}\rme^{\rmi\Delta_{yz}-\rmi\theta_{2}}+J_{z}^{*}}\right|},
\end{equation}
and $\Delta_{yz}\equiv\delta_{y}-\delta_{z}$. The restriction of
$f$ to $X^{(3)}(E,\mu_{x})$ reads, 
\begin{equation}
f|_{X^{(3)}}\left(E,\theta_{x},\theta_{2},\theta_{z}\right)=v_{1}\left(\theta_{x};E,\mu_{x}\right)-\mathcal{E}_{[011]}\left(\theta_{2},\theta_{z}\right),
\end{equation}
where $\mathcal{E}_{[011]}(\theta_{2},\theta_{z})$ equals the spectrum
of the $[11]$-SGBZ of the 2D HN model with the subscript substitutions
$x\rightarrow y$, $y\rightarrow z$, and $\theta_{[11]}\rightarrow\theta_{2}$
{[}Eq. (\ref{eq:SGBZ-E-11}){]}. As shown in Fig. \ref{fig:higher-dim-HN}(b),
similar to the case of $\mathbf{a}_{2}=\mathbf{a}_{y}$, the 3D SGBZ
constraint is also satisfied at $\mu_{x}=\gamma_{x}$. However, the
SGBZ, which reads, 
\begin{equation}
\beta_{x}=\rme^{\gamma_{x}+\rmi\theta_{x}},\beta_{2}=\rme^{\gamma_{y}+\gamma_{z}+\rmi\theta_{2}},\beta_{z}=\rme^{\mu_{z,0}(\theta_{2})+\rmi\theta_{z}},
\end{equation}
is not compatible with the case of $\mathbf{a}_{2}=\mathbf{a}_{y}$,
and the corresponding spectrum, 
\begin{equation}
E=2\rme^{\rmi\delta_{x}}\text{Re}\left(J_{x}^{*}\rme^{\rmi\theta_{y}}\right)+\mathcal{E}_{[011]}\left(\theta_{2},\theta_{z}\right),
\end{equation}
is also different. Therefore, the SGBZs in three and higher dimensions
are not uniquely determined by the major axis alone but are also influenced
by the choices of other axes.

\section{Competition of incompatible SGBZs}\label{sec:competition-SGBZ}

\begin{figure}
\centering

\includegraphics{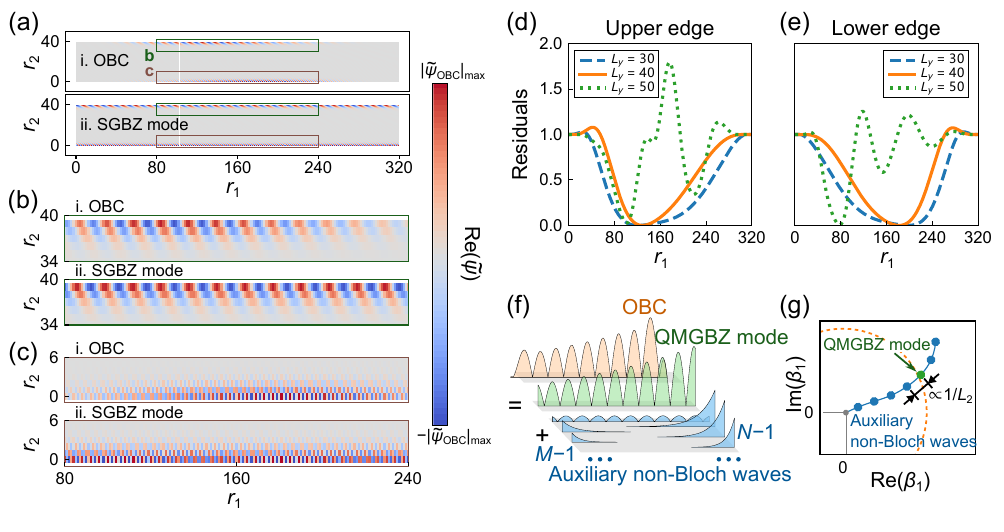}

\caption{Eigenstate distributions and the competition between incompatible
SGBZs. (a) Full view of the real part of the eigenstate in the coordinates
$r_{1}$ and $r_{2}$, scaled by $\protect\rme^{-r_{1}\gamma_{x}-r_{1}\gamma_{y}-r_{2}\gamma_{y}}$.
The OBC eigenstate and the SGBZ mode are shown in panel i and panel
ii, respectively. (b, c) Zoomed-in view of the (b) upper edge and
(c) lower edge of the eigenstate, corresponding to the green and brown
boxes marked by 'b' and 'c' in (a), respectively. The coupling terms
are the same as in Fig. \ref{fig:quasi-1D}, and the eigenenergy of
the eigenstate is $1.19+1.23\protect\rmi$. (d, e) Numerical results
of the residuals (d) $\mathcal{R}_{u}$ and (e) $\mathcal{R}_{l}$
for different $L_{y}$. (f) Quasi-1D picture of the deviation of the
QMGBZ mode from the OBC eigenstate. (g) Illustration of the effect
of the strip width. The blue dots denote the $\beta_{1}$-solutions
corresponding to the auxiliary non-Bloch waves, and the green dot
represents the QMGBZ mode.}
\label{fig:competition}
\end{figure}

In the main text, we have shown that incompatible SGBZs compete with
each other, leading to geometry-dependent bands. To understand this
competition effect, we numerically calculate the OBC eigenstate and
the corresponding SGBZ mode for the incompatible case shown in Fig.
3 of the main text. For numerical calculations, we first select an
OBC eigenstate $\psi_{\text{OBC}}$, and then calculate the SGBZ mode
in the $[11]$-SGBZ with the same eigenenergy as $\psi_{\text{OBC}}$
according to Eqs. (\ref{eq:11-SGBZ-HN}) and (\ref{eq:SGBZ-E-11}).
Figure \ref{fig:competition}(a--c) illustrates the distribution
of the OBC eigenstate $\psi_{\text{OBC}}$ and the SGBZ mode $\psi_{\text{SGBZ}}$
with eigenenergy $1.19+1.23\rmi$. For better visualization, the real-space
coordinates are transformed into the basis $\{\mathbf{a}_{[11]},\mathbf{a}_{y}\}$,
that is, $\mathbf{r}=r_{1}\mathbf{a}_{[11]}+r_{2}\mathbf{a}_{y}$,
and the modes are scaled by $\psi\left(r_{1},r_{2}\right)\rightarrow\tilde{\psi}\left(r_{1},r_{2}\right)=\rme^{-r_{1}\gamma_{x}-r_{1}\gamma_{y}-r_{2}\gamma_{y}}\psi\left(r_{1},r_{2}\right)$
to remove the common exponential factors. The real parts of $\tilde{\psi}_{\text{OBC}}\left(r_{1},r_{2}\right)$
(panel i) and $\tilde{\psi}_{\text{SGBZ}}\left(r_{1},r_{2}\right)$
(panel ii) are shown in Fig. \ref{fig:competition}(a), and the zoomed-in
views of the upper edge (green box marked by `b') and the lower edge
(brown box marked by `c') are shown in Figs. \ref{fig:competition}(b)
and \ref{fig:competition}(c), respectively. According to the numerical
results, $\tilde{\psi}_{\text{SGBZ}}$ fits well with $\tilde{\psi}_{\text{OBC}}$
in the bulk but deviates from $\tilde{\psi}_{\text{OBC}}$ near the
left and right boundaries.

Next, we change the width $L_{y}$ and compare the deviations of the
SGBZ mode from the OBC eigenstate. To describe the deviations, we
define the residuals $\mathcal{R}_{u}$ and $\mathcal{R}_{l}$ as,
\begin{align}
\mathcal{R}_{u}\left(r_{1}\right) & =\sum_{r_{2}=\frac{L_{y}}{2}}^{L_{y}}\frac{\left|\tilde{\psi}_{\text{SGBZ}}\left(r_{1},r_{2}\right)-\tilde{\psi}_{\text{OBC}}\left(r_{1},r_{2}\right)\right|^{2}}{\mathcal{N}_{u}},\label{eq:residual-upper}\\
\mathcal{R}_{l}\left(r_{1}\right) & =\sum_{r_{2}=1}^{\frac{L_{y}}{2}}\frac{\left|\tilde{\psi}_{\text{SGBZ}}\left(r_{1},r_{2}\right)-\tilde{\psi}_{\text{OBC}}\left(r_{1},r_{2}\right)\right|^{2}}{\mathcal{N}_{l}},\label{eq:residual-lower}
\end{align}
where $\mathcal{R}_{u}$ and $\mathcal{R}_{l}$ denote the residuals
at the upper edge and the lower edge, respectively, and the normalization
factors are, 
\begin{align}
\mathcal{N}_{u} & =\sum_{r_{1}=1}^{L_{[11]}}\sum_{r_{2}=\frac{L_{y}}{2}}^{L_{y}}\left|\tilde{\psi}_{\text{SGBZ}}\right|^{2}/L_{[11]},\\
\mathcal{N}_{l} & =\sum_{r_{1}=1}^{L_{[11]}}\sum_{r_{2}=1}^{\frac{L_{y}}{2}}\left|\tilde{\psi}_{\text{SGBZ}}\right|^{2}/L_{[11]}.
\end{align}
In numerical calculations, the eigenstates with $L_{y}=30$ and $50$
are compared to the eigenstate with $L_{y}=40$ {[}i.e., the eigenstate
shown in Fig. \ref{fig:competition}(a-c){]}. For each parallelogram
region, the eigenstate with the closest eigenenergy to the eigenstate
shown in Fig. \ref{fig:competition}(a) is selected. Figure \ref{fig:competition}(d)
and \ref{fig:competition}(e) show the numerical results of $\mathcal{R}_{u}$
and $\mathcal{R}_{l}$, respectively. For $L_{y}=30$ and $L_{y}=40$,
the residuals decay from the boundaries to the bulk, which is consistent
with the numerical results shown in Fig. \ref{fig:competition}(a-c).
Compared with the case of $L_{y}=40$, the residual in the case of
$L_{y}=30$ decays more rapidly from the boundaries to the bulk. In
contrast, for $L_{y}=50$, the residual is distributed in the bulk,
indicating that the $[11]$-SGBZ mode cannot describe the OBC eigenstate
in this case.

Since the SGBZ is equivalent to the QMGBZ when the width is sufficiently
large, the influence of the width can be understood through the quasi-1D
model along the major axis. For a quasi-1D strip with finite width,
supposing that the quasi-1D characteristic polynomial is $F_{L_{2}}(E,\beta_{1})$,
the number of OBC equations at the left (right) boundary equals $M$
($N$), where $-M$ and $N$ are the lowest and highest degrees of
$\beta_{1}$ in $F_{L_{2}}(E,\beta_{1})$, respectively. For each
reference energy $E$, $F_{L_{2}}(E,\beta_{1})$ has $M+N$ solutions
$\beta_{1}^{(j)}(E),j=1,2,\dots,M+N$, so there exist $M+N$ non-Bloch
waves with eigenenergy $E$. However, the QMGBZ mode lies only in
the subspace spanned by the non-Bloch waves corresponding to $\beta_{1}^{(M)}(E)$
and $\beta_{1}^{(M+1)}(E)$. Therefore, as shown in Fig. \ref{fig:competition}(f),
the non-Bloch waves corresponding to $\beta_{1}^{(j)}(E),j\neq M,M+1$,
dubbed the ``auxiliary non-Bloch waves'', are required to superpose
with the QMGBZ mode to satisfy the OBC. We suppose that the OBC mode
is expanded as the superposition of the non-Bloch waves, 
\begin{align}
\psi_{\text{OBC}}\left(r_{1},r_{2}\right) & =\sum_{j}C_{j}\psi_{\beta_{1}^{(j)}}\left(r_{1},r_{2}\right),\nonumber \\
 & =\sum_{j}C_{j}\left(\beta_{1}^{(j)}\right)^{r_{1}}\phi_{\beta_{1}^{j}}\left(r_{2}\right),
\end{align}
where $\psi_{\beta_{1}^{(j)}}(r_{1},r_{2})\equiv\langle r_{1},r_{2}|\psi_{\beta_{1}^{(j)}}\rangle$
is the real-space wavefunction of the non-Bloch wave and $\phi_{\beta_{1}^{j}}(r_{2})\equiv(\beta_{1}^{(j)})^{-1}\psi_{\beta_{1}^{j}}(1,r_{2})$
is the $\beta_{1}$-independent part. Here, the eigenenergy $E$ is
omitted for brevity. To satisfy the $M$ equations at the left boundary,
the first $M-1$ auxiliary non-Bloch waves should be comparable to
the QMGBZ mode at the left boundary, while the last $N-1$ auxiliary
non-Bloch waves should be negligible at the left boundary, that is,
$|C_{j}/C_{M}|$ is comparable to $1$ for $j<M$, and tends to $0$
for $j>M+1$. Similarly, for the right $N$ boundary equations, $|C_{j}/C_{M}|\times|\beta_{1}^{(j)}/\beta_{1}^{(M)}|^{L_{1}}$
is comparable to $1$ for $j>M+1$ and tends to $0$ for $j<M$. Therefore,
for finite $L_{1}$, the crosstalk between the left and right boundaries
can be characterized by the ratios $|\beta_{1}^{(M-1)}/\beta_{1}^{(M)}|^{L_{1}}$
and $|\beta_{1}^{(M+1)}/\beta_{1}^{(M)}|^{-L_{1}}$. When the two
ratios tend to $0$, the left (right) auxiliary non-Bloch waves do
not influence the boundary equations at the right (left) boundary.

In 1D lattices, because the difference between $|\beta_{1}^{(M-1)}|$
($|\beta_{1}^{(M+2)}|$) and $|\beta_{1}^{(M)}|$ is generally a nonzero
finite value, both ratios tend to $0$ when $L_{1}$ is large enough.
However, for a quasi-1D strip of a 2D lattice, criticality arises
when the width of the strip tends to infinity. As illustrated in Fig.
\ref{fig:competition}(g), both $|\beta_{1}^{(M-1)}|-|\beta_{1}^{(M)}|$
and $|\beta_{1}^{(M+2)}|-|\beta_{1}^{(M)}|$ tend to $0$ with the
order of $1/L_{2}$ when the width $L_{2}\rightarrow\infty$. Supposing,
\begin{equation}
|\beta_{1}^{(M-1)}|=|\beta_{1}^{(M)}|-\frac{\alpha_{L}}{L_{2}}|\beta_{1}^{(M)}|+O\left(\frac{1}{L_{2}^{2}}\right),
\end{equation}
and, 
\begin{equation}
|\beta_{1}^{(M+2)}|=|\beta_{1}^{(M)}|+\frac{\alpha_{R}}{L_{2}}|\beta_{1}^{(M)}|+O\left(\frac{1}{L_{2}^{2}}\right),
\end{equation}
and assuming that $L_{1}$ tends to infinity with $L_{1}=KL_{2}$,
the crosstalk terms are given by, 
\begin{align}
\lim_{L_{2}\rightarrow\infty}\left|\frac{\beta_{1}^{(M-1)}}{\beta_{1}^{(M)}}\right|^{L_{1}} & =\exp\left(-\alpha_{L}K\right),\label{eq:crosstalk-left}\\
\lim_{L_{2}\rightarrow\infty}\left|\frac{\beta_{1}^{(M+2)}}{\beta_{1}^{(M)}}\right|^{-L_{1}} & =\exp\left(-\alpha_{R}K\right),\label{eq:crosstalk-right}
\end{align}
which are determined by the aspect ratio $K$ rather than the size
of the system.

For compatible SGBZs, the SGBZ mode automatically satisfies the OBC
equations at the boundaries parallel to the minor axis. However, for
incompatible SGBZs, the auxiliary non-Bloch waves become relevant.
As $L_{1}$ increases, the boundaries parallel to the minor axis move
apart, reducing the crosstalk. In contrast, when $L_{2}$ increases,
the difference between $|\beta_{1}^{(M-1)}|$ (or $|\beta_{1}^{(M+2)}|$)
and $|\beta_{1}^{(M)}|$ decreases, allowing the mismatch at the boundaries
to propagate further into the bulk and increasing the crosstalk. Due
to the competition between these two effects, the influences of the
boundary terms remain non-negligible in the bulk even in the thermodynamic
limit.

\section{Details about the criterion for geometry dependence}\label{sec:details-criterion}

In the main text, we briefly discussed the criterion for geometry-dependent
or uniform bands. In this section, we will derive it in detail.

In the previous sections, we have shown that geometry-dependent bands
originate from the competition between incompatible SGBZs. Therefore,
if all SGBZs of a non-Hermitian system are compatible, the system
exhibits uniform bands. The main idea to derive the general criterion
for uniform bands is as follows: Given a basis of lattice vectors
$(\mathbf{a}_{1},\mathbf{a}_{2})$, and considering the SGBZ points
of the strip with major axis $\mathbf{a}_{1}$, for arbitrary basis
transformations, 
\begin{equation}
\begin{pmatrix}\tilde{\mathbf{a}}_{1} & \tilde{\mathbf{a}}_{2}\end{pmatrix}=\begin{pmatrix}\mathbf{a}_{1} & \mathbf{a}_{2}\end{pmatrix}\mathbf{P},
\end{equation}
where $\mathbf{P}\in\mathbb{Z}^{2\times2}$ is the transformation
matrix, if the SGBZ of the $\mathbf{a}_{1}$-strip is transformed
to the SGBZ of the $\tilde{\mathbf{a}}_{1}$-strip, then all the SGBZs
are compatible, and consequently, the system has uniform bands. Otherwise,
if the transformed points do not belong to the SGBZ of the $\tilde{\mathbf{a}}_{1}$-strip
for some matrix $\mathbf{P}$, the system exhibits geometry-dependent
bands.

Under the coordinate transformation, the momenta are transformed by,
\begin{equation}
\begin{pmatrix}\tilde{k}_{1} & \tilde{k}_{2}\end{pmatrix}=\begin{pmatrix}k_{1} & k_{2}\end{pmatrix}\mathbf{P},\label{eq:transformation-k}
\end{equation}
or equivalently $\begin{pmatrix}\ln\tilde{\beta}_{1} & \ln\tilde{\beta}_{2}\end{pmatrix}=\begin{pmatrix}\ln\beta_{1} & \ln\beta_{2}\end{pmatrix}\mathbf{P}$.
Therefore, the characteristic polynomials in the original strip and
the transformed strip are related by, 
\begin{equation}
f\left(E,\beta_{1},\beta_{2}\right)=\tilde{f}\left(E,\tilde{\beta}_{1},\tilde{\beta}_{2}\right)=\tilde{f}\left(E,\beta_{1}^{P_{11}}\beta_{2}^{P_{21}},\beta_{1}^{P_{12}}\beta_{2}^{P_{22}}\right),\label{eq:ChP-relations}
\end{equation}
where $P_{ij}$ is the matrix element of $\mathbf{P}$.

\begin{figure}
\centering

\includegraphics{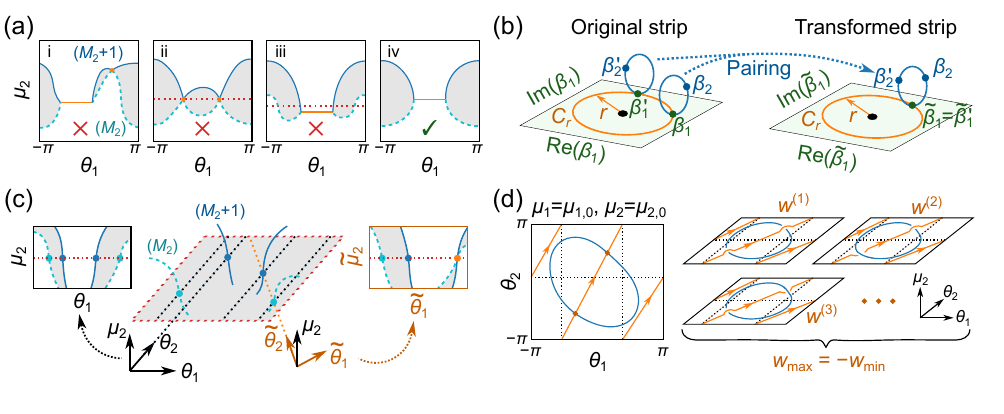}

\caption{Derivation of necessary conditions for uniform bands. (a) Cases with
and without uniform bands, where the blue and cyan curves represent
the $(M_{2}+1)$-th and $M_{2}$-th $\beta_{2}$-zeros of $f(E,\protect\rme^{\mu_{1}+\protect\rmi\theta_{1}},\beta_{2})$.
(b) Pairing of two PMGBZ points under transformations of the major
axes. (c) Illustration that the case in panel iii of (a) is equivalent
to the case in panel i through coordinate transformations. (d) Constraints
on the winding number in the $\boldsymbol{\mu}=(\mu_{1,0},\mu_{2,0})$
plane.}
\label{fig:uniform-band-cond}
\end{figure}

We first consider the necessary conditions for uniform bands. Figure
\ref{fig:uniform-band-cond}(a) shows four different cases of the
zeros of $f(E,\rme^{\mu_{1,0}+\rmi\theta_{1}},\beta_{2})$, where
$\mu_{1,0}$ is the value at which $W(E,\mu_{1})$ changes sign. The
function $\mu_{2}^{(j)}(\theta_{1};E,\mu_{1}),j=1,2,\dots,M_{2}+N_{2}$
is defined as $\mu_{2}^{(j)}=\ln|\beta_{2}^{(j)}|$, where $\beta_{2}^{(j)}$
is the $j$-th zero of $f(E,\rme^{\mu_{1,0}+\rmi\theta_{1}},\beta_{2})$,
ordered by $|\beta_{2}^{(j)}|\leq|\beta_{2}^{(k)}|,\forall j<k$.
According to the definition of the SGBZ, the intersections of $\mu_{2}^{(M_{2})}(\theta_{1};E,\mu_{1,0})$
and $\mu_{2}^{(M_{2}+1)}(\theta_{1};E,\mu_{1,0})$, marked by the
orange points or lines in Fig. \ref{fig:uniform-band-cond}(a), correspond
to the SGBZ points of the original strip. In the following part, we
will show that uniform bands are not allowed in the cases marked by
the red crosses (panels i--iii).

First, by definition, for some eigenenergy $E$, the SGBZ points are
defined as the zeros of $f(E,\cdot,\cdot)$ on the base manifold $X(E,\mu_{1,0})$,
where $\mu_{1,0}$ is the value at which $W(E,\mu_{1,0})$ changes
sign. According to the definition of $X(E,\mu_{1,0})$, the modulus
of $\beta_{1}$ must be constant ($|\beta_{1}|=\rme^{\mu_{1,0}}$).
For uniform bands, the SGBZ should be independent of the selection
of the major axis. Therefore, for a given eigenenergy $E$, both the
modulus of $\beta_{1}$ and the modulus of $\beta_{2}$ should be
constant in the SGBZ, which rules out the case of panel i in Fig.
\ref{fig:uniform-band-cond}(a). In the following text, we denote
$\mu_{2,0}$ as the constant value of $\mu_{2}\equiv\ln|\beta_{2}|$
when the system exhibits uniform bands.

Second, according to the definition of SGBZ, an SGBZ point is necessarily
a PMGBZ point. We consider the special transformations satisfying
$P_{12}=0$. For an SGBZ pair $\{(\beta_{1},\beta_{2}),(\beta_{1},\beta_{2}e^{\rmi\phi})\}$
in the original strip, the transformed points $(\beta_{1}^{P_{11}}\beta_{2}^{P_{21}},\beta_{2})$
and $(\beta_{1}^{P_{11}}\beta_{2}^{P_{21}}e^{\rmi P_{21}\phi},\beta_{2}e^{\rmi\phi})$
have different $\tilde{\beta}_{1}$ components. To ensure the transformed
points are PMGBZ points, for each point $(\beta_{1},\beta_{2})$ in
the original SGBZ, as illustrated in Fig. \ref{fig:uniform-band-cond}(b),
there must exist another SGBZ point $(\beta_{1}^{\prime},\beta_{2}^{\prime})$
paired with $(\beta_{1},\beta_{2})$, such that the transformed points
$\left(\tilde{\beta}_{1},\tilde{\beta}_{2}\right)=\left(\beta_{1}^{P_{11}}\beta_{2}^{P_{21}},\beta_{2}\right)$
and $\left(\tilde{\beta}_{1}^{\prime},\tilde{\beta}_{2}^{\prime}\right)=\left(\beta_{1}^{\prime P_{11}}\beta_{2}^{\prime P_{21}},\beta_{2}^{\prime}\right)$
form a pair of PMGBZ points with $\tilde{\beta}_{1}=\tilde{\beta}_{1}^{\prime}$,
i.e., 
\begin{align}
\beta_{1}^{P_{11}}\beta_{2}^{P_{21}} & =\beta_{1}^{\prime P_{11}}\beta_{2}^{\prime P_{21}},\label{eq:GISE-necessary-1}\\
\left|\beta_{2}\right| & =\left|\beta_{2}^{\prime}\right|,\label{eq:GISE-necessary-2}
\end{align}
Substituting Eq. (\ref{eq:GISE-necessary-2}) into (\ref{eq:GISE-necessary-1}),
the conditions above are equivalent to, 
\begin{align}
\left|\beta_{1}\right| & =\left|\beta_{1}^{\prime}\right|,\label{eq:GISE-nece-sim-1}\\
\left|\beta_{2}\right| & =\left|\beta_{2}^{\prime}\right|,\label{eq:GISE-nece-sim-2}\\
\text{Arg}\left[\left(\frac{\beta_{1}}{\beta_{1}^{\prime}}\right)^{P_{11}}\right] & =\text{Arg}\left[\left(\frac{\beta_{2}^{\prime}}{\beta_{2}}\right)^{P_{21}}\right].\label{eq:GISE-nece-sim-3}
\end{align}
It is noted that both $(\beta_{1},\beta_{2})$ and $(\beta_{1}^{\prime},\beta_{2}^{\prime})$
are located on the original SGBZ; that is, Eqs. (\ref{eq:GISE-nece-sim-1}-\ref{eq:GISE-nece-sim-3})
do not require information from the transformed SGBZ. Thus, we obtain
a necessary condition for uniform bands. For an arbitrary SGBZ point
$(\beta_{1},\beta_{2})$ and arbitrary integers $P_{11}$ and $P_{21}$,
there must exist another SGBZ point $(\beta_{1}^{\prime},\beta_{2}^{\prime})$
with the same eigenenergy $E$ that pairs with $(\beta_{1},\beta_{2})$,
satisfying Eqs. (\ref{eq:GISE-nece-sim-1}-\ref{eq:GISE-nece-sim-3}).
As $P_{21}$ runs over all integers, the point $(\beta_{1}^{\prime},\beta_{2}^{\prime})$
also changes with $P_{21}$. Consequently, for each eigenenergy $E$,
there must exist infinitely many SGBZ points, which rules out the
case of panel ii in Fig. \ref{fig:uniform-band-cond}(a).

Third, as illustrated in panel iii of Fig. \ref{fig:uniform-band-cond}(a),
if there exists a horizontal line other than $\mu_{2}=\mu_{2,0}$
that intersects both $\mu_{2}^{(M_{2})}$ and $\mu_{2}^{(M_{2}+1)}$,
the bands are also geometry-dependent. As shown in Fig. \ref{fig:uniform-band-cond}(c),
the case in panel iii can be transformed into the case in panel i
through coordinate transformations. We consider the plane with constant
$\mu_{1}$ and $\mu_{2}$, intersecting both the $M_{2}$-th and $(M_{2}+1)$-th
zeros, depicted as the gray plane in Fig. \ref{fig:uniform-band-cond}(c).
Under coordinate transformations, according to Eq. (\ref{eq:transformation-k}),
both $\boldsymbol{\mu}\equiv(\mu_{1},\mu_{2})$ and $\boldsymbol{\theta}\equiv(\theta_{1},\theta_{2})$
transform like $\mathbf{a}_{1},\mathbf{a}_{2}$, such that $\tilde{\boldsymbol{\mu}}=\boldsymbol{\mu}\mathbf{P}$
and $\tilde{\boldsymbol{\theta}}=\boldsymbol{\theta}\mathbf{P}$.
Therefore, points on the plane with constant $\boldsymbol{\mu}$ are
transformed to the plane with constant $\tilde{\boldsymbol{\mu}}$,
and the transformation of the $\boldsymbol{\theta}$ coordinates is
affine. By definition, the plots in Fig. \ref{fig:uniform-band-cond}(a)
are projections of the 3D plot in Fig. \ref{fig:uniform-band-cond}(c)
onto the $\theta_{1}$-$\mu_{2}$ plane. Thus, as illustrated by the
orange dotted line, we can adjust the direction of the major axis
via coordinate transformations so that the $M_{2}$-th and $(M_{2}+1)$-th
solutions are projected to the same point. In the transformed basis,
other than the continuum intersections at $\tilde{\mu}_{2,0}=P_{11}\mu_{1,0}+P_{21}\mu_{2,0}$,
the two curves of zeros also intersect at $\tilde{\mu}_{2}=P_{11}\mu_{1}+P_{21}\mu_{2}$,
corresponding to the case in panel i of Fig. \ref{fig:uniform-band-cond}(a).

Therefore, for systems with uniform bands, a necessary condition is
that the zeros of $f$ exhibit the pattern shown in panel iv of Fig.
\ref{fig:uniform-band-cond}(a) for each eigenenergy $E$ and in each
strip. That is, the SGBZ points with the same eigenenergy must have
a constant value of $\boldsymbol{\mu}=(\mu_{1,0},\mu_{2,0})$, and
the horizontal line in the $\mu_{2}$-$\theta_{1}$ plot must not
intersect both $\mu_{2}^{(M_{2})}$ and $\mu_{2}^{(M_{2}+1)}$ simultaneously,
except at $\mu_{2}=\mu_{2,0}$. For winding numbers, we define, 
\begin{align}
u_{1}\left(\theta_{2};E,\mu_{1},\mu_{2}\right) & \equiv\int_{-\pi}^{\pi}\frac{\rmd\theta_{1}}{2\pi\rmi}\frac{\partial\ln\left[f\left(E,\rme^{\mu_{1}+\rmi\theta_{1}},\rme^{\mu_{2}+\rmi\theta_{2}}\right)\right]}{\partial\theta_{1}},\\
u_{2}\left(\theta_{1};E,\mu_{1},\mu_{2}\right) & \equiv\int_{-\pi}^{\pi}\frac{\rmd\theta_{2}}{2\pi\rmi}\frac{\partial\ln\left[f\left(E,\rme^{\mu_{1}+\rmi\theta_{1}},\rme^{\mu_{2}+\rmi\theta_{2}}\right)\right]}{\partial\theta_{2}},
\end{align}
which are the winding numbers with constant $\boldsymbol{\mu}$ around
the $\theta_{1}$-axis and $\theta_{2}$-axis, respectively. According
to panel iv of Fig. \ref{fig:uniform-band-cond}(a), the winding number
must satisfy, 
\begin{equation}
u_{2}\left(\theta_{1};E,\mu_{1,0},\mu_{2,>}\right)\geq0,\quad u_{2}\left(\theta_{1};E,\mu_{1,0},\mu_{2,<}\right)\leq0,
\end{equation}
where $\mu_{2,>}$ and $\mu_{2,<}$ belong to $(\mu_{2,0},\mu_{2,0}+\epsilon)$
and $(\mu_{2,0}-\epsilon,\mu_{2,0})$, respectively, for an infinitesimal
positive number $\epsilon$. By the same reasoning, taking $\mathbf{a}_{2}$
as the major axis, a similar relation is also satisfied by $u_{1}$,
i.e., 
\begin{equation}
u_{1}\left(\theta_{2};E,\mu_{1,>},\mu_{2,0}\right)\geq0,\quad u_{1}\left(\theta_{2};E,\mu_{1,<},\mu_{2,0}\right)\leq0.
\end{equation}
However, to obtain the sufficient conditions for uniform bands, the
requirements on $u_{1}$ and $u_{2}$ must be generalized to arbitrary
winding loops. As illustrated in Fig. \ref{fig:uniform-band-cond}(d),
for an arbitrary winding loop (orange lines) in the plane $\boldsymbol{\mu}=(\mu_{1,0},\mu_{2,0})$,
if the winding loop intersects the zeros of $f$ (blue curve), the
winding number around the loop is ill-defined. Nevertheless, the winding
loops can be deformed along the $\mu_{2}$ direction (or equivalently
$\mu_{1}$) to avoid these intersections. Different deformations yield
different winding numbers $w^{(1)},w^{(2)},\dots$. Among all possible
deformations, we select the maximum and minimum winding numbers, denoted
as $w_{\text{max}}$ and $w_{\text{min}}$, respectively. We will
demonstrate that the uniformity of bands requires $w_{\text{max}}=-w_{\text{min}}$
for all possible winding loops, and this condition is also sufficient
for uniform bands.

\begin{figure}
\centering

\includegraphics{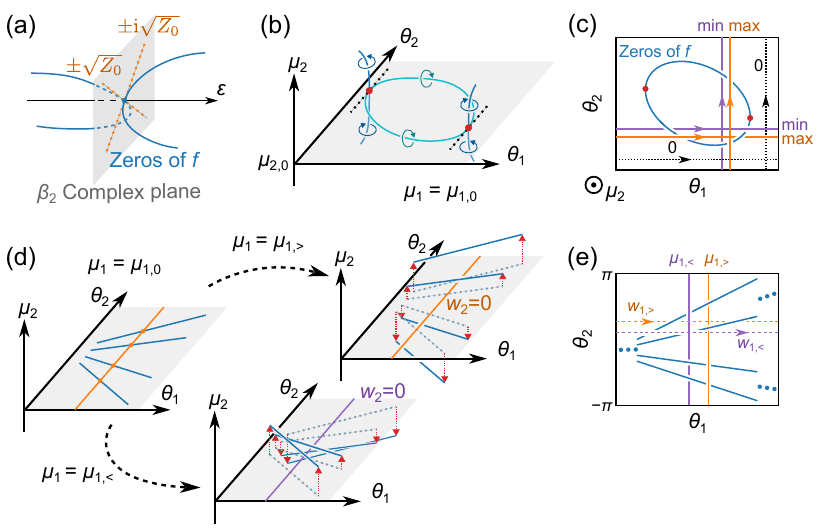}

\caption{Properties of the zeros of $f$ and sufficient conditions for uniform
bands. (a) Zeros of $f$ near the point where $\partial f/\partial\beta_{2}=0$.
(b) Zeros of $f$ near the plane $\boldsymbol{\mu}=(\mu_{1,0},\mu_{2,0})$.
(c) Maximum and minimum winding numbers near the winding loops in
the plane $\boldsymbol{\mu}=(\mu_{1,0},\mu_{2,0})$. (d) Movement
of the zeros as $\mu_{1}$ increases or decreases. (e) Limits of the
winding loops as $\mu_{1,>}\to\mu_{1,0}+0^{+}$ and $\mu_{1,<}\to\mu_{1,0}-0^{+}$.}
\label{fig:ChP-zeros}
\end{figure}

To understand this criterion, we first provide an overview of the
properties of the zeros of $f$. As previously discussed, $f(E,\beta_{1},\beta_{2})$
is a polynomial in $E$ and a Laurent polynomial in $\beta_{1}$ and
$\beta_{2}$. For generality, we only require $f$ to be holomorphic
on $\mathbb{C}\backslash\{0\}$, which includes both Laurent polynomials
and Laurent series.

Treating $E$ as a parameter, the solution set of $f(E,\beta_{1},\beta_{2})=0$
consists of algebraic curves. When $\partial f/\partial\beta_{2}\neq0$,
within the zero set, $\beta_{2}$ is locally a holomorphic function
of $\beta_{1}$, and its derivative is given by, 
\begin{equation}
\frac{\rmd\beta_{2}}{\rmd\beta_{1}}=-\frac{\partial f/\partial\beta_{1}}{\partial f/\partial\beta_{2}}.
\end{equation}
When $\partial f/\partial\beta_{2}=0$, $\beta_{2}$ cannot be considered
as a local function of $\beta_{1}$. Assuming $\partial^{j}f/\partial\beta_{2}^{j}=0$
for $j=1,2,\dots,m-1$, the increment of $\beta_{1}$ is related to
the increment of $\beta_{2}$ by, 
\begin{equation}
\frac{\Delta\beta_{2}^{m}}{m!}\frac{\partial^{m}f}{\partial\beta_{2}^{m}}+\Delta\beta_{1}\frac{\partial f}{\partial\beta_{1}}=o\left(\Delta\beta_{2}^{m}\right).
\end{equation}
Geometrically, the points satisfying $\partial f/\partial\beta_{2}=0$
form a nexus of $m$ complex curves. Taking $m=2$ as an example,
if $\Delta\beta_{1}=\varepsilon\hat{n}$, where $\hat{n}$ is a unit
complex number and $\varepsilon$ is real, the increment of $\beta_{2}$
is given by, 
\begin{align*}
\Delta\beta_{2} & =\pm i\sqrt{2\varepsilon\hat{n}\frac{\partial f/\partial\beta_{1}}{\partial^{2}f/\partial\beta_{2}^{2}}},\\
 & =\begin{cases}
\pm i\sqrt{\left|\varepsilon\right|}\sqrt{2\hat{n}\frac{\partial f/\partial\beta_{1}}{\partial^{2}f/\partial\beta_{2}^{2}}}, & \epsilon\geq0,\\
\pm\sqrt{\left|\varepsilon\right|}\sqrt{2\hat{n}\frac{\partial f/\partial\beta_{1}}{\partial^{2}f/\partial\beta_{2}^{2}}}, & \epsilon<0.
\end{cases}
\end{align*}
As illustrated in Fig. \ref{fig:ChP-zeros}(a), as $\varepsilon$
approaches $0$ from both positive and negative directions, the zero
curves converge toward two orthogonal lines in the complex $\beta_{2}$
plane, aligned with $\pm\rmi\sqrt{Z_{0}}$ and $\pm\sqrt{Z_{0}}$,
respectively, where $Z_{0}\equiv2\hat{n}\left(\partial f/\partial\beta_{1}\right)/\left(\partial^{2}f/\partial\beta_{2}^{2}\right)$.

For uniform bands, as discussed in the main text, the SGBZ points
corresponding to a given reference energy $E$ must maintain constant
values of $\mu_{1}\equiv\ln|\beta_{1}|$ and $\mu_{2}\equiv\ln|\beta_{2}|$.
Consequently, we examine the zeros of $f$ situated in the $(\mu_{1},\mu_{2})=(\mu_{1,0},\mu_{2,0})$
plane. When $\beta_{2}$ is a holomorphic function of $\beta_{1}$,
the mapping $(\mu_{1}+\rmi\theta_{1})\mapsto(\mu_{2}+\rmi\theta_{2})$
is also holomorphic, as it constitutes a composition of holomorphic
functions, i.e., 
\begin{equation}
\mu_{1}+\rmi\theta_{1}\overset{\exp}{\mapsto}\beta_{1}\mapsto\beta_{2}\overset{\ln}{\mapsto}\mu_{2}+\rmi\theta_{2}.
\end{equation}
According to the Riemann extension theorem, when a portion of the
set of zeros lies in the plane $(\mu_{1},\mu_{2})=(\mu_{1,0},\mu_{2,0})$,
the entire branch of the curve is also contained in that plane unless
$\partial f/\partial\beta_{2}=0$. Figure \ref{fig:ChP-zeros}(b)
shows the zeros of $f$ near the plane $(\mu_{1},\mu_{2})=(\mu_{1,0},\mu_{2,0})$
in the 3D hyperplane with $\mu_{1}=\mu_{1,0}$, where the cyan and
blue lines represent zeros within and outside the plane, respectively.
Based on the preceding discussion, the nexus points (red dots) occur
where $\partial\theta_{2}/\partial\theta_{1}$ diverges.

Next, we examine the winding numbers of $f$ in the 4D space $(\beta_{1},\beta_{2})\in\mathbb{C}^{2}$.
Similar to the winding numbers on $X(E,\mu_{1})$, the winding numbers
in $\mathbb{C}^{2}$ are also topologically invariant. When the winding
loop passes through a zero of $f$, the winding number changes according
to the winding number of an infinitesimal loop around that zero. Through
direct calculation, we can determine the winding number for each infinitesimal
winding loop around the zero curve. Assuming $(\beta_{1,0},\beta_{2,0})$
is a zero of $f(E,\cdot,\cdot)$, in the neighborhood of $(\beta_{1,0},\beta_{2,0})$
we have, 
\begin{equation}
f\left(E,\beta_{1,0}+\Delta\beta_{1},\beta_{2,0}+\Delta\beta_{2}\right)=\frac{\partial f}{\partial\beta_{1}}\Delta\beta_{1}+\frac{\partial f}{\partial\beta_{2}}\Delta\beta_{2}+O\left(|\Delta\beta_{1}|^{2}+|\Delta\beta_{2}|^{2}\right).
\end{equation}
For infinitesimal loops oriented perpendicular to the $\beta_{1}$-plane,
i.e., 
\begin{equation}
\beta_{1}=\beta_{1,0},\quad\beta_{2}=\beta_{2,0}\rme^{\Delta\mu_{2}+\rmi\Delta\theta_{2}},
\end{equation}
the characteristic polynomial reads, 
\begin{equation}
f\left(E,\beta_{1,0},\beta_{2,0}\rme^{\Delta\mu_{2}+\rmi\Delta\theta_{2}}\right)=\beta_{2,0}\frac{\partial f}{\partial\beta_{2}}\left(\Delta\mu_{2}+\rmi\Delta\theta_{2}\right).
\end{equation}
By direct calculation, when the winding loop takes the form $\Delta\mu_{2}=\epsilon\cos t$,
$\Delta\theta_{2}=\epsilon\sin t$, where $\epsilon>0$ is an infinitesimal
radius and $t\in[0,2\pi]$ is the parameter, the winding number equals
$+1$. Otherwise, the winding number equals $-1$. In Fig. \ref{fig:ChP-zeros}(b),
the positive directions of the winding loops are marked around each
curve of zeros. Therefore, moving the winding loop across an intersection
point changes the winding number by $\pm1$, where the sign is determined
by the direction in which the winding loop passes the intersection
point.

Based on the above discussion, we first consider the necessity. That
is, the relation $w_{\text{max}}=-w_{\text{min}}$ holds for arbitrary
closed loops when a system exhibits uniform bands. According to the
positive directions marked in Fig. \ref{fig:ChP-zeros}(b), for winding
loops parallel to the $\theta_{2}$-axis, the winding loop with maximum
(minimum) winding number braids above (below) the intersection points.
For loops parallel to the $\theta_{1}$-axis, the loop with maximum
(minimum) winding number braids above (below) the intersection point
when $\rmd\theta_{2}/\rmd\theta_{1}<0$, and below (above) the intersection
point when $\rmd\theta_{2}/\rmd\theta_{1}>0$. For both the winding
loops in $\theta_{1}$ and $\theta_{2}$ directions, the braiding
directions of the winding loops with maximum and minimum winding numbers
are opposite to each other at every intersection point. Figure \ref{fig:ChP-zeros}(c)
illustrates the loops for maximum and minimum winding numbers in the
top view of the plane $\boldsymbol{\mu}=(\mu_{1,0},\mu_{2,0})$. As
illustrated by the black dotted lines, when the winding loop does
not intersect the zeros of $f$, the winding number equals $0$. When
intersection points increase, $w_{\text{max}}$ increases by $1$,
and $w_{\text{min}}$ decreases by $1$ simultaneously, so that the
sum of $w_{\text{max}}$ and $w_{\text{min}}$ remains $0$. Therefore,
for uniform bands, $w_{\text{max}}=-w_{\text{min}}$ must be satisfied
for winding loops in $\theta_{1}$ and $\theta_{2}$ directions. Furthermore,
to ensure the above discussion holds for arbitrary strips, $w_{\text{max}}=-w_{\text{min}}$
must hold for arbitrary winding loops.

Finally, we consider the sufficiency. If there exists a plane $\boldsymbol{\mu}=(\mu_{1,0},\mu_{2,0})$
that contains a continuum of zeros of $f$, and $w_{\text{max}}=-w_{\text{min}}$
holds for arbitrary winding loops, we will show that the SGBZ constraint
is satisfied, i.e., $W(E,\mu_{1,>})$ and $W(E,\mu_{1,<})$ have opposite
signs for $\mu_{1,<}<\mu_{1,0}<\mu_{1,>}$. To compute $W(E,\mu_{1,>})$
and $W(E,\mu_{1,<})$, the winding numbers on the base manifolds $X(E,\mu_{1,>})$
and $X(E,\mu_{1,<})$ are required. By definition, $X(E,\mu_{1})$
is defined by the radius function $\ln|\beta_{2}|=\mu_{2,0}(\theta_{1};E,\mu_{1})$
satisfying $\mu_{2}^{(M_{2})}(\theta_{1};E,\mu_{1})\leq\mu_{2,0}(\theta_{1};E,\mu_{1})\leq\mu_{2}^{(M_{2}+1)}(\theta_{1};E,\mu_{1})$.
Although the radius function $\mu_{2,0}(\theta_{1};E,\mu_{1,0})$
is constant, $\mu_{2,0}(\theta_{1};E,\mu_{1,>})$ and $\mu_{2,0}(\theta_{1};E,\mu_{1,<})$
are not necessarily constant. However, when $\mu_{1,>}$ and $\mu_{1,<}$
are close to $\mu_{1,0}$, the strip winding numbers can be calculated
using the winding loops in the neighborhood of the plane $\boldsymbol{\mu}=(\mu_{1,0},\mu_{2,0})$.

As illustrated in Fig. \ref{fig:ChP-zeros}(d), consider a winding
loop with constant $\theta_{1}$. Assuming that the maximum and minimum
winding numbers are $w_{\text{max}}=-w_{\text{min}}=w_{0}$, the winding
loop should have $2w_{0}$ intersections with the zeros of $f$. Because
the map $\mu_{1}+\rmi\theta_{1}\mapsto\mu_{2}+\rmi\theta_{2}$ is
a holomorphic function when $\partial f/\partial\beta_{2}\neq0$,
the Cauchy-Riemann equations hold, i.e., $\partial\mu_{2}/\partial\mu_{1}=\partial\theta_{2}/\partial\theta_{1}$,
$\partial\mu_{2}/\partial\theta_{1}=-\partial\theta_{2}/\partial\mu_{1}$.
For the zeros of $f$ on the plane $\boldsymbol{\mu}=(\mu_{1,0},\mu_{2,0})$,
$\partial\theta_{2}/\partial\mu_{1}=-\partial\mu_{2}/\partial\theta_{1}=0$.
Therefore, when $\mu_{1}$ increases or decreases, the zeros move
in the direction normal to the plane. For a small increment in $\mu_{1}$,
the increment in $\mu_{2}$ is determined by the derivative $\partial\theta_{2}/\partial\theta_{1}$,
i.e., 
\begin{equation}
\Delta\mu_{2}=\frac{\partial\theta_{2}}{\partial\theta_{1}}\Delta\mu_{1}+O\left(\Delta\mu_{1}^{2}\right).
\end{equation}
Therefore, for positive $\Delta\mu_{1}$, the zeros with larger $\partial\theta_{2}/\partial\theta_{1}$
have larger values of $\mu_{2}$, and vice versa. According to the
Cauchy argument principle, for a fixed value of $\theta_{1}$, the
winding number around the loop parallel to the $\theta_{2}$ axis
with $\mu_{2}=\mu_{2,0}(\theta_{1};E,\mu_{1})$ is equal to $0$.
As illustrated in Fig. \ref{fig:ChP-zeros}(d), for $\mu_{1}=\mu_{1,>}$,
$\mu_{2,0}(\theta_{1};E,\mu_{1,>})$ lies above the zeros with the
$w_{0}$ lowest $\partial\theta_{2}/\partial\theta_{1}$ and below
the zeros with the $w_{0}$ highest $\partial\theta_{2}/\partial\theta_{1}$.

Next, as illustrated by the orange and purple solid lines in Fig.
\ref{fig:ChP-zeros}(e), we topologically deform the winding loops
to the plane $\boldsymbol{\mu}=(\mu_{1,0},\mu_{2,0})$, except at
the intersection points. Then, by sweeping $\theta_{1}$ over $[-\pi,\pi]$,
the two families of winding loops form two closed surfaces, which
are the topological deformations of $X(E,\mu_{1,>})$ and $X(E,\mu_{1,<})$
when $\mu_{1,>}$ and $\mu_{1,<}$ are sufficiently close to $\mu_{1,0}$.
According to the discussions above, at each intersection point, the
relative positions between the winding loop and the zero of $f$ are
opposite for the cases of $\mu_{1,>}$ and $\mu_{1,<}$. We define
the winding loop to be on the ``positive side'' of an intersection
point when its direction matches the positive direction shown in Fig.
\ref{fig:ChP-zeros}(b), and on the ``negative side'' when its direction
is opposite to the positive direction. For a winding loop with $2w_{0}^{\prime}$
intersection points, where at $p$ points the loop braids on the positive
side, the winding number of the loop takes the form $w=p+n_{0}$,
where $n_{0}$ is a constant independent of $p$. When $p$ ranges
from $0$ to $2w_{0}^{\prime}$, the minimum and maximum winding numbers
are $w_{\text{min}}=n_{0}$ and $w_{\text{max}}=2w_{0}^{\prime}+n_{0}$,
respectively. From $w_{\text{max}}=-w_{\text{min}}$, it follows that
$n_{0}=-w_{0}^{\prime}$. Therefore, for the two winding loops with
$p$ positive sides and $2w_{0}^{\prime}-p$ positive sides, respectively,
the winding numbers are $p-w_{0}^{\prime}$ and $w_{0}^{\prime}-p$,
which are opposites. As illustrated by the orange and purple dashed
lines in Fig. \ref{fig:ChP-zeros}(e), the winding loops for $\mu_{1,>}$
and $\mu_{1,<}$ lie on opposite sides at every intersection point,
so the winding numbers $w_{1,>}$ and $w_{1,<}$ are opposite. That
is, the strip winding numbers $W(E,\mu_{1,>})$ and $W(E,\mu_{1,<})$
have opposite signs.

With the discussions above, we have shown that the SGBZ constraint
is satisfied at $\mu_{1}=\mu_{1,0}$ if there are infinitely many
zeros of $f$ located on the plane $\boldsymbol{\mu}=\boldsymbol{\mu}_{0}\equiv(\mu_{1,0},\mu_{2,0})$,
and the condition $w_{\text{max}}=-w_{\text{min}}$ is satisfied for
an arbitrary winding loop on the plane. Under coordinate transformations,
because the zeros are mapped to the plane $\tilde{\boldsymbol{\mu}}=\boldsymbol{\mu}_{0}P$,
which is constant in $\tilde{\boldsymbol{\mu}}$, and the condition
$w_{\text{max}}=-w_{\text{min}}$ still holds, the SGBZ constraint
is also satisfied at $\tilde{\mu}_{1}=P_{11}\mu_{1,0}+P_{21}\mu_{2,0}$.
Therefore, the condition in Fig. \ref{fig:uniform-band-cond}(d) is
also sufficient for uniform bands.

\begin{figure}
\centering

\includegraphics{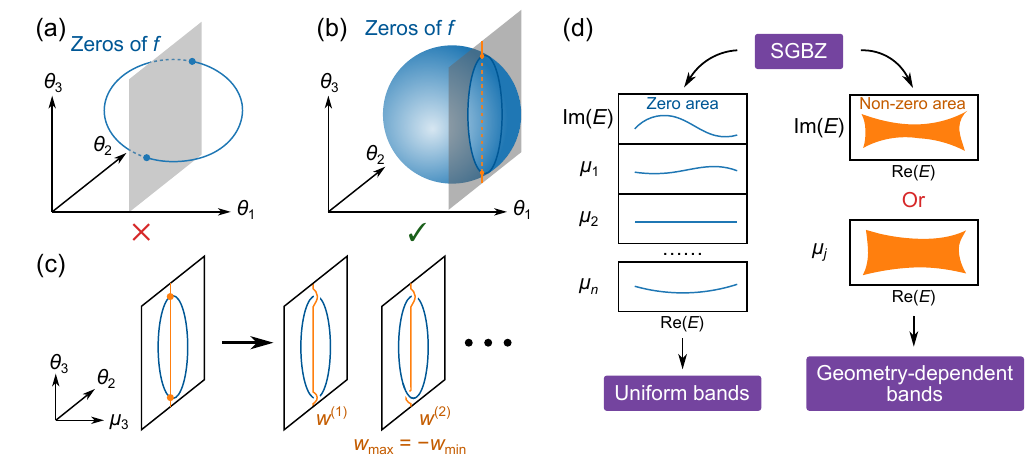}

\caption{General criterion in $n$-dimensional lattices. (a, b) Cases where
the codimension of the zeros of $f$ is (a) greater than $1$ and
(b) equal to $1$. Uniform bands are permitted only in case (b). (c)
Conditions on winding numbers. For uniform bands, $w_{\text{max}}=-w_{\text{min}}$
holds for all possible winding loops in the plane $\boldsymbol{\mu}=\boldsymbol{\mu}_{0}$.
(d) Illustration of the criterion for uniform or geometry-dependent
bands. For an arbitrary SGBZ, if the spectrum and all the $\text{Re}(E)$-$\mu_{j}$
plots have zero area, the bands are uniform. Otherwise, if the spectrum
or at least one of the $\text{Re}(E)$-$\mu_{j}$, $j=1,2,\dots,n$
plots has non-zero area, the bands are geometry-dependent.}
\label{fig:high-dim-criterion}
\end{figure}

The discussions above can be extended to general $n$-dimensional
lattices. First, by reordering the axes, all SGBZ points corresponding
to the same eigenenergy must lie in the plane $\boldsymbol{\mu}=\boldsymbol{\mu}_{0}\equiv(\mu_{1,0},\mu_{2,0},\dots,\mu_{n,0})$.
Next, generalizing the results for $n=2$, the zeros of $f(E,\boldsymbol{\beta})$
form a subspace of codimension $1$ for uniform bands. For instance,
in 2D uniform bands, the zeros of $f(E,\beta_{1},\beta_{2})$ at fixed
$E$ form 1D curves. As shown in Fig. \ref{fig:high-dim-criterion}(a),
if the codimension of the zero set exceeds $1$, there must exist
a subspace {[}gray surface in Fig. \ref{fig:high-dim-criterion}(a){]}
where the codimension of the zero set relative to that subspace is
also greater than $1$. Since each such subspace corresponds to an
$(n-1)$-dimensional subsystem in a specific coordinate system, the
dimensionality conditions for the zero set must also hold within the
subspace. Thus, by recursively applying the results for 2D uniform
bands, we conclude that the codimension of the zero set for a given
eigenenergy cannot exceed $1$. As depicted in Fig. \ref{fig:high-dim-criterion}(b),
for 3D lattices, the zero set forms a closed surface in the 3D subspace
defined by $\boldsymbol{\mu}=\boldsymbol{\mu}_{0}$. Moreover, as
illustrated in Fig. \ref{fig:high-dim-criterion}(c), the winding
numbers for any closed loop in the $\boldsymbol{\mu}=\boldsymbol{\mu}_{0}$
space must satisfy $w_{\text{max}}=-w_{\text{min}}$, where $w_{\text{max}}$
and $w_{\text{min}}$ are the maximum and minimum winding numbers
under all possible perturbations at the intersections.

The criterion for uniform or geometry-dependent bands also appears
in the spectrum and the imaginary momentum spectrum. Consider a non-Hermitian
system with uniform bands. For the spectrum, since the codimension
of the SGBZ points for a given eigenenergy is at most $1$, when $\boldsymbol{\beta}$
traverses the entire SGBZ, the spectrum has at most one dimension,
resulting in zero area. For the imaginary momentum spectrum {[}plots
of $\text{Re}(E)$-$\mu_{j}$, or equivalently $\text{Re}(E)$-$\text{Im}(-k_{j})${]},
because all SGBZ points for the same eigenenergy have constant $\boldsymbol{\mu}$,
the $\text{Re}(E)$-$\mu_{j}$ plots also exhibit zero area. In summary,
as illustrated in Fig. \ref{fig:high-dim-criterion}(d), the system
exhibits uniform bands if and only if both the spectrum and all imaginary
momentum spectra have zero area.

\section{Uniform and geometry-dependent bands for 2D HN model}

\begin{figure}
\centering

\includegraphics{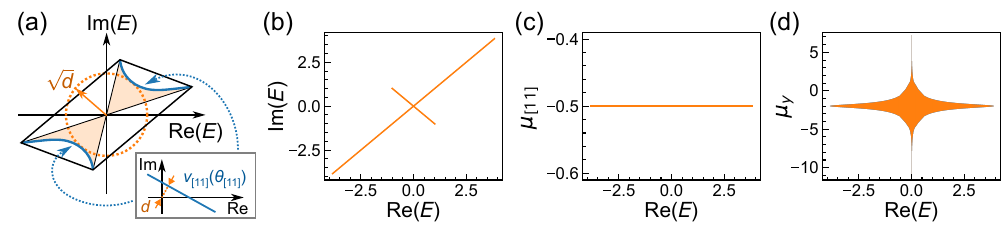}

\caption{Characterization of the geometric dependence of the 2D HN model via
the $[11]$-SGBZ. (a) Spectrum of the $[11]$-SGBZ. (b--d) The (b)
spectrum, (c) $\mu_{[11]}$, and (d) $\mu_{y}$ for the $[11]$-SGBZ
of the model satisfying $|J_{x}|=|J_{y}|$ and $\sin\Delta_{xy}\protect\neq0$.}
\label{fig:example-uniform-HN}
\end{figure}

In Sec. \ref{sec:SGBZs-HN}, we calculated three SGBZs for the 2D
HN model. As discussed in Sec. \ref{sec:details-criterion}, the geometry
dependence can be determined using any of the SGBZs. Here, we take
the $[11]$-SGBZ as an example. In Sec. \ref{sec:SGBZs-HN}, we derived
that the $[11]$-SGBZ is given by, 
\begin{equation}
\begin{cases}
\tilde{\beta}_{[11]}=\rme^{\gamma_{x}+\gamma_{y}+\rmi\theta_{[11]}},\\
\tilde{\beta}_{y}=\rme^{\gamma_{y}+\rmi\theta_{y}}\sqrt{\left|\frac{J_{x}^{*}\rme^{\rmi\Delta_{xy}+\rmi\theta_{[11]}}+J_{y}}{J_{x}\rme^{\rmi\Delta_{xy}-\rmi\theta_{[11]}}+J_{y}^{*}}\right|},
\end{cases}\label{eq:11-SGBZ-HN-1}
\end{equation}
and the corresponding spectrum reads, 
\begin{equation}
E\left(\theta_{[11]},\theta_{y}\right)=2\rme^{\rmi\bar{\varphi}(\theta_{[11]})}\sqrt{\left|v_{[11]}\left(\theta_{[11]};\gamma_{x}+\gamma_{y}\right)\right|}\cos\left(\theta_{y}-\frac{\Delta\varphi(\theta_{[11]};\gamma_{x}+\gamma_{y})}{2}\right),
\end{equation}
where $v_{[11]}$, $\bar{\varphi}$, and $\Delta\varphi$ are defined
in Eqs. (\ref{eq:v-11}--\ref{eq:Delta-varphi}). As illustrated
in Fig. \ref{fig:example-uniform-HN}(a), the spectrum of the $[11]$-SGBZ
is formed by the segments connecting two branches of square roots
of $v_{[11]}(\theta_{[11]};\gamma_{x}+\gamma_{y})$. By definition,
$v_{[11]}(\theta_{[11]};\gamma_{x}+\gamma_{y})$ reads, 
\begin{equation}
v_{[11]}\left(\theta_{[11]};\gamma_{x}+\gamma_{y}\right)=\rme^{\rmi\delta_{x}}\rme^{\rmi\delta_{y}}\left(2\text{Re}\left(J_{x}^{*}J_{y}^{*}\rme^{\rmi\theta_{[11]}}\right)+\rme^{\rmi\Delta_{xy}}\left|J_{x}\right|^{2}+\rme^{-\rmi\Delta_{xy}}\left|J_{y}\right|^{2}\right).
\end{equation}
When $\theta_{[11]}$ traverses $[-\pi,\pi]$, the minimum value of
$|v_{[11]}|$ is given by, 
\begin{equation}
d\equiv\min_{\theta_{[11]}}v_{[11]}\left(\theta_{[11]};\gamma_{x}+\gamma_{y}\right)=\left|\left(\left|J_{x}\right|^{2}-\left|J_{y}\right|^{2}\right)\sin\Delta_{xy}\right|.
\end{equation}

If the bands of the 2D HN model are uniform, the spectrum of an arbitrary
SGBZ should have zero area. As shown in Fig. \ref{fig:example-uniform-HN}(a),
the area is zero if and only if $\sqrt{d}=0$, which occurs when $|J_{x}|=|J_{y}|$
or $\sin\Delta_{xy}=0$. When $\sin\Delta_{xy}=0$, the $[11]$-SGBZ
becomes 
\begin{equation}
\begin{cases}
\tilde{\beta}_{[11]}=\rme^{\gamma_{x}+\gamma_{y}+\rmi\theta_{[11]}},\\
\tilde{\beta}_{y}=\rme^{\gamma_{y}+\rmi\theta_{y}}\sqrt{\left|\frac{J_{x}^{*}\rme^{\rmi\theta_{[11]}}\cos\Delta_{xy}+J_{y}}{J_{x}\rme^{-\rmi\theta_{[11]}}\cos\Delta_{xy}+J_{y}^{*}}\right|}=\rme^{\gamma_{y}+\rmi\theta_{y}}.
\end{cases}
\end{equation}
In this case, $\mu_{[11]}\equiv\ln|\tilde{\beta}_{[11]}|=\gamma_{x}+\gamma_{y}$
and $\mu_{y}\equiv\ln|\tilde{\beta}_{y}|=\gamma_{y}$ are both constant,
satisfying the criterion for uniform bands. In contrast, when $|J_{x}|=|J_{y}|$
but $\sin\Delta_{xy}\neq0$, $\mu_{y}$ is not constant for fixed
eigenvalue $E$, which violates the criterion for uniform bands. Figures
\ref{fig:example-uniform-HN}(b-d) illustrate the spectrum, $\mu_{[11]}$,
and $\mu_{y}$ of the $[11]$-SGBZ, where the parameters are $\gamma_{x}=1.5$,
$\gamma_{y}=-2$, $\delta_{x}=\pi/3$, $\delta_{y}=\pi/6$, $J_{x}=\sqrt{2}$,
and $J_{y}=1+\rmi$. As illustrated by the plots, while the area of
the spectrum is zero, the area of the $\mu_{y}-\text{Re}(E)$ plot
is non-zero. In fact, it is shown in Sec. \ref{sec:SGBZs-HN} that
the $[11]$-SGBZ is incompatible with the $[10]$-SGBZ or $[01]$-SGBZ
when $\sin\Delta_{xy}\neq0$. Therefore, $\sin\Delta_{xy}=0$ is necessary
for uniform bands in the 2D HN model.

Next, we will show that $\sin\Delta_{xy}=0$ is also sufficient for
uniform bands. When $\sin\Delta_{xy}=0$, we consider the real-space
Hamiltonian, 
\begin{align}
H_{\text{HN}} & =\sum_{r_{x},r_{y}}J_{x1}c_{r_{x}+1,r_{y}}^{\dagger}c_{r_{x},r_{y}}+J_{x2}c_{r_{x}-1,r_{y}}^{\dagger}c_{r_{x},r_{y}}+J_{y1}c_{r_{x},r_{y}+1}^{\dagger}c_{r_{x},r_{y}}+J_{y2}c_{r_{x},r_{y}-1}^{\dagger}c_{r_{x},r_{y}},\nonumber \\
 & =\sum_{r_{x},r_{y}}\rme^{\rmi\delta_{x}}\left[\left(\rme^{\gamma_{x}}J_{x}c_{r_{x}+1,r_{y}}^{\dagger}c_{r_{x},r_{y}}+\rme^{-\gamma_{x}}J_{x}^{*}c_{r_{x}-1,r_{y}}^{\dagger}c_{r_{x},r_{y}}\right)\pm\left(\rme^{\gamma_{y}}J_{y}c_{r_{x},r_{y}+1}^{\dagger}c_{r_{x},r_{y}}+\rme^{-\gamma_{y}}J_{y}^{*}c_{r_{x},r_{y}-1}^{\dagger}c_{r_{x},r_{y}}\right)\right],
\end{align}
where the sign of ``$\pm$'' depends on whether $\Delta_{xy}=0$
or $\Delta_{xy}=\pi$. In the non-unitary basis $|r_{x},r_{y}\rangle\equiv\rme^{\gamma_{x}r_{x}+\gamma_{y}r_{y}}c_{r_{x},r_{y}}^{\dagger}|0\rangle$,
$H_{\text{HN}}$ can be written in matrix form as, 
\begin{equation}
H_{r_{x},r_{y}}^{r_{x}^{\prime},r_{y}^{\prime}}=\rme^{\rmi\delta_{x}}\left(J_{x}\delta_{r_{x}+1,r_{y}}^{r_{x}^{\prime},r_{y}^{\prime}}+J_{x}^{*}\delta_{r_{x}-1,r_{y}}^{r_{x}^{\prime},r_{y}^{\prime}}\pm J_{y}\delta_{r_{x},r_{y}+1}^{r_{x}^{\prime},r_{y}^{\prime}}\pm J_{y}^{*}\delta_{r_{x},r_{y}-1}^{r_{x}^{\prime},r_{y}^{\prime}}\right),
\end{equation}
where $H_{r_{x},r_{y}}^{r_{x}^{\prime},r_{y}^{\prime}}$ is defined
as, 
\begin{equation}
H_{\text{HN}}\left|r_{x},r_{y}\right>=H_{r_{x},r_{y}}^{r_{x}^{\prime},r_{y}^{\prime}}\left|r_{x}^{\prime},r_{y}^{\prime}\right>,
\end{equation}
and the delta function is defined as, 
\begin{align}
\delta_{r_{x},r_{y}}^{r_{x}^{\prime},r_{y}^{\prime}}=\begin{cases}
1, & r_{x}^{\prime}=r_{x},r_{y}^{\prime}=r_{y},\\
0, & \text{Otherwise}.
\end{cases}
\end{align}
It is noted that the part, 
\begin{equation}
\tilde{H}_{r_{x},r_{y}}^{r_{x}^{\prime},r_{y}^{\prime}}\equiv J_{x}\delta_{r_{x}+1,r_{y}}^{r_{x}^{\prime},r_{y}^{\prime}}+J_{x}^{*}\delta_{r_{x}-1,r_{y}}^{r_{x}^{\prime},r_{y}^{\prime}}\pm J_{y}\delta_{r_{x},r_{y}+1}^{r_{x}^{\prime},r_{y}^{\prime}}\pm J_{y}^{*}\delta_{r_{x},r_{y}-1}^{r_{x}^{\prime},r_{y}^{\prime}},
\end{equation}
is a Hermitian matrix, so that the eigenvalues and eigenstates of
$H_{r_{x},r_{y}}^{r_{x}^{\prime},r_{y}^{\prime}}$ are equivalent
to those of the Hermitian matrix $\tilde{H}_{r_{x},r_{y}}^{r_{x}^{\prime},r_{y}^{\prime}}$
except for a common phase factor $\exp(\rmi\delta_{x})$ in the eigenvalues.
Because the bands of a Hermitian system are independent of geometries,
the bands of the 2D HN model with $\sin\Delta_{xy}=0$ are also independent
of geometries. Therefore, for the 2D HN model, the condition $\sin\Delta_{xy}=0$
is the sufficient and necessary condition for uniform bands.

We can also test our criterion in the $[10]$-SGBZ or $[01]$-SGBZ.
Because $\mu_{x}=\gamma_{x}$ and $\mu_{y}=\gamma_{y}$ are constant,
the only thing we need to check is the zero spectral area. According
to Eq. (\ref{eq:SGBZ-xy-E}), the sufficient and necessary condition
for the zero spectral area is $\sin\Delta_{xy}=0$, which is the same
as the result we obtained from the $[11]$-SGBZ.

\bibliography{2D-skin-effect}